\documentclass[11pt]{article}

\usepackage[margin=1in]{geometry}
\usepackage{amsmath,amssymb}
\usepackage{booktabs}
\usepackage{graphicx}
\usepackage{caption}
\usepackage{subcaption}
\usepackage{pgffor}       
\usepackage{setspace}
\usepackage{microtype}
\usepackage[authoryear,round]{natbib}
\usepackage[colorlinks=true,linkcolor=black,citecolor=black,urlcolor=blue]{hyperref}

\makeatletter
\newenvironment{tablenotes}{%
  \footnotesize
  \par\medskip
  \list{}{\labelwidth\z@ \itemindent\z@ \leftmargin\z@ \listparindent\z@
          \parsep\z@ \itemsep 2\p@ \topsep 2\p@ \partopsep\z@}%
  \item\relax
}{\endlist\par\medskip}
\makeatother

\graphicspath{{./}}

\title{\bf Administrative Boundary Reforms and Firm Pollution in China:\\[0.3em]
Institutional Path Divergence and Spatial Reallocation of Pollution}
\author{Yuting Chen\thanks{Yokohama National University, Yokohama, Japan. Email: chen-yuting-bh@ynu.ac.jp}\\
\and
Xuancheng Qin\thanks{Yokohama National University, Yokohama, Japan. Email: qin-xuancheng-gn@ynu.ac.jp}}
\date{\today}

\begin{document}
\maketitle

\begin{abstract}
\noindent
Administrative boundary reforms in China alter the assignment of local
authority and growth incentives without changing physical geography. We study
two reforms with opposite institutional content---\emph{county-to-district}
(centralization: a county absorbed into a municipality as an urban district)
and \emph{county-to-city} (empowerment: a county upgraded to a county-level
city)---using a firm-year pollution panel (1998--2013) and a border-pair
research design that compares firms on opposite sides of the same county
border. Centralizing reform lowers local NOx \emph{intensity} by
$14.8\%$ through cleaner production, without shrinking output or employment;
empowering reform raises COD, dust, and wastewater levels by $87$--$95\%$
through dirty growth. The local abatement is not net abatement: NOx rises in
same-prefecture neighbor counties ($+0.77$, $p<0.01$), reallocation follows
administrative boundaries rather than geographic distance, is confirmed in
satellite ambient data, and is not driven by wind or water transport. The
reallocation operates through incumbent neighbor firms becoming dirtier, not
through firm relocation or new entry. A back-of-the-envelope welfare
calculation yields a statistically insignificant welfare gain from the only
significant abatement (NOx, \textyen 47 million/yr, CI crossing zero) and a
damage of -\textyen 2.3 billion/yr from county-to-city reform. These
findings show that boundary reforms reshape the spatial configuration of
pollution across administrative boundaries rather than reducing it.
\medskip

\medskip

\end{abstract}

\setstretch{1.15}

\section{Introduction}\label{sec:introduction}

The administrative boundaries of local jurisdictions shape environmental
outcomes in two ways that point in opposite directions. Boundaries define the
scope of authority over regulation and the scope of responsibility for
pollution: a broader jurisdiction internalizes externalities, while a
fragmented one lets pollution spill across borders. But boundaries also define
the incentive to grow: jurisdictions that compete for mobile firms can race to
the bottom in environmental standards, and jurisdictions endowed with greater
development autonomy have stronger incentives to attract polluting industry.
Which force dominates---the internalization of externalities through
centralization, or the intensification of growth incentives through
empowerment---is a central, unresolved question in environmental federalism.

China's county-level administrative boundary reforms offer a clean setting in
which to separate these two forces, because two of the most important reforms
change the \emph{assignment of authority} between a county and its
prefecture-level city in opposite directions while leaving physical geography
untouched. In a \emph{county-to-district} reform (chedi she qu), a county is
absorbed into the prefecture-level city and converted into an urban district,
shifting planning, fiscal, and regulatory authority upward---an act of
\emph{administrative centralization}. In a \emph{county-to-city} reform (chedi
she shi), a county is upgraded to a county-level city and endowed with greater
autonomy over land, industrial development, and local public finance---an act
of \emph{local growth empowerment}. Because both reforms leave the territory,
population, and firms in place, they provide a rare experiment in which the
\emph{governance boundary} changes while the physical boundary does not.

We combine a firm-year pollution panel covering 1998--2013 with a
\emph{border-pair} research design that compares firms on opposite sides of the
same county border before and after reform. This design absorbs border-pair
$\times$ year shocks and isolates the local treatment effect from all shocks
common to the two sides of the border. We further assemble satellite ambient
pollution data (PM$_{2.5}$ and NO$_2$), wind fields, and river network
topology to trace whether local emission changes are offset by changes in
adjacent counties.

Three sets of findings emerge. First, the two reforms have \emph{opposite}
effects on firm pollution. Centralizing county-to-district reform lowers local
NOx \emph{intensity} by 14.8 log points ($p<0.01$) and SO$_2$ intensity by 14.9
log points, through cleaner production rather than scale contraction: output is
unchanged and employment rises. Empowering county-to-city reform raises COD,
dust, and wastewater \emph{levels} by 87--95 log points and COD and dust
\emph{intensity}, through dirty growth. The institutional path divergence
$\beta_{\text{district}}-\beta_{\text{countycity}}$ is therefore opposite-signed
across pollutants: the same umbrella of administrative boundary adjustment
delivers abatement or pollution depending on the direction of the authority
shift.

Second, the local abatement is not net abatement. NOx rises in
same-prefecture neighbor counties by 77 log points ($p<0.01$) following
county-to-district reform, and this reallocation follows \emph{administrative}
boundaries rather than geographic distance: it is concentrated in
same-prefecture counties 20--50 km away, while cross-prefecture counties
immediately across the border experience a decline. The pattern is confirmed
in satellite ambient data (local NO$_2$ falls; adjacent NO$_2$ and PM$_{2.5}$
rise) and is not driven by wind or water transport. The reallocation operates
through \emph{incumbent} neighbor firms becoming dirtier---same output, fewer
workers, more NOx---rather than through firm relocation or new entry.

Third, the welfare consequences are muted. A back-of-the-envelope
monetization using China-specific health-damage parameters yields a
statistically insignificant welfare gain from the only significant abatement
(NOx, +\textyen 47 million per year, with a confidence interval crossing
zero), while county-to-city reform generates roughly -\textyen 2.3 billion
per year in damages. Neither reform measurably reduces output. Taken together,
the evidence suggests that county-level boundary reforms reshape the
\emph{spatial configuration} of pollution across administrative boundaries
rather than reducing it.

We contribute to three literatures. Relative to work on Chinese
county-to-district reform (\citet{liu2023centralization},
\citet{liuwei2022urbanization}), we study the \emph{institutional contrast}
between centralizing and empowering reforms, so the estimand is the
\emph{path divergence} $\beta_{\text{district}}-\beta_{\text{countycity}}$ rather
than the average effect of a single reform. Relative to the literature on
administrative consolidation and environmental outcomes, we show that
within-jurisdiction abatement need not survive at the regional level, making
spatial reallocation a first-order identification and welfare concern rather
than an afterthought (cf.\ \citet{butts2021geocoded}). Relative to the environmental
federalism literature, the setting provides direct evidence on whether
centralization or growth empowerment dominates for pollution, and we decompose
the responses into scale, selection, and clean-production margins to identify
the channel.

The remainder of the paper proceeds as follows. Section~\ref{sec:background}
describes the institutional background. Section~\ref{sec:data} presents the
data and measurement. Section~\ref{sec:framework} lays out the empirical
framework. Section~\ref{sec:results} reports the main results,
Section~\ref{sec:spillover} the spatial reallocation analysis, and
Section~\ref{sec:mechanisms} the mechanisms. Section~\ref{sec:land} examines
land development and exposure, and Section~\ref{sec:welfare} the welfare
accounting. Section~\ref{sec:robustness} discusses robustness and
Section~\ref{sec:conclusion} concludes.

\section{Institutional Background}\label{sec:background}

\subsection{County-level administrative status in China}
China's territorial administration has five levels: the central government,
province, prefecture-level city, county, and township. The county (xian) is the
basic unit of territorial governance for land allocation, fiscal management,
and environmental regulation. A given county-level territory can hold one of
three administrative statuses. A \emph{county} ($\sim$1{,}600) is subordinate
to a prefecture-level city and enjoys limited urban-development authority. A
\emph{county-level city} (xianji shi) is a county that has been upgraded to
city status while remaining at the county administrative level; it retains
its own planning, land, and fiscal autonomy but gains an urban label and
associated growth mandates. An \emph{urban district} (shixia qu) is an
administrative division directly under the prefecture-level city government,
governed as part of the urban core, with substantially less independent
authority over land and fiscal matters.

Administrative status is not a purely bureaucratic label: it determines the
fiscal contract between the county and the prefecture, the allocation of land
quotas, the locus of approval authority for industrial projects, and the
hierarchical position of the local environmental protection bureau. Changes in
status therefore shift the \emph{governance boundary} and the \emph{growth
incentives} faced by local officials, even though the physical territory and
population are unchanged.

\subsection{Three boundary reforms}
We study three reforms that convert a county-level territory from one status to
another.

\paragraph{County-to-district (chedi she qu).} A county is absorbed into the
prefecture-level city and converted into an urban district. The reform shifts
authority \emph{upward}: planning, land, and fiscal decisions are centralized
in the prefecture government, and the district becomes part of the integrated
urban administration. We interpret this reform as \emph{administrative
centralization}. In our reform roster, 99 such reforms occur, concentrated in
the 2000s and 2010s.

\paragraph{County-to-city (chedi she shi).} A county is upgraded to a
county-level city. The territory retains county-level authority but acquires
urban status, typically accompanied by greater autonomy over land conversion,
industrial development, and local public finance, together with explicit
urbanization and growth mandates. We interpret this reform as \emph{local
growth empowerment}. In our reform roster, 168 such reforms occur, but they are
concentrated in the early 1990s (71 of the 77 reforms in the analytic sample
occur in 1993--1997, before the firm panel begins); this timing constrains
identification and is addressed in Section~\ref{sec:framework}.

\paragraph{City-to-district (chedi she qu, from county-level city).} A
county-level city is converted into an urban district. Institutionally this
resembles county-to-district---an upward shift of authority---and we treat it
as a robustness arm rather than a third primary treatment. In our reform
roster, 135 such reforms occur.

\subsection{Why boundary reforms affect firm pollution}
The three reforms change the \emph{assignment of authority and incentives}
between the prefecture and the county without changing physical geography.
This generates two opposing mechanisms with opposite predictions for firm
pollution.

First, \emph{centralization} (county-to-district, city-to-district) can
internalize environmental externalities and strengthen enforcement. When a
county becomes a district, its environmental bureau is folded into the
prefecture-level regulatory hierarchy, weakening the county's incentive to
compete for polluting investment through lax enforcement and shifting authority
toward a government with a broader, more urbanized constituency. If regulation
binds, incumbent firms should reduce emission \emph{intensity} through cleaner
production, with no necessary change in scale. This is the
\emph{boundary-internalization} hypothesis (H1).

Second, \emph{empowerment} (county-to-city) strengthens local growth
incentives. Upgraded cities receive greater autonomy over land and industrial
development and face urbanization mandates; local officials have stronger
incentives to attract polluting manufacturing and expand industrial output.
This should raise pollution through a combination of dirty entry and
intensity-increasing expansion---a \emph{dirty growth} channel (H2).

Third, because both reforms operate on the \emph{governance boundary} rather
than on the physical geography, the resulting changes in local pollution need
not correspond to changes in \emph{regional} pollution. If abatement in the
reformed territory induces polluting activity to relocate to adjacent
counties---particularly those sharing the same prefecture, which face the same
upper-level government---then local abatement is offset spatially. This is the
\emph{spatial reallocation} hypothesis (H3), and it motivates the spillover
design of Section~\ref{sec:spillover}.

\subsection{Related literature}
Our institutional contrast is closest to two studies of county-to-district
reform in China. \citet{liu2023centralization} find that county-to-district reform reduces
firm SO$_2$ emissions and raises COD removal, attributing the effect to
total-factor-productivity and output channels rather than to enforcement.
\citet{liuwei2022urbanization} link
county-to-district reform to reduced pollution through a fiscal channel---a
decline in county fiscal revenue and infrastructure spending and an increase in
livelihood spending. Our contribution relative to this work is twofold. First,
we study the \emph{institutional contrast} between centralizing and empowering
reforms (county-to-district versus county-to-city), so the estimand is the
\emph{path divergence} $\beta_{\text{district}}-\beta_{\text{countycity}}$,
not the average effect of a single reform. Second, we test whether local
abatement reflects net abatement or \emph{spatial reallocation} across
administrative boundaries, using border-pair and neighbor-spillover designs
complemented by satellite ambient data and wind/water falsifications.

The identification concerns are shared with a broader literature on
administrative consolidation and environmental outcomes. Recent work on
environmental vertical management reform and on city--county mergers
documented that centralizing environmental authority reduces firm pollution
within the jurisdiction; our border-pair design isolates whether such within-
jurisdiction abatement survives at the regional level. Methodologically, we
follow \citet{butts2021geocoded} in treating spatial spillovers as a first-order
identification threat in difference-in-differences designs with administrative
treatment boundaries.

\section{Data and Measurement}\label{sec:data}

\subsection{Data sources}
The analysis combines three administrative data assets. First, a \emph{firm
pollution panel} (1998--2013; 694{,}835 firm-year observations) constructed
from the Chinese Annual Survey of Industrial Firms matched to the
Environmental Statistics Database, recording six pollutants---SO$_2$, COD,
NOx, industrial dust, wastewater, and waste gas---together with firm location
(county), industry, output, sales, assets, and employment. Emissions of SO$_2$,
COD, NOx, and dust are in kilograms; wastewater is in tons; waste gas is a
volume measure (10{,}000 standard m$^3$) and is excluded from monetary damage
calculations. Second, an \emph{administrative reform roster} covering three
types of boundary adjustment: county-to-district (99 reforms, centralization),
county-to-city (168 reforms, empowerment), and city-to-district (135 reforms).
Because 71 of 77 county-to-city reforms in the analytic sample occurred in
1993--1997, before the panel begins, county-to-city effects are identified off
a small set of late reforms and are reported with the appropriate caveat.
Third, \emph{border-pair and spatial data}: a border-pair file fixed at the
year-2000 boundary, distance bands, same- vs.\ cross-prefecture adjacency,
wind direction (MERRA2), and river network topology (MERIT-Basins). Ambient
validation uses satellite PM$_{2.5}$ (1998--2024), satellite tropospheric NO$_2$
(1996--2017), and satellite-derived river water quality (2000--2023).

\subsection{Outcomes}
We study three outcome families. \emph{Emission levels} are $\ln(1+E)$.
\emph{Emission intensity} scales emissions by employment or sales,
$\ln(1+E/\text{employment})$ and $\ln(1+E/\text{sales})$, separating the
\emph{scale} margin from the \emph{clean-production} margin. \emph{Damage
proxies} weight emissions by population exposure and, for the welfare
exercise, by monetized marginal damage coefficients.

\subsection{Estimation samples}
Three complementary samples are used. The \emph{firm panel} (firm $\times$
year) supports the baseline and border-pair designs. The \emph{county panel}
(county $\times$ year, 2{,}687 counties) supports neighbor spillover and
land-supply regressions. The \emph{border-pair panel} (firm $\times$ border
pair $\times$ year) identifies local treatment effects by comparing firms on
opposite sides of the same border.

\section{Empirical Framework}\label{sec:framework}

\subsection{Baseline difference-in-differences}
The baseline specification is
\begin{equation}
Y_{ict} = \beta_1\,\text{District}_{ct} + \beta_2\,\text{CountyCity}_{ct}
+ \alpha_i + \lambda_t + \eta_{p(i)\times t} + \varepsilon_{ict},
\label{eq:baseline}
\end{equation}
where $\text{District}_{ct}$ ($\text{CountyCity}_{ct}$) equals one in the
years after county $c$ undergoes a county-to-district (county-to-city) reform.
Firm fixed effects $\alpha_i$ absorb time-invariant firm heterogeneity;
prefecture-by-year effects $\eta_{p(i)\times t}$ absorb prefecture-level
shocks. The parameter of primary interest is the \emph{institutional path
divergence} $\beta_1-\beta_2$. Standard errors are clustered at the original
county. Because treatment timing is staggered, the two-way fixed-effects
specification serves as a benchmark only.

\subsection{Border-pair design}
The border-pair design is the identification workhorse. It compares firms on
the treated side of a county border with firms on the untreated side of the
\emph{same} border:
\begin{equation}
Y_{ibt} = \tau\,(\text{TreatedSide}_i \times \text{Post}_t)
+ \alpha_i + \lambda_{b(i)\times t} + \varepsilon_{ibt},
\label{eq:border}
\end{equation}
absorbing border-pair-by-year shocks and clustering at the border-pair level.

\subsection{Event studies and modern estimators}
For each reform type we estimate a dynamic specification with a binned window
$[-5,+5]$ (endpoints pooled) and $t=-1$ omitted:
\begin{equation}
Y_{ict} = \sum_{k\neq -1}\theta_k\,\mathbb{1}[t-T_c=k]\cdot\text{Treated}_{c}
+ \alpha_i + \lambda_{b(i)\times t} + \varepsilon_{ict},
\label{eq:event}
\end{equation}
reporting the joint pre-trend test $\theta_{-5}=\dots=\theta_{-2}=0$. For
staggered designs we complement the event study with the Sun--Abraham
interaction-weighted estimator, Borusyak et al.\ imputation, the
de Chaisemartin--D'Haultf\oe uille dynamic estimator, and stacked DID by
treatment cohort.

\subsection{Spatial spillover design}
To test whether local abatement is offset by spatial reallocation, we regress
outcomes of \emph{never-treated} counties on exposure to reformed neighbors:
\begin{equation}
Y_{jct} = \sum_{d}\gamma_d\,\text{Ring}^d_{jct}
+ \delta_1\,\text{SamePref}_{jct} + \delta_2\,\text{CrossPref}_{jct}
+ \alpha_j + \eta_{p(j)\times t} + \varepsilon_{jct},
\label{eq:spillover}
\end{equation}
where $\text{Ring}^d$ indicates distance band $d$ (0--20, 20--50, 50--100\,km)
and $\text{SamePref}$ ($\text{CrossPref}$) indicates adjacency within (across)
a prefecture. The distance-band $\times$ prefecture interaction isolates
whether reallocation follows \emph{administrative} boundaries or
\emph{geographic} distance. We also aggregate each treated county with its
fixed set of pre-reform neighbors to test whether \emph{total} regional
pollution changes.

\subsection{Mechanisms}
Mechanisms are organized around the identity
\[
\Delta\ln(\text{Total emissions}) =
\Delta\ln(\text{firm count/employment})
+ \Delta\ln(\text{emission intensity}),
\]
decomposed into three margins: (i) \emph{scale and selection} (entry, exit,
employment, sales); (ii) \emph{clean production} (emission intensity, industry
heterogeneity); and (iii) \emph{regulatory/enforcement} channels. As
Section~\ref{sec:mechanisms} documents, the enforcement channel is not
testable in our data window, so the mechanism analysis is identified off the
first two margins.

\section{Main Results}\label{sec:results}

\subsection{Emission levels}
Table~\ref{tab:levels} reports border-pair DID estimates for emission levels.
County-to-district reform reduces NOx by 29.4 log points ($-0.294$, $p<0.01$)
with no significant effect on SO$_2$, COD, dust, or wastewater. County-to-city
reform raises COD, dust, and wastewater by 86.9, 94.5, and 92.6 log points,
respectively. The mirror-image pattern across the two reform types is the
paper's central fact: the same umbrella of administrative boundary adjustment
produces \emph{opposite-signed} effects depending on whether authority shifts
upward or local growth autonomy is enhanced. The institutional path divergence
$\beta_1-\beta_2$ is negative for NOx and positive for COD, dust, and
wastewater.

\subsection{Emission intensity}
Table~\ref{tab:intensity} scales emissions by employment and sales. The
county-to-district effect is concentrated in the \emph{clean-production}
margin: NOx per worker falls by 14.8 log points ($-0.148$, $p<0.01$), SO$_2$ per
worker by 14.9 log points ($-0.149$, $p<0.01$), and wastewater per worker by
12.9 log points ($-0.129$, $p<0.10$); sales-denominated intensities point in
the same direction. County-to-city reform instead \emph{raises} intensity: COD
per worker ($+0.275$, $p<0.05$), dust per worker ($+0.644$, $p<0.10$), COD per
sales ($+0.045$, $p<0.05$), and dust per sales ($+0.151$, $p<0.05$). The
intensity results discipline interpretation: county-to-district reform does
\emph{not} shrink scale (employment \emph{rises}; Section~\ref{sec:mechanisms});
it makes incumbent production cleaner. County-to-city reform makes production
dirtier at the margin.

\subsection{Pre-trends and modern estimators}
Table~\ref{tab:borderpretrend} reports border-pair joint pre-trend tests: the
clean-production margin (NOx per worker) passes ($F{=}1.25$, $p{=}0.289$), while
NOx levels are marginal ($p{=}0.073$); we anchor the local-abatement claim on
the intensity margin. Table~\ref{tab:estimators} compares estimators for
levels: the county-to-district NOx effect is largest and most precise in the
border-pair design. The stacked county-to-city level specification fails
parallel trends for SO$_2$, COD, dust, and wastewater (Table~\ref{tab:pretrend},
joint $F{=}30$--$307$), so county-to-city level effects are reported only from
the border-pair design. For the border-pair design, the correct pre-trend test
is the joint test on pre-period coefficients \emph{within} the pair-by-year
structure, not the standard staggered-DID estimator, which discards the
within-pair comparison that identifies the design.

\begin{figure}[p]
\caption{County-to-district reform: border-pair event studies, emission levels}
\label{fig:district_border}
\centering
  \begin{subfigure}[b]{0.32\textwidth}
    \includegraphics[width=\textwidth]{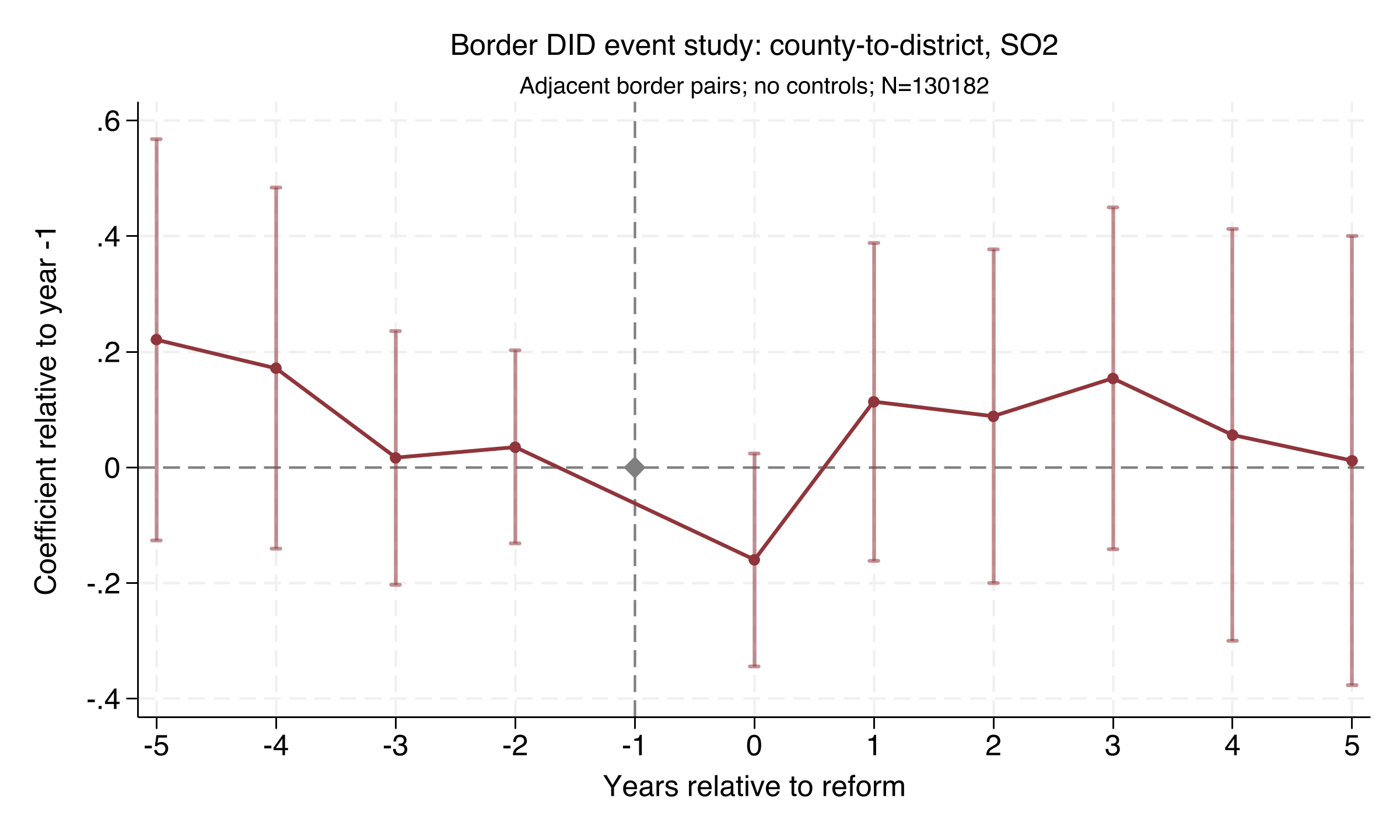}
    \caption{SO$_2$}
  \end{subfigure}\hfill
  \begin{subfigure}[b]{0.32\textwidth}
    \includegraphics[width=\textwidth]{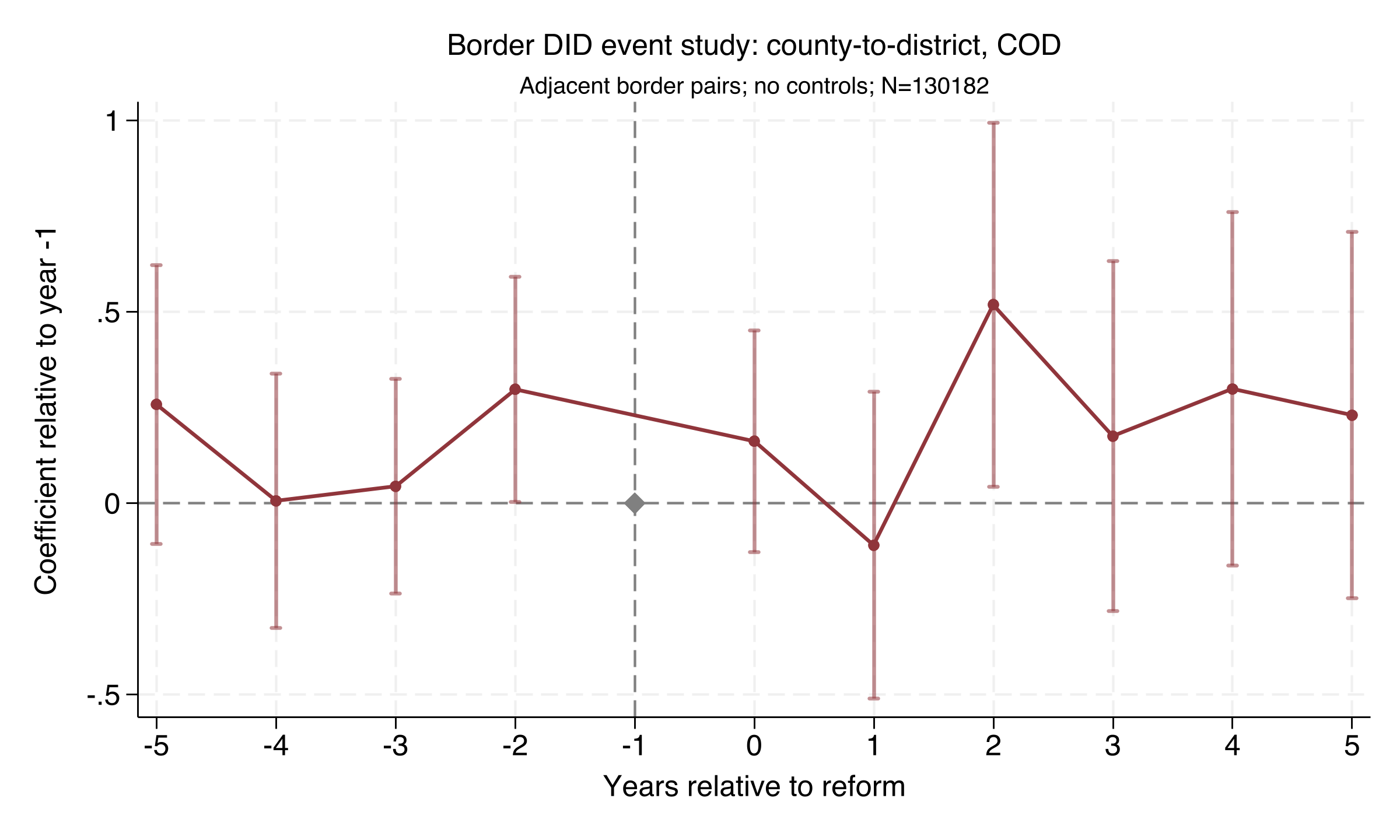}
    \caption{COD}
  \end{subfigure}\hfill
  \begin{subfigure}[b]{0.32\textwidth}
    \includegraphics[width=\textwidth]{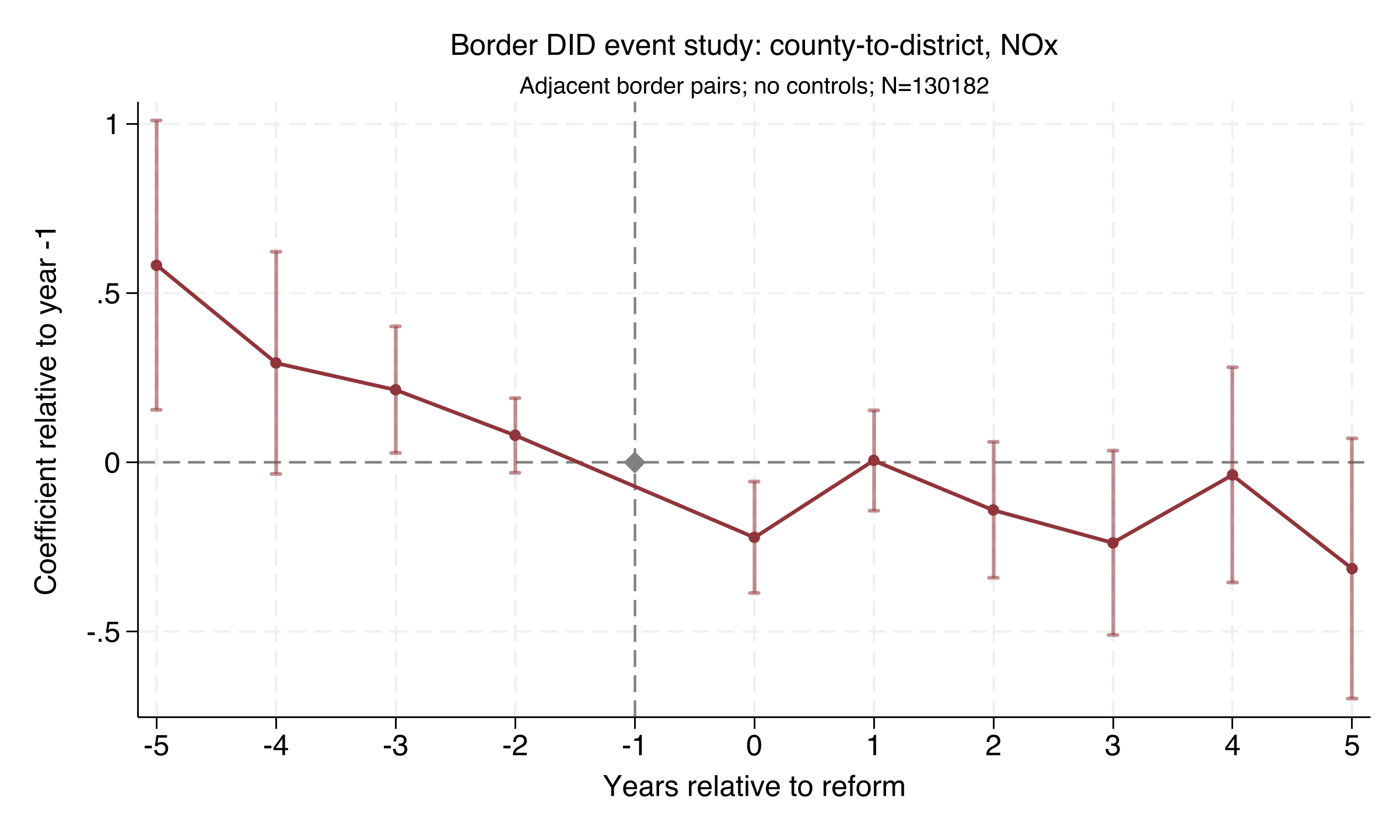}
    \caption{NOx}
  \end{subfigure}\hfill
  \begin{subfigure}[b]{0.32\textwidth}
    \includegraphics[width=\textwidth]{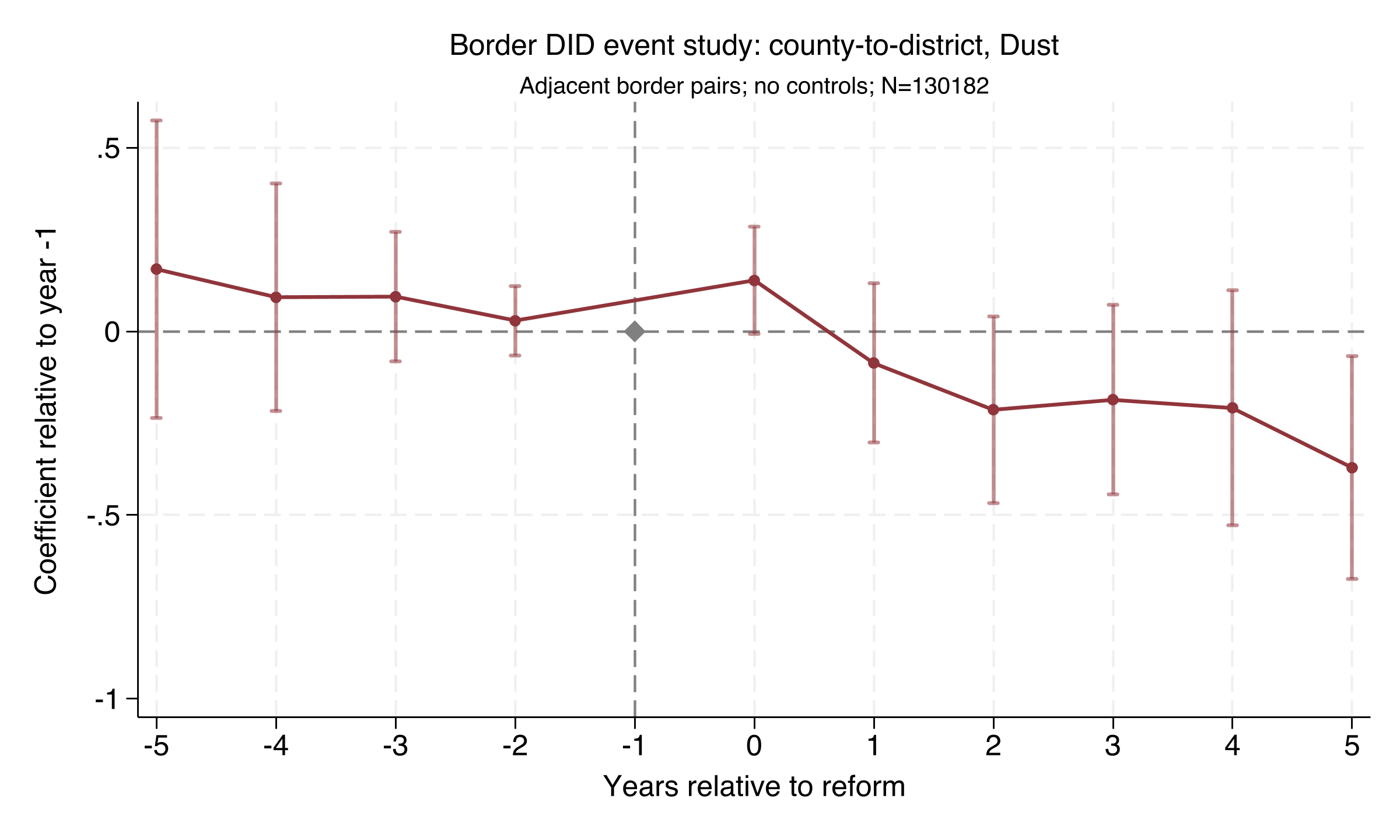}
    \caption{Dust}
  \end{subfigure}\hfill
  \begin{subfigure}[b]{0.32\textwidth}
    \includegraphics[width=\textwidth]{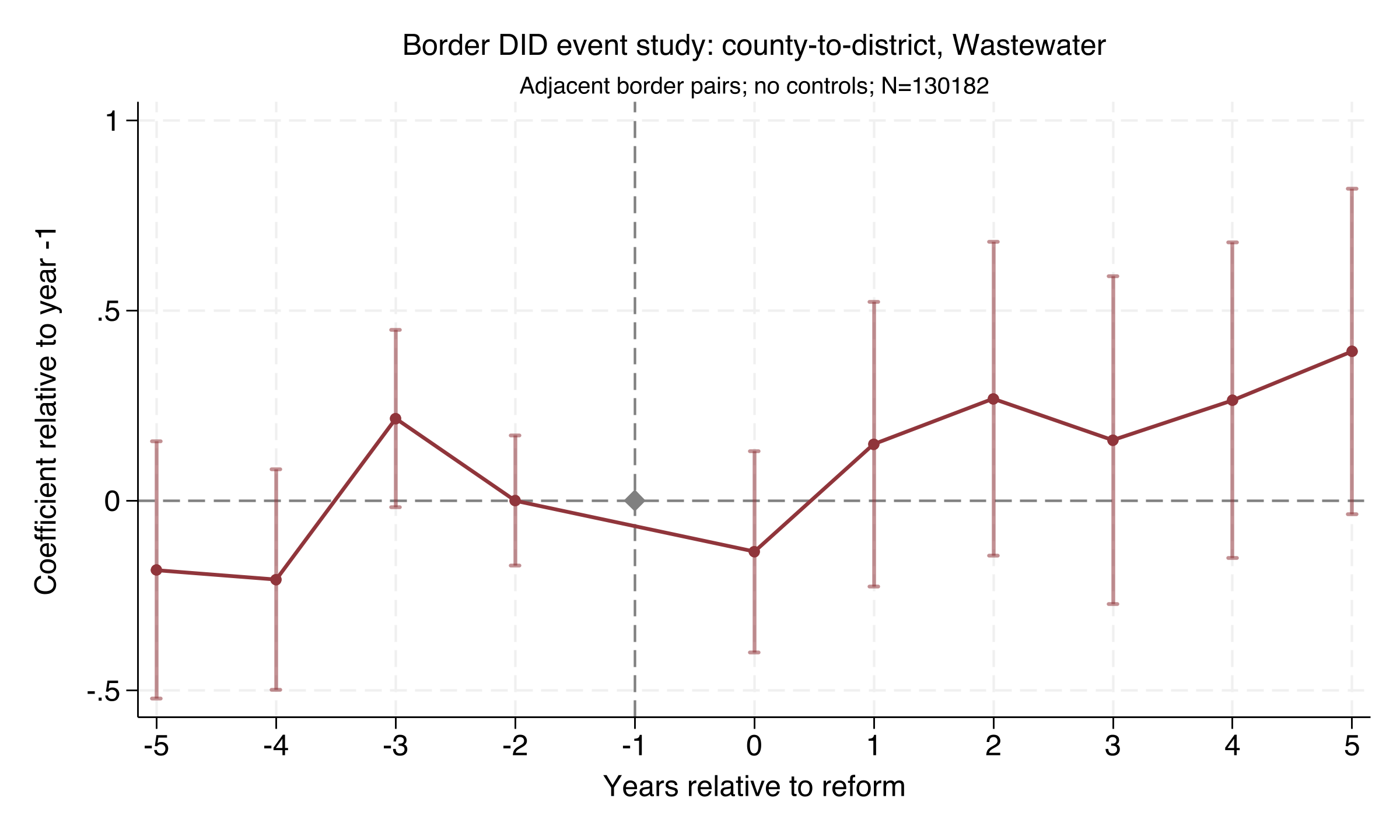}
    \caption{Wastewater}
  \end{subfigure}\hfill
  \begin{subfigure}[b]{0.32\textwidth}
    \includegraphics[width=\textwidth]{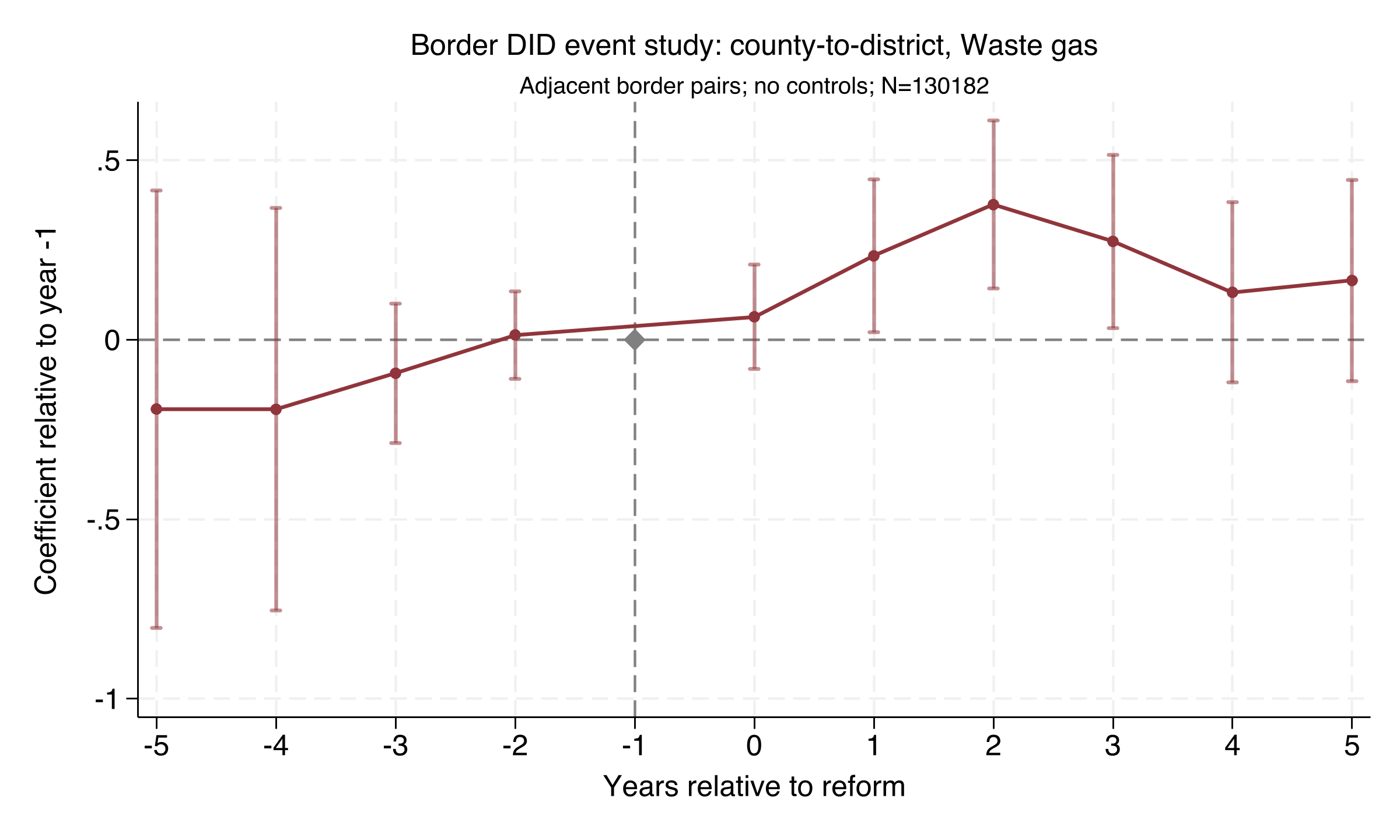}
    \caption{Waste~gas}
  \end{subfigure}\hfill
\end{figure}

\begin{figure}[p]
\caption{County-to-city reform: border-pair event studies, emission levels}
\label{fig:countycity_border}
\centering
  \begin{subfigure}[b]{0.32\textwidth}
    \includegraphics[width=\textwidth]{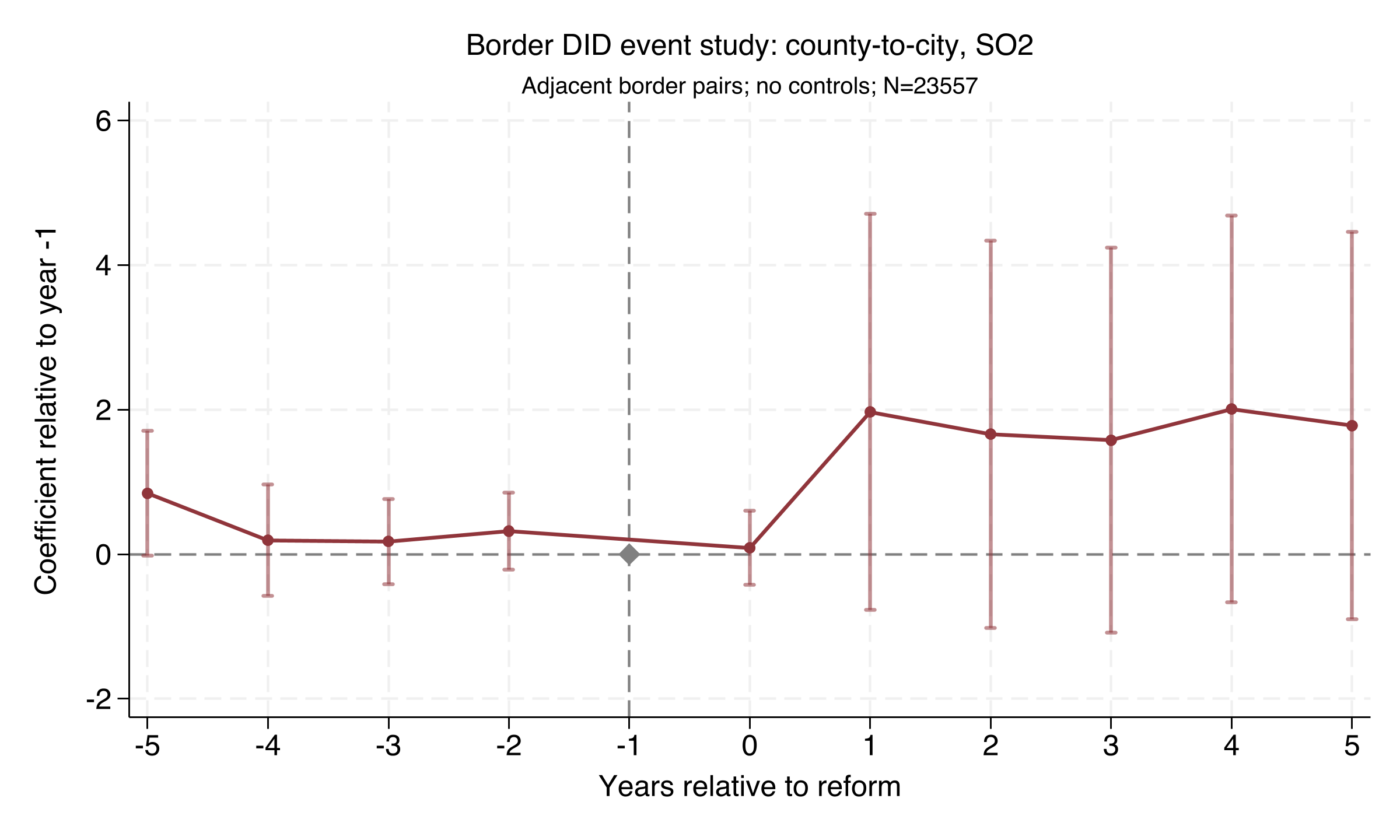}
    \caption{SO$_2$}
  \end{subfigure}\hfill
  \begin{subfigure}[b]{0.32\textwidth}
    \includegraphics[width=\textwidth]{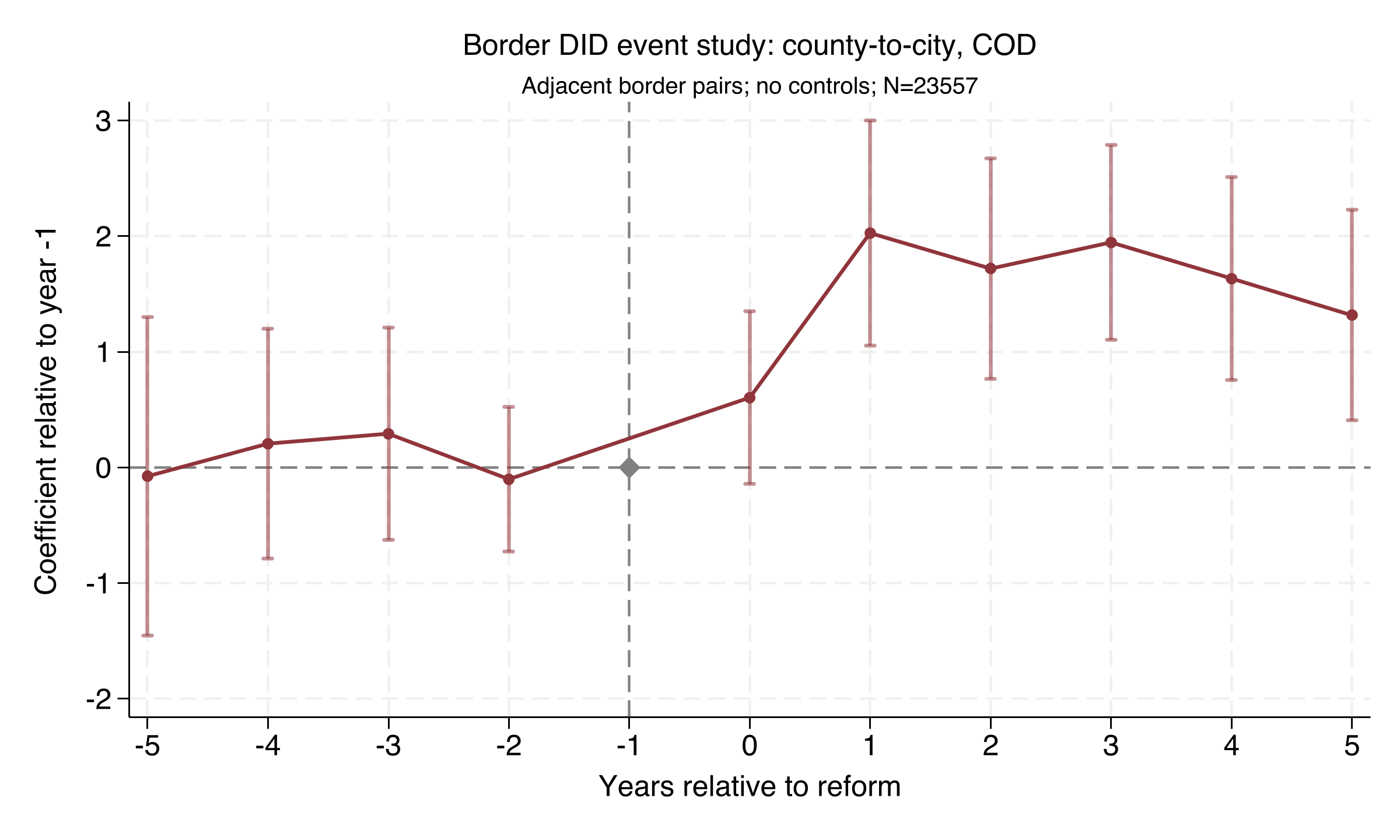}
    \caption{COD}
  \end{subfigure}\hfill
  \begin{subfigure}[b]{0.32\textwidth}
    \includegraphics[width=\textwidth]{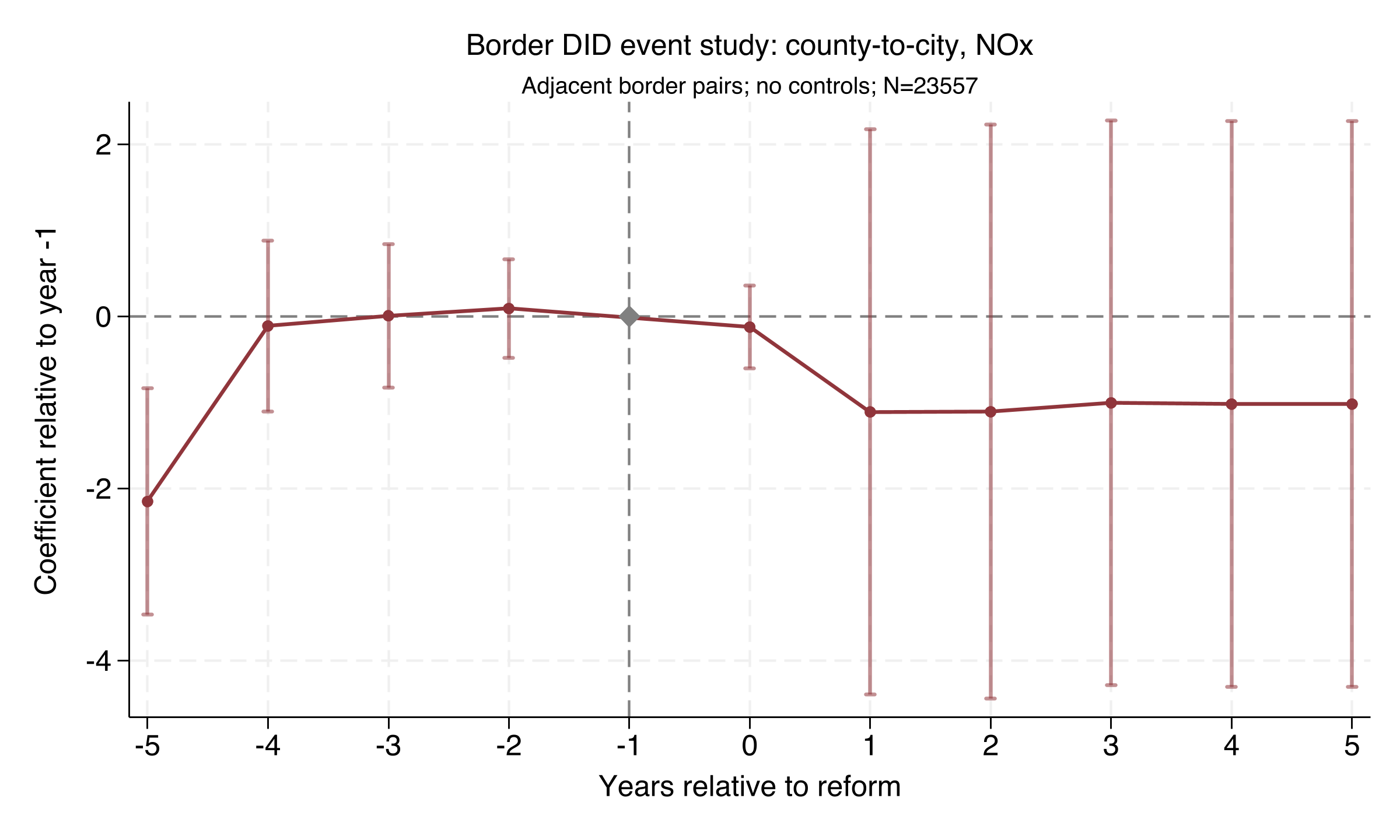}
    \caption{NOx}
  \end{subfigure}\hfill
  \begin{subfigure}[b]{0.32\textwidth}
    \includegraphics[width=\textwidth]{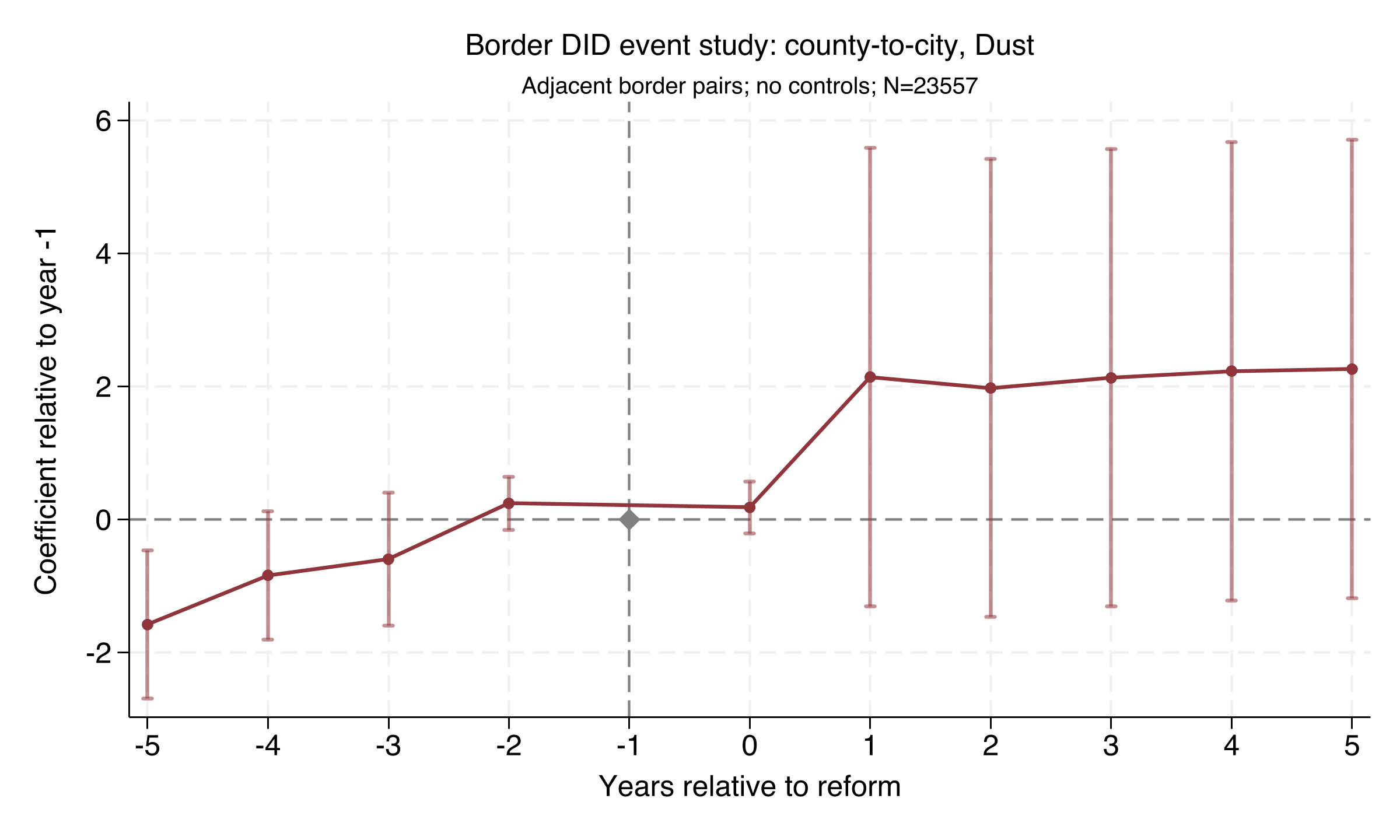}
    \caption{Dust}
  \end{subfigure}\hfill
  \begin{subfigure}[b]{0.32\textwidth}
    \includegraphics[width=\textwidth]{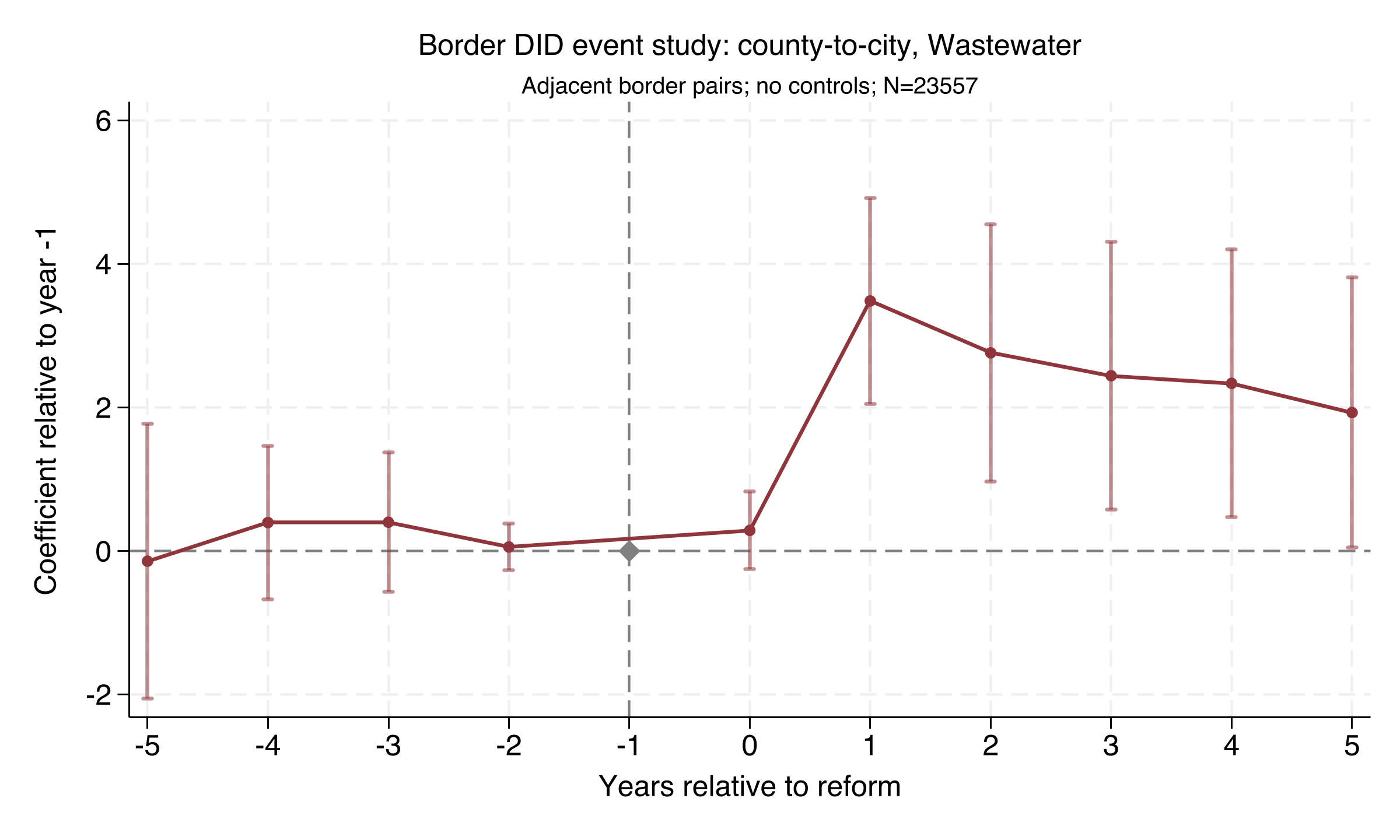}
    \caption{Wastewater}
  \end{subfigure}\hfill
  \begin{subfigure}[b]{0.32\textwidth}
    \includegraphics[width=\textwidth]{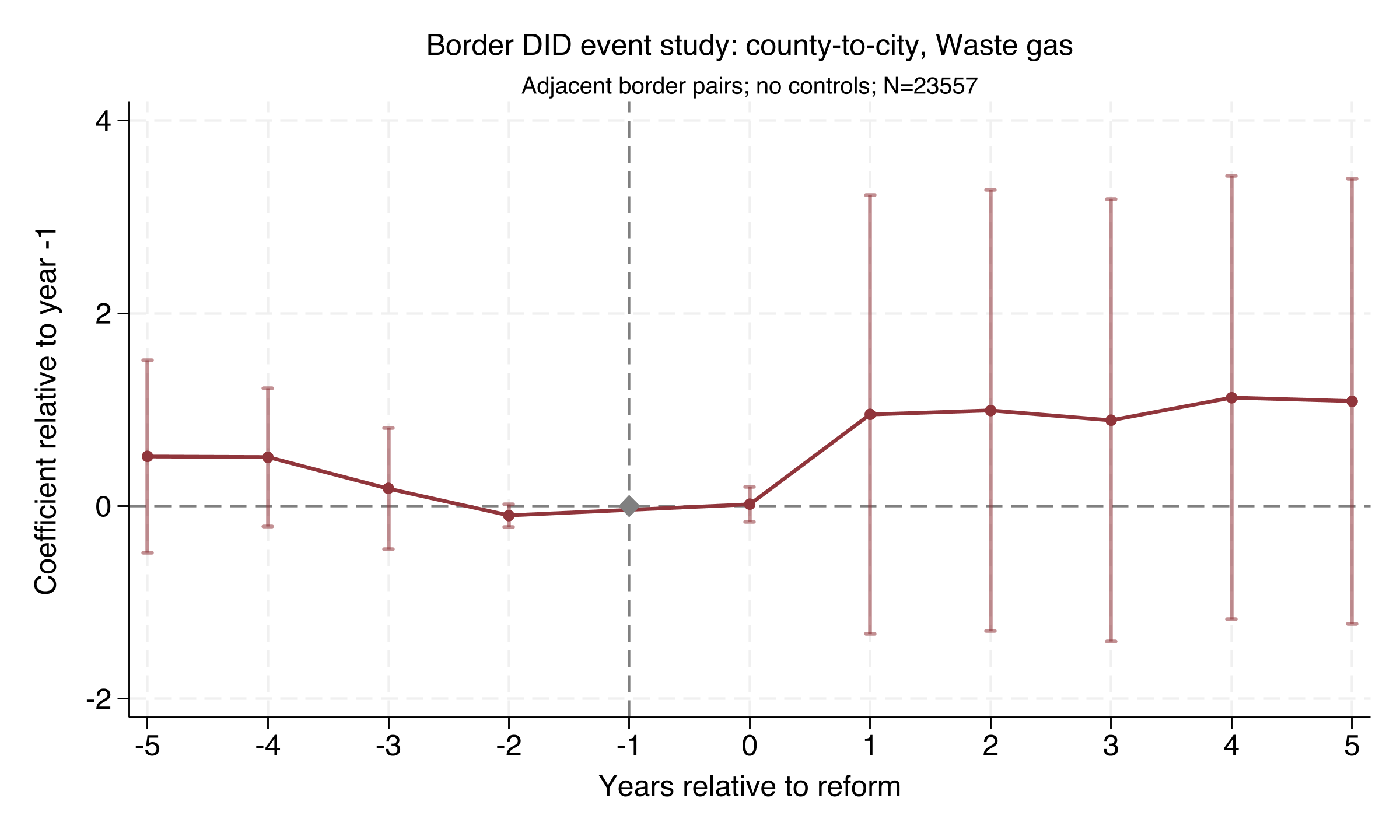}
    \caption{Waste~gas}
  \end{subfigure}\hfill
\end{figure}

Figure~\ref{fig:district_border} and Figure~\ref{fig:countycity_border} plot
the border-pair event studies.

\section{Spatial Reallocation versus Net Abatement}\label{sec:spillover}

\subsection{Neighbor spillovers along administrative boundaries}
Table~\ref{tab:spillover} reports neighbor spillovers. A reformed county's
\emph{same-prefecture} neighbors experience a significant NOx \emph{increase}
($+0.773$, $p{=}0.004$); cross-prefecture neighbors do not ($+0.034$, n.s.).
The distance-band $\times$ prefecture interaction sharpens the pattern: the
positive spillover is concentrated in same-prefecture counties 20--50\,km away
($+0.890$, $p{<}0.05$), while cross-prefecture counties immediately across the
border (0--20\,km) experience a significant \emph{decline} ($-2.215$,
$p{<}0.001$). Reallocation therefore follows \textbf{administrative
boundaries, not geographic distance}.

\subsection{Ambient pollution validates the reallocation}
Table~\ref{tab:ambient} reports satellite ambient data. In the 0--20\,km ring,
ambient NO$_2$ rises by 1.184 ($p{<}0.01$) and PM$_{2.5}$ by 1.514 ($p{<}0.01$);
locally, NO$_2$ falls after reform (event-time coefficient at $t{=}1$ of
$-0.589$, $p{<}0.05$). Two independent data chains---firm emissions and
ambient concentration---point to the same conclusion: pollution falls locally
and rises in adjacent counties.

\subsection{Falsification: wind and water}
\emph{Wind.} Table~\ref{tab:wind} assigns each neighbor to downwind, upwind,
and crosswind cones ($\pm45^{\circ}$) from MERRA2. Downwind effects are zero
and crosswind placebo effects are zero; the apparent upwind effect vanishes
once fixed compass sectors are controlled for---an artifact of bimodal monsoon
winds under vector averaging. The spillover is not wind-driven.
\emph{Water.} Table~\ref{tab:water} uses satellite river water quality and
precise reach-level hydrological distance. There is no downstream deterioration
for COD or ammonia; the county-level ``far-downstream positive'' result
disappears at the reach level. The reallocation is specific to air (NOx).

\subsection{Aggregate regional pollution}
Aggregating each treated county with its fixed pre-reform neighbors, regional
NOx is initially estimated to rise ($+0.50$, $p{<}0.05$). However, treated
regions exhibit a positive \emph{selection} pre-trend, and the aggregate result
disappears after matching on pre-treatment NOx levels and trends
(Figure~\ref{fig:region_modern}). We report the aggregate result in the
appendix and anchor the reallocation claim on the within-neighbor,
administrative-boundary comparison, which is identified off never-treated
counties and is immune to the selection of treated counties.

\section{Mechanisms}\label{sec:mechanisms}

\subsection{Scale and selection}
Table~\ref{tab:mechanisms} reports firm-count, entry, exit, and scale
responses. County-to-district reform reduces firm count ($-0.196$, $p{<}0.01$)
and exits ($-0.208$, $p{<}0.01$) while employment rises ($+0.052$, $p{<}0.01$)
and sales fall modestly ($-0.071$, $p{<}0.10$)---restructuring toward a
cleaner composition rather than broad contraction. County-to-city reform
generates firm churn: entry $+0.231$ ($p{<}0.01$) and exit $+0.492$ ($p{<}0.01$)
both rise, with no significant net change in the firm count.

\subsection{Clean production and industry heterogeneity}
Table~\ref{tab:heterogeneity} reports industry heterogeneity. County-to-district
intensity reductions are concentrated in high-polluting industries (DID
$\times$ high-polluting industry $=-0.153$, $p{<}0.01$ for COD per worker),
while county-to-city intensity \emph{increases} are concentrated there
($+0.129$, $p{<}0.05$). The composition of abatement and of dirty growth are
both concentrated where pollution is concentrated.

\subsection{The relocation margin within neighbors}
Within same-prefecture neighbor counties, the NOx increase is attributable to
\emph{incumbent firms becoming dirtier}, not to new entry or relocation.
Table~\ref{tab:entry} shows that true entry and true polluting entry---measured
from business-registration founding years, which corrects the left-censoring of
the 1998 panel start (92.4\% of ``1998 entrants'' were pre-existing firms)---are
insignificant. By contrast, incumbent neighbor firms raise NOx ($+0.820$,
$p{<}0.01$) while holding output constant ($+0.005$, n.s.) and \emph{reducing}
employment ($-0.253$, $p{<}0.10$). The reallocation operates through a
deterioration of the clean-production margin of incumbent neighbor
firms---same output, fewer workers, more NOx.

\subsection{The enforcement channel is not testable}
The regulatory channel cannot be identified in the 1998--2013 panel:
environmental penalty records begin in 2011, and the central environmental
inspection (2016) and environmental vertical-management reforms (2016--2018)
postdate the panel. The ``11th Five-Year Plan'' total-control targets
(2006--2010) provide a within-window regulatory proxy; Table~\ref{tab:totalcontrol}
shows null or \emph{opposite-signed} interaction effects (stricter targets
associated with \emph{smaller} reform-induced abatement, consistent with
targets assigned to initially dirtier counties). We report this as a null
result and anchor the mechanism on scale, selection, and clean production.

\section{Land Development and Exposure}\label{sec:land}

\subsection{Land supply}
Table~\ref{tab:land} reports land-supply effects. County-to-district reform
significantly increases land supply: industrial land area $+0.628$ ($p{<}0.01$),
residential area $+0.558$ ($p{<}0.01$), total area $+0.546$ ($p{<}0.01$),
parcels $+0.522$ ($p{<}0.01$), and industrial land value $+1.309$ ($p{<}0.01$).
A stacked-DID specification with cohort-specific controls confirms the result
and passes pre-trend tests (industrial area $+0.261$, $p{<}0.10$; residential
$+0.240$, $p{<}0.05$; parcels $+0.380$, $p{<}0.01$; value $+0.756$, $p{<}0.05$).
The reform is associated with intensified local development, consistent with a
``reform $\to$ development $\to$ pollution reallocation'' chain.

\subsection{Exposure-weighted damages}
Exposure-weighted damage outcomes (emissions $\times$ county population,
downwind population, downstream population, with seasonal splits) are uniformly
insignificant for county-to-district reform. Reforms do not produce detectable
population-weighted damage reductions---again consistent with local abatement
being offset by reallocation. The only clean exposure finding is that local
NOx abatement is concentrated among \emph{near-urban firms} (within 10\,km of
built-up area): NOx falls by $0.279$ ($p{<}0.10$) near urban areas versus
$-0.031$ (n.s.) far from them (Table~\ref{tab:land}, Panel C).

\section{Welfare Accounting}\label{sec:welfare}

We monetize the emission changes in a back-of-the-envelope welfare calculation.
Emissions in kilograms are converted to tons; marginal damages (yuan per ton,
2003 prices) are constructed from China-specific health-damage parameters---the
GEMM concentration--response function \citep{burnett2018global} applied to
age-specific NCD+LRI mortality from GBD, population-weighted PM$_{2.5}$
exposure, and a value of a statistical life of approximately \textyen 1~million
\citep{worldbank2007cost}; COD damages derive from \citet{worldbank2007cost}
aggregate water-pollution damages.

Table~\ref{tab:welfare} reports the results. For county-to-district reform, the
only significant abatement is NOx, which monetizes to \textyen 47 million
per year---but the 95\% CI spans -\textyen 3.3 billion to +\textyen 6.0
billion and crosses zero, so the welfare gain is statistically
indistinguishable from zero. For county-to-city reform, the significant
increases in COD and dust monetize to a damage of -\textyen 2.3 billion per
year. Neither reform produces a significant output loss, so a unit abatement
cost cannot be computed---a finding in itself: the observed abatement is not
achieved at a measurable cost to local output. These magnitudes are sensitivity
calculations: the air-pollutant marginal damages rely on generic intake
fractions rather than China-specific source--receptor estimates.

\section{Robustness}\label{sec:robustness}

The findings are subjected to the following checks. (i) \emph{Estimator
robustness}: Sun--Abraham, Borusyak imputation, and de Chaisemartin--
D'Haultf\oe uille dynamic estimators confirm the neighbor spillover results;
stacked DID confirms the land-supply results. (ii) \emph{Clustering}: standard
errors are clustered at the border-pair level, the original county, and two-way
county $\times$ prefecture for regulatory interactions, with cluster counts
reported. (iii) \emph{Pre-trends}: joint pre-trend tests reported throughout
(Tables~\ref{tab:pretrend} and \ref{tab:borderpretrend}); the county-to-district
intensity margin and the stacked land-supply specification pass; county-to-city
stacked levels and the aggregate regional specification fail and are handled
accordingly. (iv) \emph{Directional falsifications}: wind-cone and
river-downstream falsifications rule out physical transportation
(Tables~\ref{tab:wind} and \ref{tab:water}). (v) \emph{Measurement of entry}:
true entry from business-registration founding years corrects panel
left-censoring (Table~\ref{tab:entry}).

\section{Conclusion}\label{sec:conclusion}

Administrative boundary reforms do not have a uniform effect on firm
pollution; they reshape pollution along two institutional margins.
Centralizing reform (county-to-district) lowers local NOx \emph{intensity}
($-0.148^{***}$) through cleaner production, without shrinking output or
employment. Empowering reform (county-to-city) raises COD, dust, and wastewater
levels ($+0.87^{**}$ to $+0.95^{*}$) and COD/dust intensity through dirty
growth. The local abatement is offset by increases in same-prefecture neighbor
counties ($+0.77^{***}$), which follow administrative boundaries rather than
geographic distance, are validated in satellite ambient data, and are not
driven by wind or water transport; the reallocation operates through incumbent
neighbor firms becoming dirtier, not through relocation or new entry.
Exposure-weighted damages do not fall, and the monetized welfare gain from the
only significant abatement is statistically indistinguishable from zero, while
county-to-city reform generates roughly \textyen 2.3 billion per year in
monetized damages. The net effect of these reforms on regional pollution is, at
best, a spatial reshuffling of emissions.

\bibliographystyle{plainnat}
\bibliography{references}

\clearpage
\appendix
\section{Tables}

\begin{table}[t]
\caption{Border-pair difference-in-differences: emission levels}
\label{tab:levels}
\centering
\small
\begin{tabular}{lcccccc}
\toprule
 & SO$_2$ & COD & NOx & Dust & Wastewater & Waste gas \\
\midrule
\multicolumn{7}{l}{\emph{Panel A. County-to-district}} \\
TreatedSide $\times$ Post &
$-0.162$ & $0.030$ & $-0.294^{***}$ & $0.026$ & $-0.066$ & $0.118$ \\
 & $(0.111)$ & $(0.122)$ & $(0.094)$ & $(0.094)$ & $(0.129)$ & $(0.089)$ \\
\addlinespace
\multicolumn{7}{l}{\emph{Panel B. County-to-city}} \\
TreatedSide $\times$ Post &
$0.250$ & $0.869^{**}$ & $0.224$ & $0.945^{*}$ & $0.926^{**}$ & $0.080$ \\
 & $(0.365)$ & $(0.340)$ & $(0.382)$ & $(0.570)$ & $(0.433)$ & $(0.293)$ \\
\midrule
Firm FE & \multicolumn{6}{c}{Yes} \\
Border-pair $\times$ year FE & \multicolumn{6}{c}{Yes} \\
\midrule
Observations (Panel A / B) & \multicolumn{6}{c}{297,319 / 132,000} \\
\bottomrule
\end{tabular}
\begin{tablenotes}
\footnotesize
\item \emph{Notes:} Dependent variable is $\ln(1+E)$ for each pollutant. Standard errors in parentheses, clustered by border pair. Significance: $^{*}\,p{<}0.10$, $^{**}\,p{<}0.05$, $^{***}\,p{<}0.01$.
\end{tablenotes}
\end{table}

\begin{table}[t]
\caption{Border-pair difference-in-differences: emission intensity}
\label{tab:intensity}
\centering
\small
\begin{tabular}{lcccccc}
\toprule
 & SO$_2$ & COD & NOx & Dust & Wastewater & Waste gas \\
\midrule
\multicolumn{7}{l}{\emph{Panel A. County-to-district, per worker}} \\
TreatedSide $\times$ Post &
$-0.149^{***}$ & $-0.007$ & $-0.148^{***}$ & $-0.041$ & $-0.129^{*}$ & $-0.058$ \\
 & $(0.057)$ & $(0.068)$ & $(0.042)$ & $(0.049)$ & $(0.072)$ & $(0.047)$ \\
\addlinespace
\multicolumn{7}{l}{\emph{Panel B. County-to-district, per sales}} \\
TreatedSide $\times$ Post &
$-0.019$ & $-0.005$ & $-0.015^{*}$ & $-0.012$ & $0.020$ & $-0.016^{*}$ \\
 & $(0.017)$ & $(0.017)$ & $(0.008)$ & $(0.019)$ & $(0.031)$ & $(0.008)$ \\
\addlinespace
\multicolumn{7}{l}{\emph{Panel C. County-to-city, per worker}} \\
TreatedSide $\times$ Post &
$0.001$ & $0.275^{**}$ & $-0.003$ & $0.644^{*}$ & $0.253$ & $0.237$ \\
 & $(0.200)$ & $(0.140)$ & $(0.185)$ & $(0.328)$ & $(0.218)$ & $(0.159)$ \\
\addlinespace
\multicolumn{7}{l}{\emph{Panel D. County-to-city, per sales}} \\
TreatedSide $\times$ Post &
$0.033$ & $0.045^{**}$ & $0.019$ & $0.151^{**}$ & $0.059$ & $0.015$ \\
 & $(0.075)$ & $(0.021)$ & $(0.032)$ & $(0.065)$ & $(0.040)$ & $(0.033)$ \\
\midrule
Firm FE & \multicolumn{6}{c}{Yes} \\
Border-pair $\times$ year FE & \multicolumn{6}{c}{Yes} \\
\bottomrule
\end{tabular}
\begin{tablenotes}
\footnotesize
\item \emph{Notes:} Intensity is $\ln(1+E/\text{employment})$ (Panels A, C) and $\ln(1+E/\text{sales})$ (Panels B, D). Standard errors clustered by border pair. $^{*}\,p{<}0.10$, $^{**}\,p{<}0.05$, $^{***}\,p{<}0.01$.
\end{tablenotes}
\end{table}

\begin{table}[t]
\caption{Emission levels across estimators}
\label{tab:estimators}
\centering
\small
\begin{tabular}{lcccccc}
\toprule
 & \multicolumn{3}{c}{County-to-district} & \multicolumn{3}{c}{County-to-city} \\
\cmidrule(lr){2-4}\cmidrule(lr){5-7}
 & NOx & SO$_2$ & Waste gas & COD & Dust & Wastewater \\
\midrule
Two-way FE & $-0.280^{*}$ & $0.061$ & $0.318^{**}$ & $0.726$ & $-0.530$ & $0.439$ \\
 & $(0.149)$ & $(0.178)$ & $(0.151)$ & $(0.672)$ & $(0.542)$ & $(0.660)$ \\
Stacked DID & $-0.191$ & $0.009$ & $0.253^{*}$ & $0.886$ & $-0.541$ & $0.478$ \\
 & $(0.135)$ & $(0.144)$ & $(0.131)$ & $(0.745)$ & $(0.477)$ & $(0.560)$ \\
Border-pair DID & $-0.294^{***}$ & $-0.162$ & $0.118$ & $0.869^{**}$ & $0.945^{*}$ & $0.926^{**}$ \\
 & $(0.094)$ & $(0.111)$ & $(0.089)$ & $(0.340)$ & $(0.570)$ & $(0.433)$ \\
\bottomrule
\end{tabular}
\begin{tablenotes}
\footnotesize
\item \emph{Notes:} Dependent variable $\ln(1+E)$. Two-way FE absorbs firm and year FE with prefecture-by-year trends and clusters by county; stacked DID clusters by stack-county; border-pair DID clusters by border pair. The stacked county-to-city level specification fails parallel trends for SO$_2$, COD, dust, and wastewater (joint $F=30$--$307$, $p\approx0$); county-to-city level effects are therefore anchored on the border-pair design.
\end{tablenotes}
\end{table}

\begin{table}[t]
\caption{Spatial spillover of NOx (never-treated counties)}
\label{tab:spillover}
\centering
\small
\begin{tabular}{lcc}
\toprule
 & (1) & (2) \\
\midrule
Same-prefecture neighbor & $0.773^{***}$ & \\
 & $(0.267)$ & \\
Cross-prefecture neighbor & $0.034$ & \\
 & $(0.202)$ & \\
Same-pref $\times$ 0--20\,km & & $-0.159$ \\
 & & $(0.542)$ \\
Same-pref $\times$ 20--50\,km & & $0.890^{**}$ \\
 & & $(0.409)$ \\
Same-pref $\times$ 50--100\,km & & $0.670$ \\
 & & $(0.436)$ \\
Cross-pref $\times$ 0--20\,km & & $-2.215^{***}$ \\
 & & $(0.382)$ \\
Cross-pref $\times$ 20--50\,km & & $0.197$ \\
 & & $(0.275)$ \\
Cross-pref $\times$ 50--100\,km & & $0.123$ \\
 & & $(0.124)$ \\
\midrule
County FE & Yes & Yes \\
Prefecture $\times$ year FE & Yes & Yes \\
Observations & 32,409 & 32,409 \\
\bottomrule
\end{tabular}
\begin{tablenotes}
\footnotesize
\item \emph{Notes:} Dependent variable is $\ln(1+\text{total NOx})$. Sample restricted to counties never treated by county-to-district reform. Standard errors clustered by county. $^{*}\,p{<}0.10$, $^{**}\,p{<}0.05$, $^{***}\,p{<}0.01$.
\end{tablenotes}
\end{table}

\begin{table}[t]
\caption{Ambient pollution: neighbor spillovers and local dynamics}
\label{tab:ambient}
\centering
\small
\begin{tabular}{lccccc}
\toprule
 & \multicolumn{2}{c}{0--20\,km ring} & \multicolumn{2}{c}{Same-prefecture} & Local \\
\cmidrule(lr){2-3}\cmidrule(lr){4-5}
 & NO$_2$ & PM$_{2.5}$ & NO$_2$ & PM$_{2.5}$ & NO$_2$, $t{=}1$ \\
\midrule
Coefficient & $1.184^{***}$ & $1.514^{***}$ & $0.482^{**}$ & $0.301$ & $-0.589^{**}$ \\
 & $(0.434)$ & $(0.522)$ & $(0.245)$ & $(0.425)$ & $(0.270)$ \\
\midrule
County FE & Yes & Yes & Yes & Yes & Yes \\
Pref. $\times$ year FE & Yes & Yes & Yes & Yes & Yes \\
\bottomrule
\end{tabular}
\begin{tablenotes}
\footnotesize
\item \emph{Notes:} ``0--20\,km ring'' reports the effect of being within 20\,km of a reformed county on ambient NO$_2$ (satellite tropospheric column) and PM$_{2.5}$; 20--50 and 50--100\,km bands are insignificant. ``Local'' reports the event-time coefficient at $t=1$ for NO$_2$ in the reformed county itself (local decline). $^{*}\,p{<}0.10$, $^{**}\,p{<}0.05$, $^{***}\,p{<}0.01$.
\end{tablenotes}
\end{table}

\begin{table}[t]
\caption{Mechanisms: scale and selection}
\label{tab:mechanisms}
\centering
\small
\begin{tabular}{lccccc}
\toprule
 & Firm count & Entry & Exit & Employment & Sales \\
\midrule
County-to-district & $-0.196^{***}$ & $0.045$ & $-0.208^{***}$ & $0.052^{***}$ & $-0.071^{*}$ \\
 & $(0.048)$ & $(0.050)$ & $(0.057)$ & $(0.019)$ & $(0.037)$ \\
County-to-city & $0.103$ & $0.231^{***}$ & $0.492^{***}$ & $0.120$ & $0.029$ \\
 & $(0.084)$ & $(0.075)$ & $(0.108)$ & $(0.080)$ & $(0.088)$ \\
\midrule
Level & County & County & County & Firm & Firm \\
\bottomrule
\end{tabular}
\begin{tablenotes}
\footnotesize
\item \emph{Notes:} Firm-count, entry, and exit are county-level outcomes ($\ln(1{+}x)$); employment and sales are firm-level outcomes ($\ln x$). Standard errors clustered by county (county level) and border pair (firm level). $^{*}\,p{<}0.10$, $^{**}\,p{<}0.05$, $^{***}\,p{<}0.01$.
\end{tablenotes}
\end{table}

\begin{table}[t]
\caption{Industry heterogeneity: DID $\times$ high-polluting industry}
\label{tab:heterogeneity}
\centering
\small
\begin{tabular}{lcccc}
\toprule
 & COD/worker & NOx/worker & COD level & Wastewater \\
\midrule
\multicolumn{5}{l}{\emph{Panel A. County-to-city}} \\
DID (low-polluting ref.) & $0.246^{*}$ & $0.025$ & $0.846^{**}$ & $0.965^{**}$ \\
 & $(0.141)$ & $(0.185)$ & $(0.336)$ & $(0.431)$ \\
DID $\times$ HighPolluting & $0.129^{**}$ & $-0.124^{**}$ & $0.100$ & $-0.162^{*}$ \\
 & $(0.065)$ & $(0.058)$ & $(0.097)$ & $(0.091)$ \\
\addlinespace
\multicolumn{5}{l}{\emph{Panel B. County-to-district}} \\
DID (low-polluting ref.) & $0.029$ & $-0.143^{***}$ & $0.097$ & $-0.062$ \\
 & $(0.071)$ & $(0.044)$ & $(0.123)$ & $(0.132)$ \\
DID $\times$ HighPolluting & $-0.153^{***}$ & $-0.020$ & $-0.291^{***}$ & $-0.019$ \\
 & $(0.038)$ & $(0.030)$ & $(0.058)$ & $(0.057)$ \\
\bottomrule
\end{tabular}
\begin{tablenotes}
\footnotesize
\item \emph{Notes:} High-polluting industries are the top quartile of baseline industry-average COD intensity. Firm FE and border-pair $\times$ year FE absorbed. $^{*}\,p{<}0.10$, $^{**}\,p{<}0.05$, $^{***}\,p{<}0.01$.
\end{tablenotes}
\end{table}

\begin{table}[t]
\caption{Land supply and urban-proximity heterogeneity}
\label{tab:land}
\centering
\small
\begin{tabular}{lccccc}
\toprule
 & Ind.\ area & Res.\ area & All area & Ind.\ parcels & Ind.\ price \\
\midrule
\multicolumn{6}{l}{\emph{Panel A. Land supply, two-way FE (county-year)}} \\
County-to-district & $0.628^{***}$ & $0.558^{***}$ & $0.546^{***}$ & $0.522^{***}$ & $1.309^{***}$ \\
 & $(0.144)$ & $(0.110)$ & $(0.118)$ & $(0.131)$ & $(0.277)$ \\
\addlinespace
\multicolumn{6}{l}{\emph{Panel B. Land supply, stacked DID}} \\
Treated $\times$ Post & $0.261^{*}$ & $0.240^{**}$ & $0.191$ & $0.380^{***}$ & $0.756^{**}$ \\
 & $(0.145)$ & $(0.118)$ & $(0.125)$ & $(0.146)$ & $(0.331)$ \\
\addlinespace
\multicolumn{6}{l}{\emph{Panel C. NOx by urban proximity (border-pair DID)}} \\
 & \multicolumn{2}{c}{Near urban (${\leq}10$\,km)} & \multicolumn{2}{c}{Far urban (${>}10$\,km)} & \\
County-to-district & \multicolumn{2}{c}{$-0.279^{*}$} & \multicolumn{2}{c}{$-0.031$} & \\
 & \multicolumn{2}{c}{$(0.148)$} & \multicolumn{2}{c}{$(0.034)$} & \\
\bottomrule
\end{tabular}
\begin{tablenotes}
\footnotesize
\item \emph{Notes:} Panels A--B: county-year panel of land transactions, 2000--2013, zero-filled balanced panel ($N=37{,}600$); outcomes are $\ln(1{+}\text{area/parcels/price})$ by land use. Panel B stacks by reform cohort with never-/not-yet-treated controls and passes pre-trend tests. Panel C: firm-level NOx $\ln(1{+}E)$ by distance to built-up area (GAIA). $^{*}\,p{<}0.10$, $^{**}\,p{<}0.05$, $^{***}\,p{<}0.01$.
\end{tablenotes}
\end{table}

\begin{table}[t]
\caption{Back-of-the-envelope welfare accounting (2003 yuan)}
\label{tab:welfare}
\centering
\small
\begin{tabular}{lcccc}
\toprule
 & Environmental benefit & \multicolumn{2}{c}{95\% CI} & Output change \\
 & (significant terms) & lower & upper & \\
\midrule
County-to-district (border) & $+4{,}701$ & $-334{,}353$ & $+599{,}912$ & $+256{,}191$ \\
 & & & & (n.s.) \\
County-to-city (border) & $-232{,}944$ & $-1{,}052{,}357$ & $+290{,}528$ & $+53{,}761$ \\
 & & & & (n.s.) \\
\bottomrule
\end{tabular}
\begin{tablenotes}
\footnotesize
\item \emph{Notes:} Annual environmental benefit/damage in 10,000 yuan, from $\Delta \text{tons} \times \text{marginal damage}$. Marginal damages use China-specific health-damage parameters (GEMM concentration--response with age-specific NCD+LRI mortality, population-weighted PM$_{2.5}$ exposure, VSL $\approx$ \textyen 1\,M) and World Bank \& SEPA (2007) for COD; dust uses PM$_{2.5}$ mass fraction 0.25. County-to-district's only significant abatement (NOx) monetizes to +\textyen 47 million/yr with a CI crossing zero; county-to-city's COD and dust increases monetize to -\textyen 2.33 billion/yr. Neither reform shows a significant output change, so unit abatement cost is unidentified. Treated as sensitivity calculations; see notes.
\end{tablenotes}
\end{table}

\begin{table}[t]
\caption{Wind falsification: downwind vs.\ upwind vs.\ crosswind}
\label{tab:wind}
\centering
\small
\begin{tabular}{lcccc}
\toprule
 & \multicolumn{2}{c}{Heating season} & \multicolumn{2}{c}{Non-heating} \\
\cmidrule(lr){2-3}\cmidrule(lr){4-5}
 & PM$_{2.5}$ & NOx (firm) & PM$_{2.5}$ & NOx (firm) \\
\midrule
Downwind & $-0.058$ & $0.028$ & $-0.047$ & \\
 & $(0.166)$ & $(0.125)$ & $(0.152)$ & \\
Upwind & $0.451^{***}$ & $0.110$ & $0.336^{**}$ & \\
 & $(0.166)$ & $(0.118)$ & $(0.145)$ & \\
Crosswind (placebo) & $0.110$ & $0.163$ & $0.095$ & \\
 & $(0.138)$ & $(0.104)$ & $(0.123)$ & \\
Same-prefecture & & $0.758^{***}$ & & \\
 & & $(0.268)$ & & \\
\midrule
\multicolumn{5}{l}{\emph{Controlling for fixed compass sectors, the upwind coefficient loses significance}} \\
Upwind (heat), PM$_{2.5}$ & \multicolumn{4}{c}{$0.210$ (n.s.) \quad vs.\ fixed N $0.538^{*}$, E $0.920^{***}$ (not shown), S $1.341^{***}$} \\
\bottomrule
\end{tabular}
\begin{tablenotes}
\footnotesize
\item \emph{Notes:} Wind cones $\pm45^{\circ}$, $\leq100$\,km, from MERRA2. The ``upwind positive'' result is not robust to fixed compass sectors and reflects bimodal monsoon vector-average noise. Downwind $\approx 0$ and crosswind $\approx 0$ reject atmospheric transport. $^{*}\,p{<}0.10$, $^{**}\,p{<}0.05$, $^{***}\,p{<}0.01$.
\end{tablenotes}
\end{table}

\begin{table}[t]
\caption{Water falsification: no downstream deterioration}
\label{tab:water}
\centering
\small
\begin{tabular}{lcccc}
\toprule
 & \multicolumn{2}{c}{County-year} & \multicolumn{2}{c}{Reach-year (county-to-city)} \\
\cmidrule(lr){2-3}\cmidrule(lr){4-5}
 & CODMn & NH$_4$-N & CODMn & NH$_4$-N \\
\midrule
Downstream & $0.0033$ & $0.0002$ & & \\
 & $(0.0031)$ & $(0.0005)$ & & \\
Upstream (placebo) & $-0.0049$ & $0.0006$ & & \\
 & $(0.0043)$ & $(0.0008)$ & & \\
Downstream near (${\leq}100$\,km) & & & $-0.0230$ & $-0.0035^{*}$ \\
 & & & $(0.0197)$ & $(0.0018)$ \\
Downstream far (100--500\,km) & & & $-0.0095$ & $0.0021$ \\
 & & & $(0.0108)$ & $(0.0016)$ \\
\bottomrule
\end{tabular}
\begin{tablenotes}
\footnotesize
\item \emph{Notes:} Satellite-derived river water quality, 2000--2023 (212,803 reaches/month). County-year columns use county-to-district exposure; reach-year columns use county-to-city exposure with precise hydrological distance. No downstream deterioration; the county-level ``far-downstream positive'' result does not survive the reach-level specification. $^{*}\,p{<}0.10$, $^{**}\,p{<}0.05$, $^{***}\,p{<}0.01$.
\end{tablenotes}
\end{table}

\begin{table}[t]
\caption{Extensive margin: true entry vs.\ left-censored entry}
\label{tab:entry}
\centering
\small
\begin{tabular}{lccc}
\toprule
 & Same-prefecture & Cross-prefecture & Own reform \\
\midrule
\multicolumn{4}{l}{\emph{Panel A. True entry (founding year)}} \\
True entry count & $0.027$ & $-0.030$ & $-0.093$ \\
 & $(0.070)$ & $(0.053)$ & $(0.090)$ \\
True polluting entry & $0.024$ & $-0.026$ & $-0.086$ \\
 & $(0.069)$ & $(0.052)$ & $(0.093)$ \\
\addlinespace
\multicolumn{4}{l}{\emph{Panel B. Left-censored entry (panel first-year)}} \\
Entry count (old) & $-0.918$ & $1.077^{**}$ & \\
 & $(0.544)$ & $(0.455)$ & \\
Polluting entry (old) & $-0.909$ & $0.939^{**}$ & \\
 & $(0.513)$ & $(0.440)$ & \\
\bottomrule
\end{tabular}
\begin{tablenotes}
\footnotesize
\item \emph{Notes:} True entry uses business-registration founding years, correcting left-censoring of the 1998 panel start (92.4\% of ``1998 entrants'' were pre-existing firms). The old proxy gives directionally wrong signals. $^{*}\,p{<}0.10$, $^{**}\,p{<}0.05$, $^{***}\,p{<}0.01$.
\end{tablenotes}
\end{table}

\begin{table}[t]
\caption{Pre-trend tests: stacked level event studies}
\label{tab:pretrend}
\centering
\small
\begin{tabular}{lcccc}
\toprule
 & \multicolumn{2}{c}{County-to-district} & \multicolumn{2}{c}{County-to-city} \\
\cmidrule(lr){2-3}\cmidrule(lr){4-5}
 & $F$ & $p$ & $F$ & $p$ \\
\midrule
SO$_2$ & $0.731$ & $0.570$ & $62.447$ & $0.000$ \\
COD & $0.368$ & $0.831$ & $61.316$ & $0.000$ \\
NOx & $2.340$ & $0.053$ & $1.499$ & $0.200$ \\
Dust & $1.439$ & $0.218$ & $202.139$ & $0.000$ \\
Wastewater & $0.546$ & $0.702$ & $306.823$ & $0.000$ \\
Waste gas & $2.503$ & $0.040$ & $30.083$ & $0.000$ \\
\bottomrule
\end{tabular}
\begin{tablenotes}
\footnotesize
\item \emph{Notes:} Joint test of pre-period coefficients $\theta_{-5}=\dots=\theta_{-2}=0$ (4-period test), no controls. County-to-city stacked levels fail for SO$_2$, COD, dust, wastewater, and waste gas; county-to-district passes except marginal NOx and waste gas.
\end{tablenotes}
\end{table}

\begin{table}[t]
\caption{``11th Five-Year Plan'' total-control interaction}
\label{tab:totalcontrol}
\centering
\small
\begin{tabular}{lcc}
\toprule
 & $\ln(\text{SO}_2)$ & $\ln(\text{COD})$ \\
\midrule
\multicolumn{3}{l}{\emph{Panel A. County-to-district}} \\
Reform (DID) & $-0.405^{**}$ & $-0.005$ \\
 & $(0.155)$ & $(0.180)$ \\
Reform $\times$ target (per 10\,pp) & $0.212^{***}$ & $0.034$ \\
 & $(0.051)$ & $(0.036)$ \\
\addlinespace
\multicolumn{3}{l}{\emph{Panel B. County-to-city}} \\
Reform (DID) & $0.311^{*}$ & $0.233$ \\
 & $(0.181)$ & $(0.468)$ \\
Reform $\times$ target (per 10\,pp) & $-0.053$ & $0.849$ \\
 & $(0.080)$ & $(0.514)$ \\
\bottomrule
\end{tabular}
\begin{tablenotes}
\footnotesize
\item \emph{Notes:} Targets are prefecture-level ``11th Five-Year Plan'' reduction targets (county-level when available). Two-way clustering by border pair $\times$ prefecture. The opposite-signed interaction for county-to-district SO$_2$ suggests stricter targets are associated with \emph{smaller} reform-induced abatement, consistent with targets being assigned to initially dirtier counties; reported as a null/opposite-signed result.
\end{tablenotes}
\end{table}

\begin{table}[t]
\caption{Border-pair joint pre-trend tests}
\label{tab:borderpretrend}
\centering
\small
\begin{tabular}{lcc}
\toprule
Outcome & $F(4,493)$ & $p$ \\
\midrule
NOx per worker & $1.25$ & $0.289$ \\
NOx level & $2.15$ & $0.073$ \\
Output & $2.27$ & $0.061$ \\
Employment & $1.11$ & $0.351$ \\
\bottomrule
\end{tabular}
\begin{tablenotes}
\footnotesize
\item \emph{Notes:} Joint test $e_{-5}=\dots=e_{-2}=0$ within the border-pair structure (pair $\times$ year FE retained). The clean-production margin (NOx per worker) passes; the NOx level margin is marginal.
\end{tablenotes}
\end{table}

\clearpage
\section{Figures}

\begin{figure}[t]
\caption{Aggregate regional NOx: modern-DID comparison}
\label{fig:region_modern}
\centering
\includegraphics[width=0.85\textwidth]{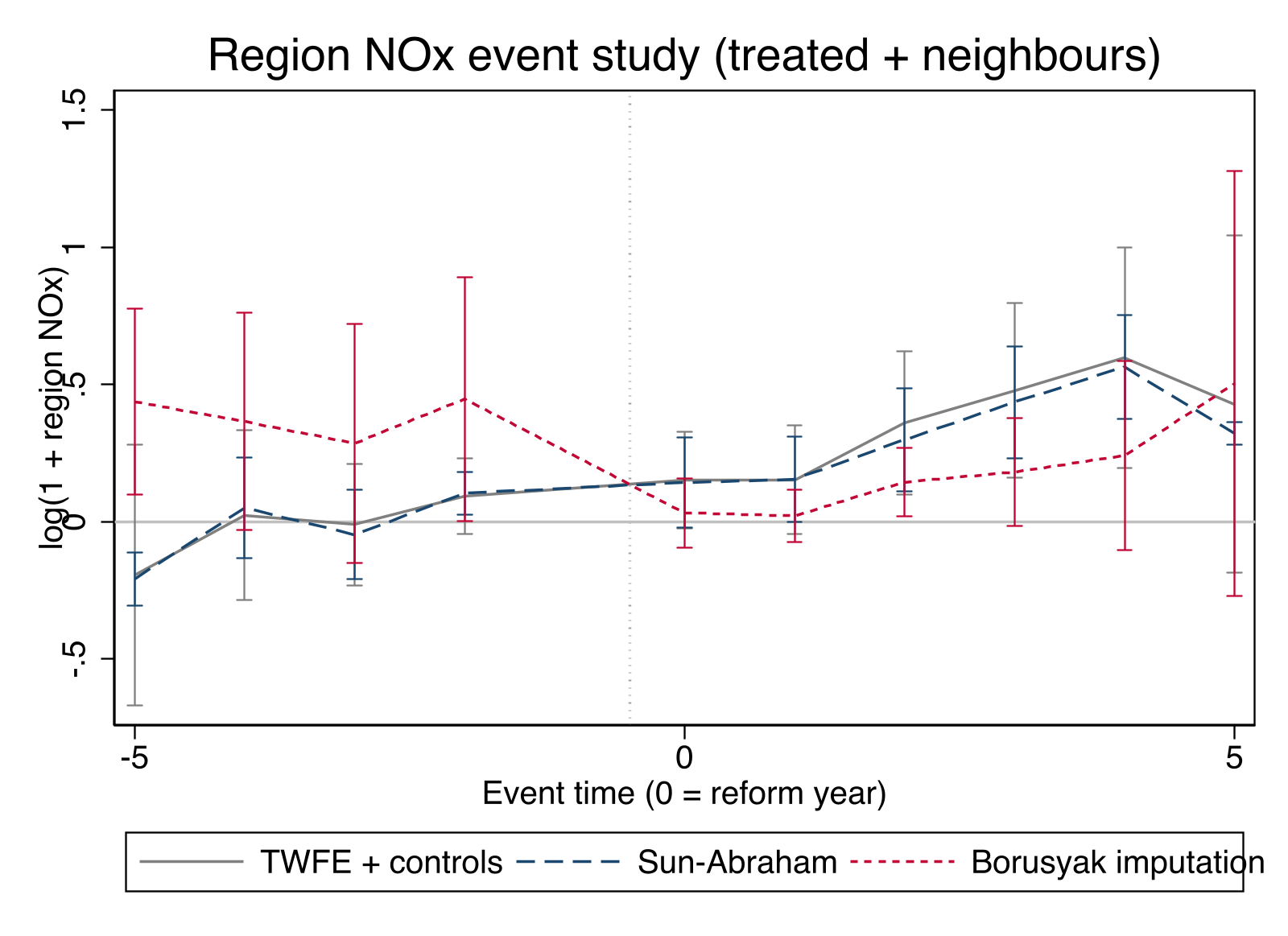}
\begin{minipage}{0.85\textwidth}
\footnotesize
\emph{Notes:} The naive TWFE aggregate region estimate ($+0.50^{**}$) is driven by a positive selection pre-trend: treated regions are on an urbanization trajectory before reform. Adding a never-treated control group and imputation estimators (Borusyak) reveals the pre-trend; matching on pre-treatment NOx levels and trends eliminates the aggregate effect. Reported in the appendix; the reallocation claim is anchored on the within-neighbor comparison (Table~\ref{tab:spillover}).
\end{minipage}
\end{figure}

\begin{figure}[t]
\caption{Neighbor spillover by distance band (ambient/emission)}
\label{fig:spillover}
\centering
\begin{subfigure}[b]{0.45\textwidth}
  \includegraphics[width=\textwidth]{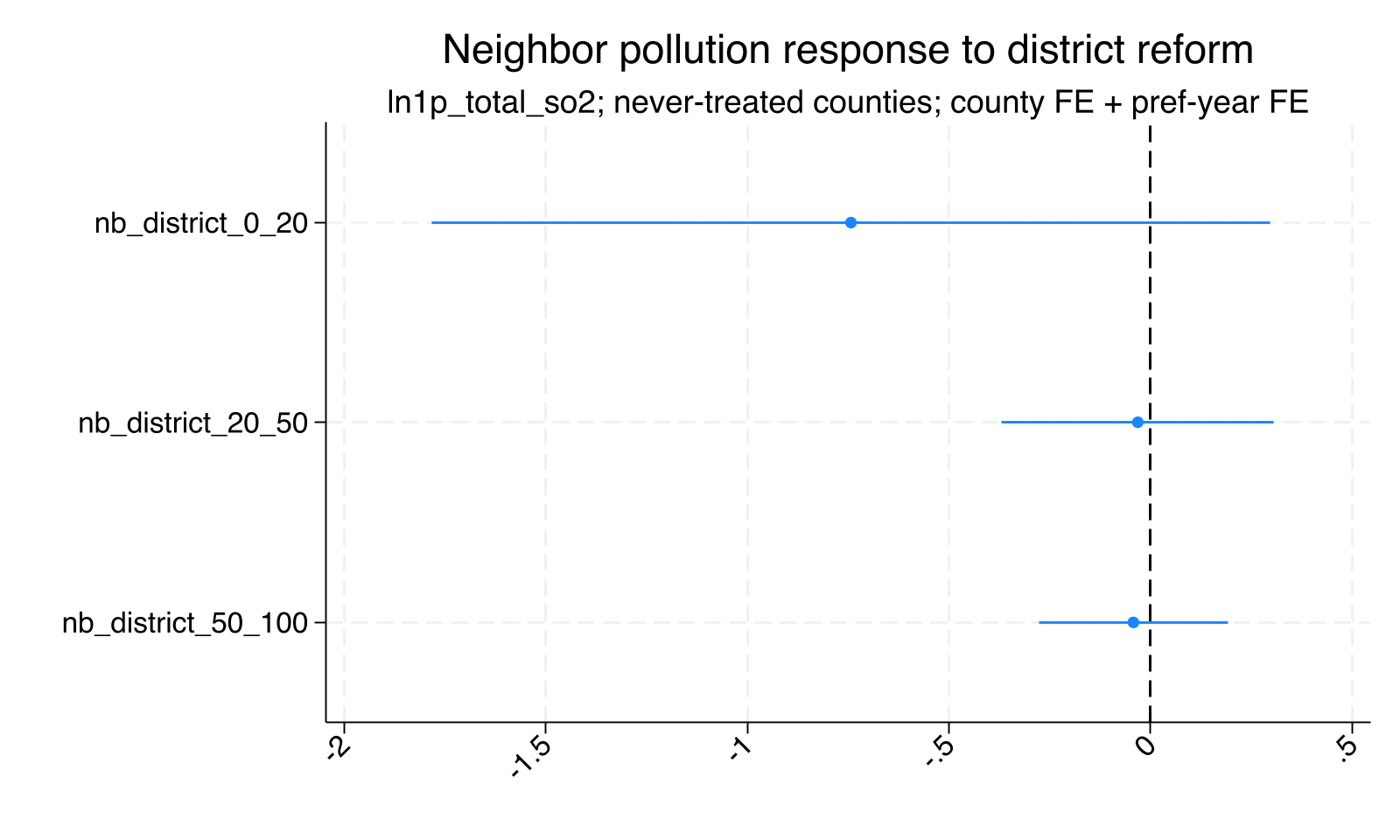}
  \caption{SO$_2$}
\end{subfigure}\hfill
\begin{subfigure}[b]{0.45\textwidth}
  \includegraphics[width=\textwidth]{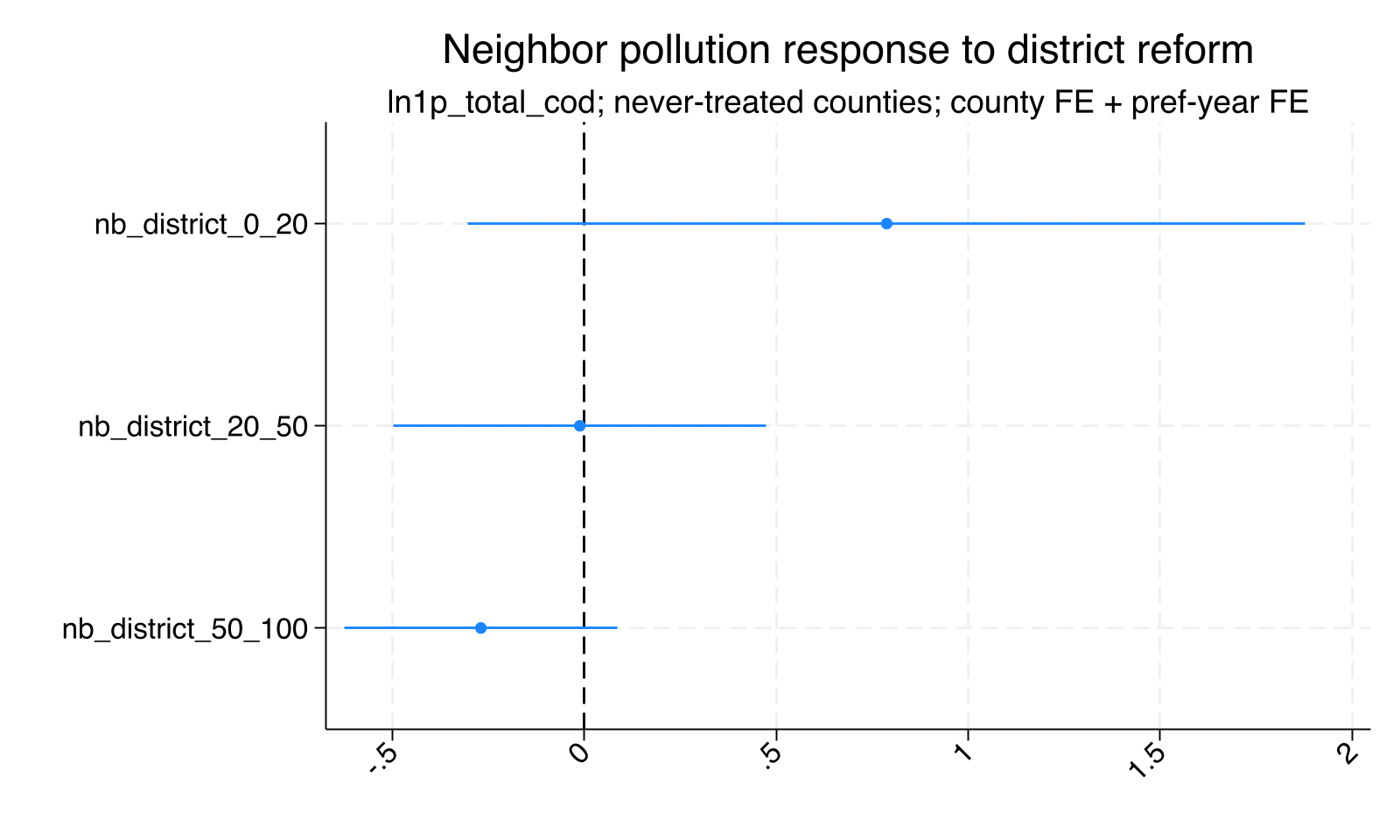}
  \caption{COD}
\end{subfigure}
\end{figure}

\begin{figure}[p]
\caption{County-to-district reform: border-pair event studies with controls}
\label{fig:district_border_ctrl}
\centering
  \begin{subfigure}[b]{0.32\textwidth}
    \includegraphics[width=\textwidth]{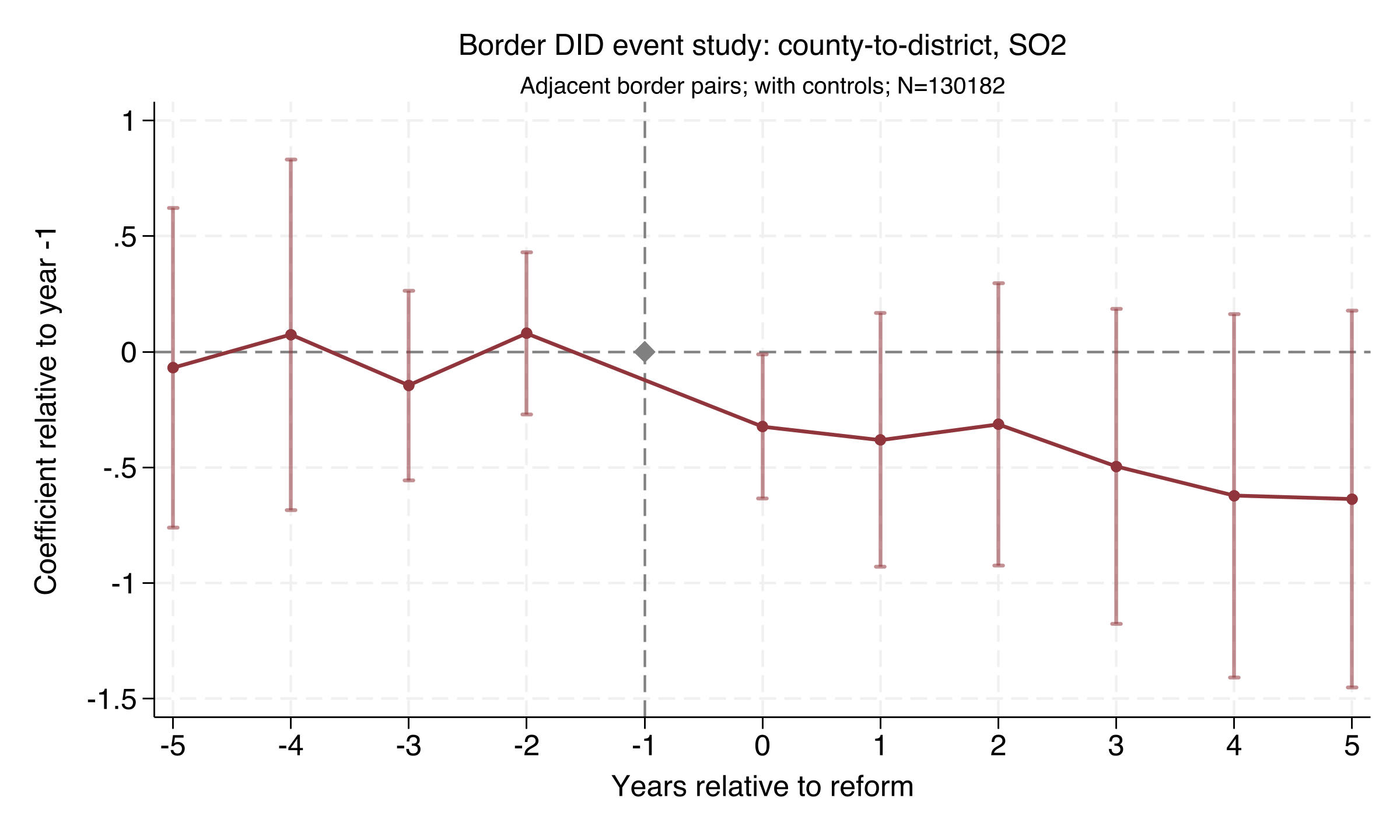}
    \caption{SO$_2$}
  \end{subfigure}\hfill
  \begin{subfigure}[b]{0.32\textwidth}
    \includegraphics[width=\textwidth]{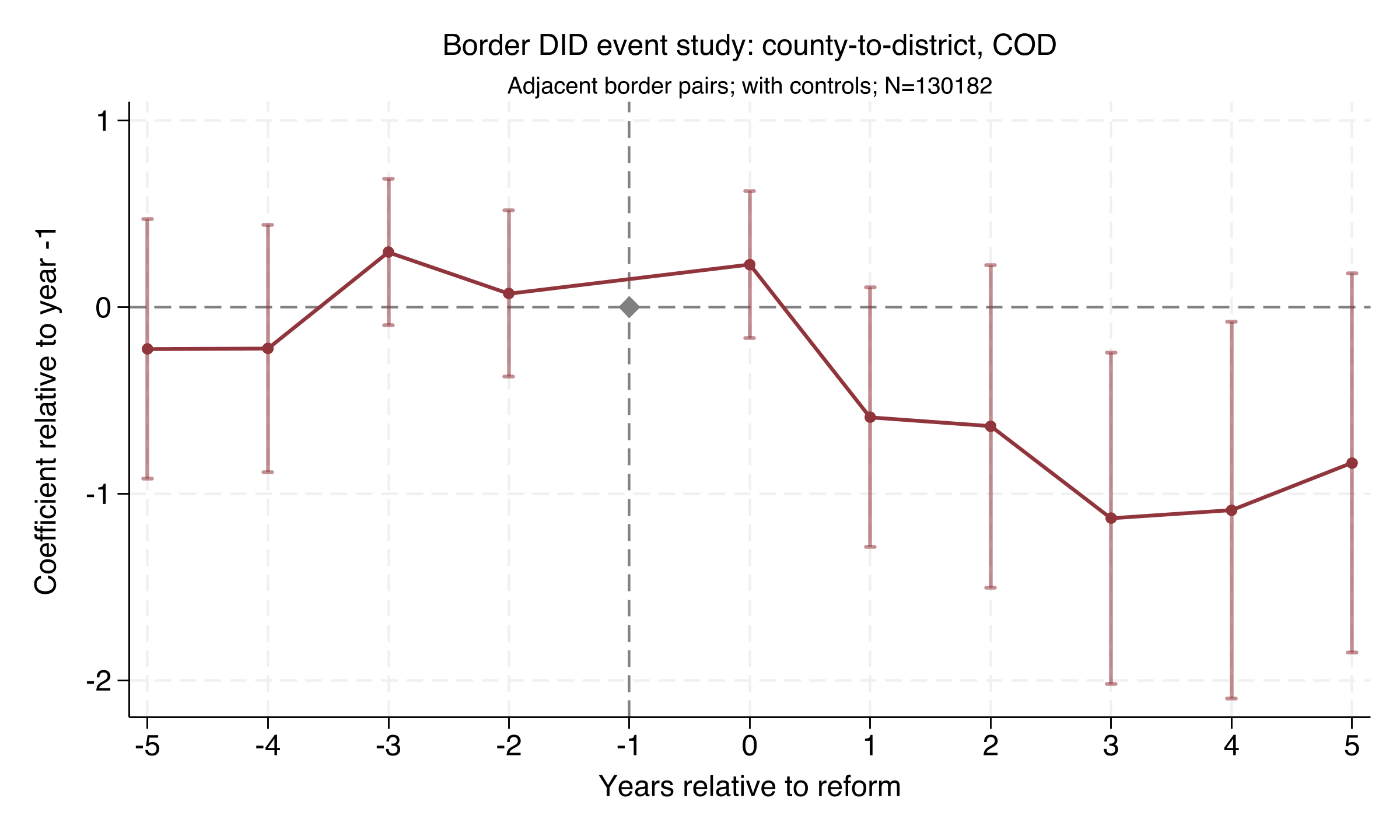}
    \caption{COD}
  \end{subfigure}\hfill
  \begin{subfigure}[b]{0.32\textwidth}
    \includegraphics[width=\textwidth]{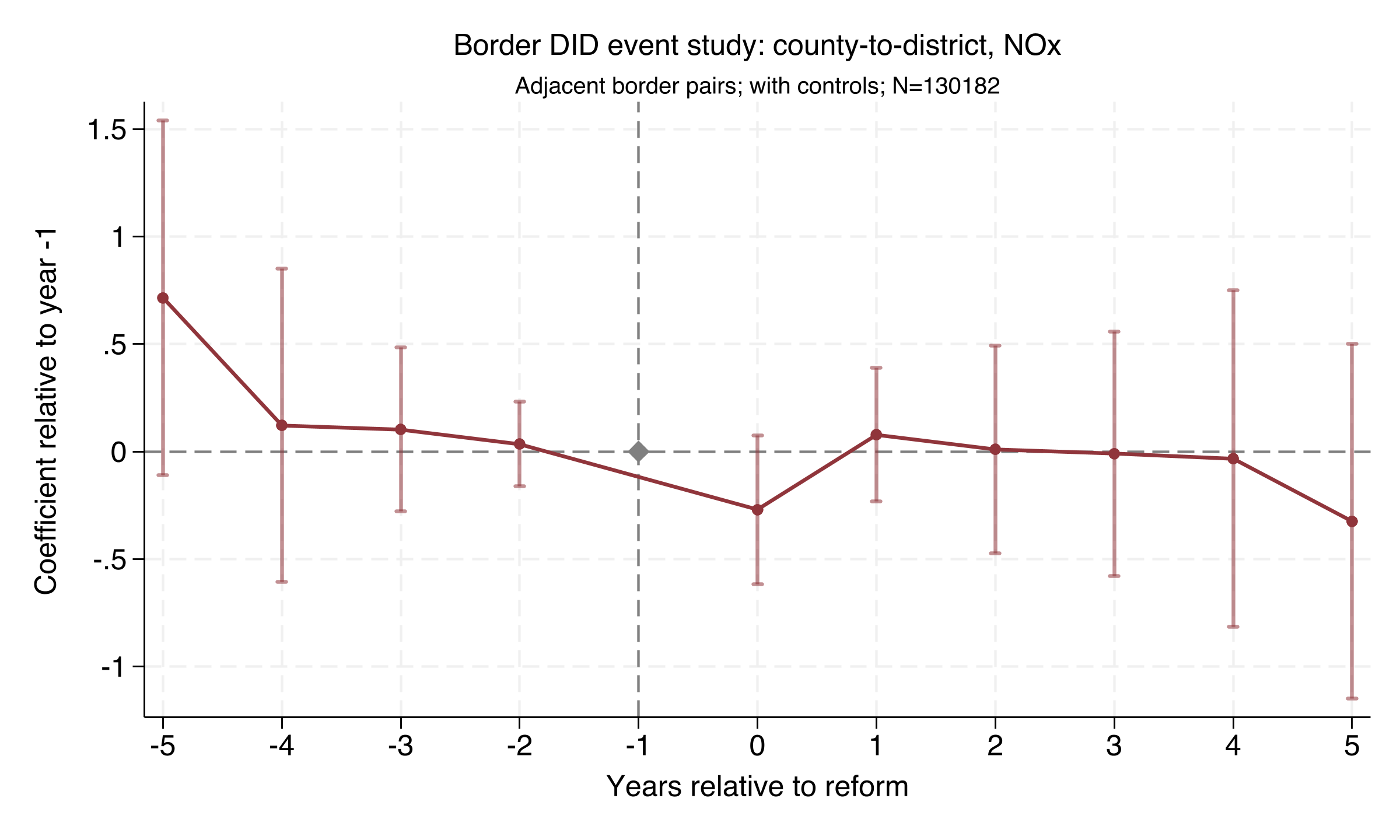}
    \caption{NOx}
  \end{subfigure}\hfill
  \begin{subfigure}[b]{0.32\textwidth}
    \includegraphics[width=\textwidth]{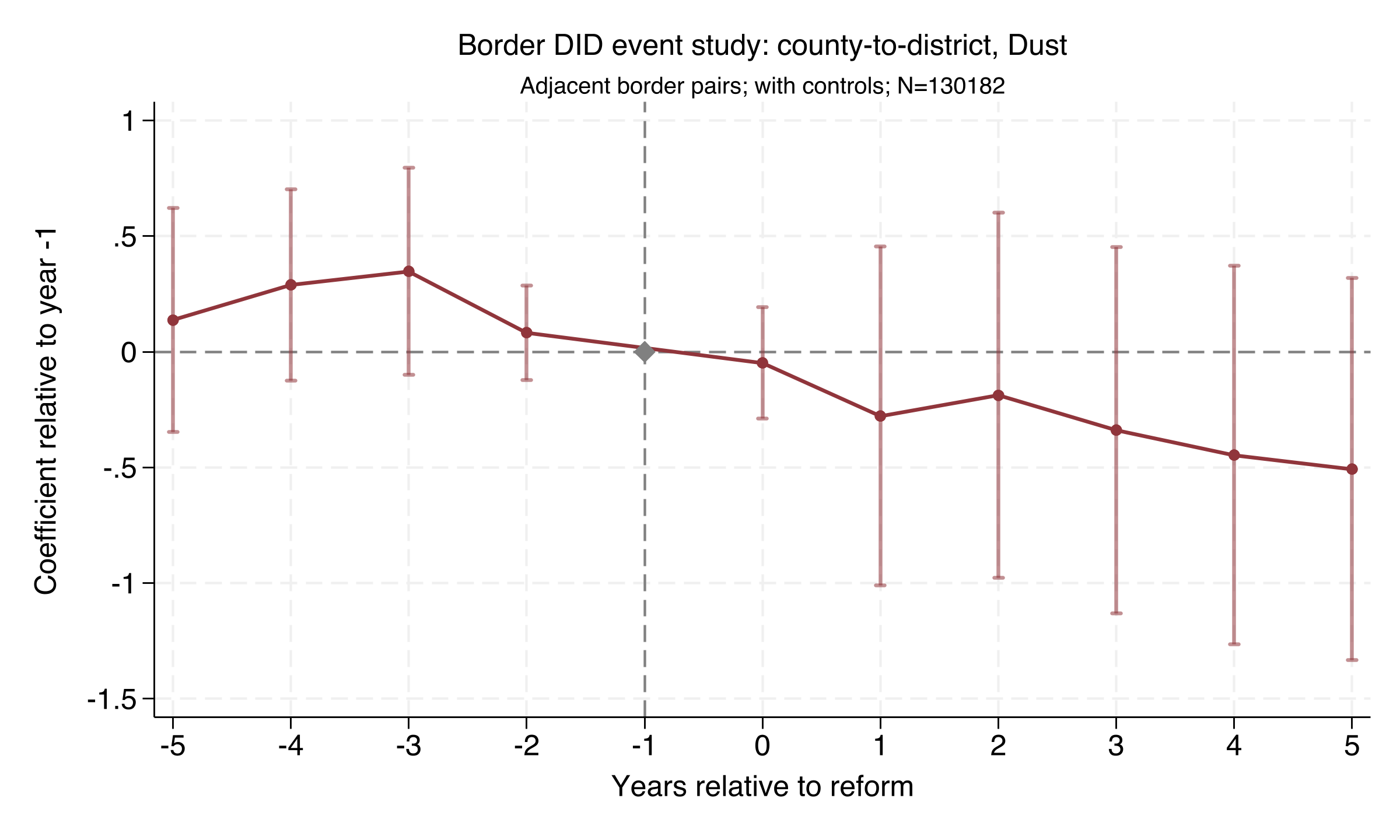}
    \caption{Dust}
  \end{subfigure}\hfill
  \begin{subfigure}[b]{0.32\textwidth}
    \includegraphics[width=\textwidth]{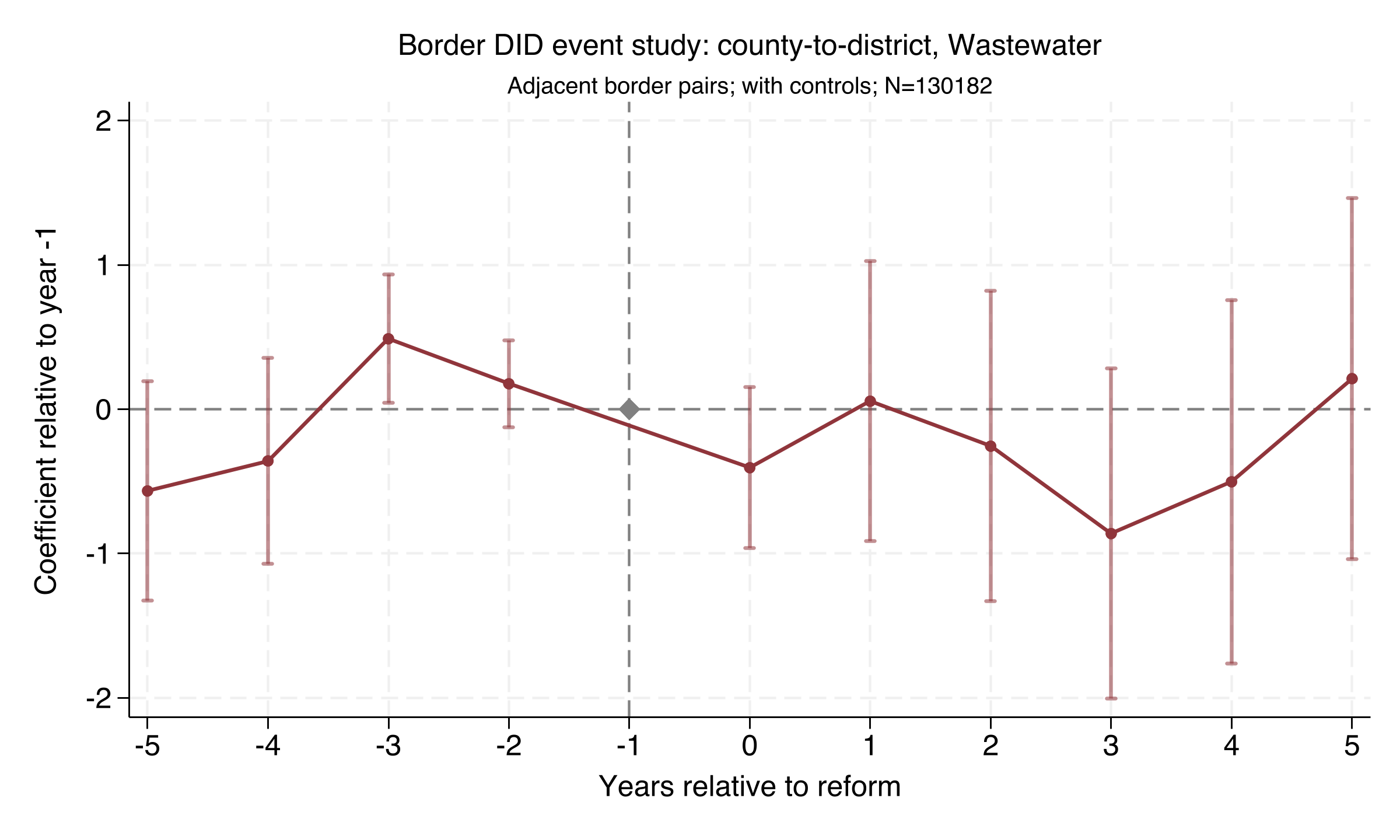}
    \caption{Wastewater}
  \end{subfigure}\hfill
  \begin{subfigure}[b]{0.32\textwidth}
    \includegraphics[width=\textwidth]{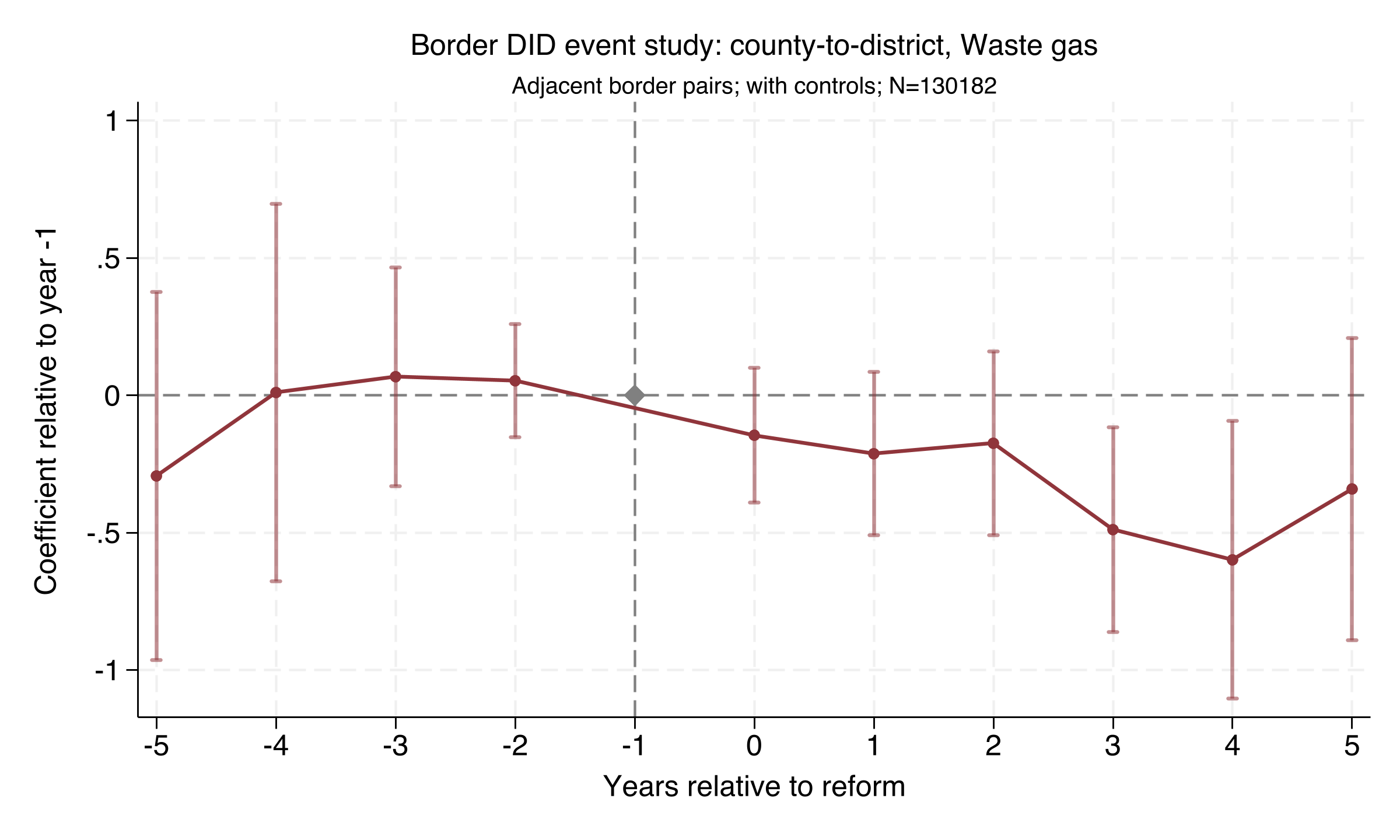}
    \caption{Waste~gas}
  \end{subfigure}\hfill
\end{figure}

\begin{figure}[p]
\caption{County-to-city reform: border-pair event studies with controls}
\label{fig:countycity_border_ctrl}
\centering
  \begin{subfigure}[b]{0.32\textwidth}
    \includegraphics[width=\textwidth]{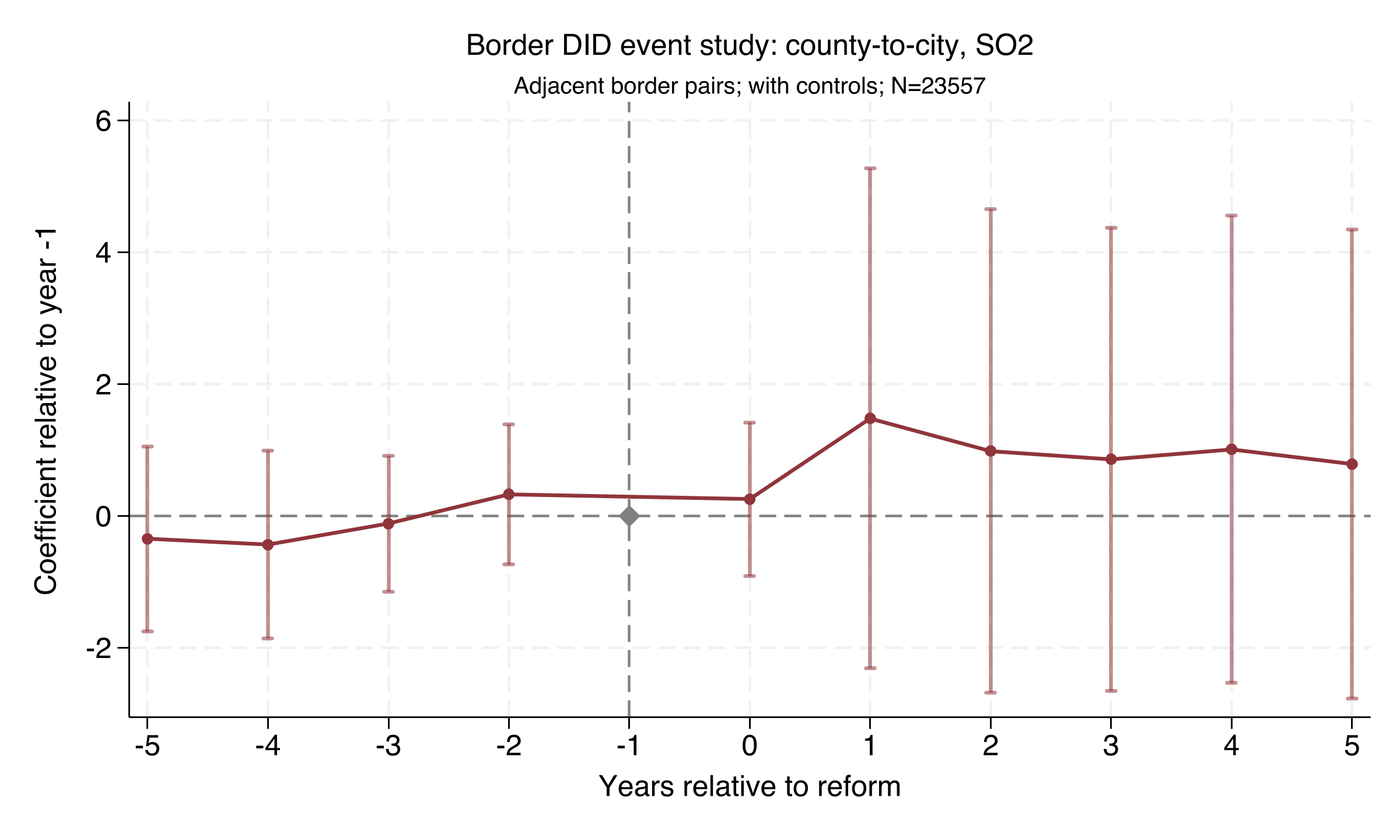}
    \caption{SO$_2$}
  \end{subfigure}\hfill
  \begin{subfigure}[b]{0.32\textwidth}
    \includegraphics[width=\textwidth]{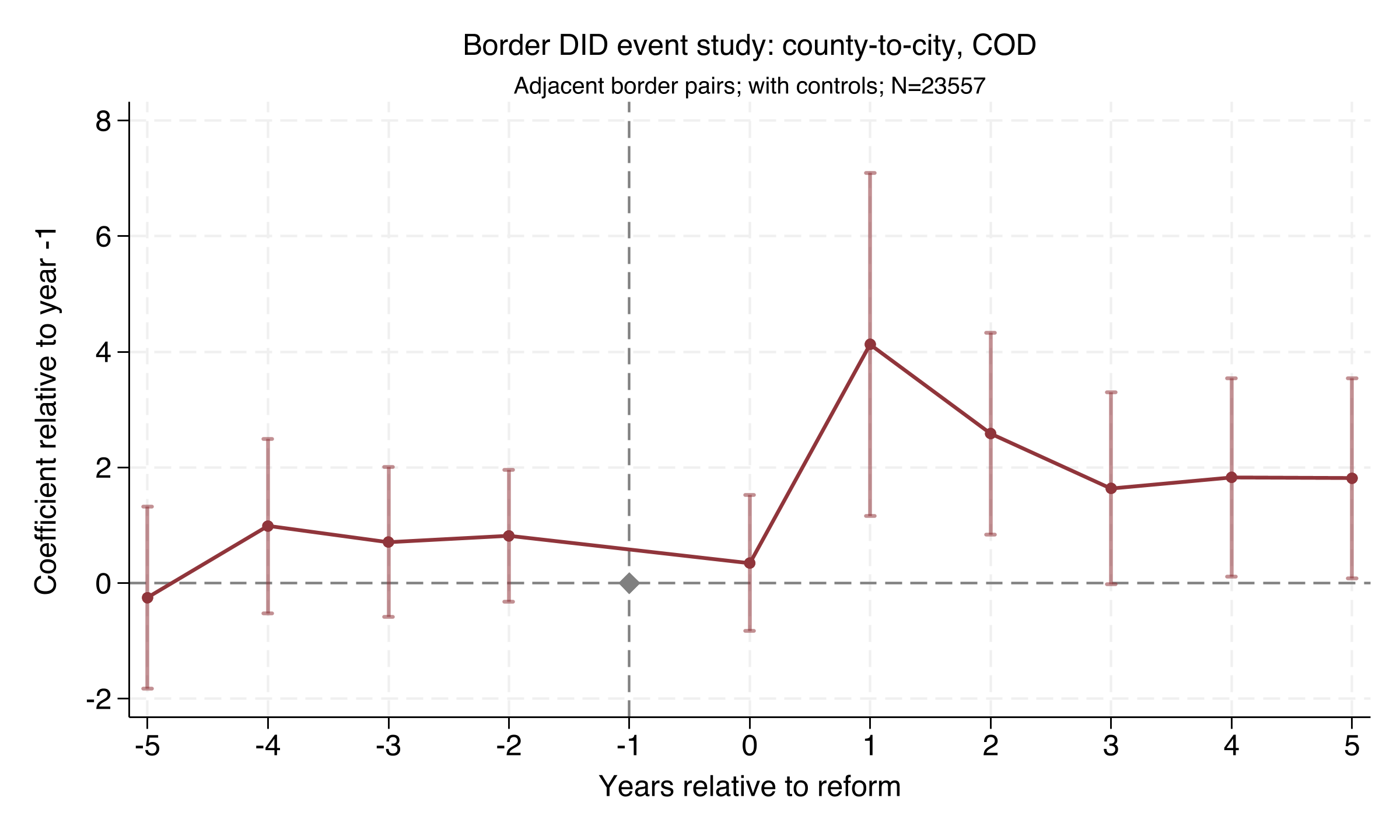}
    \caption{COD}
  \end{subfigure}\hfill
  \begin{subfigure}[b]{0.32\textwidth}
    \includegraphics[width=\textwidth]{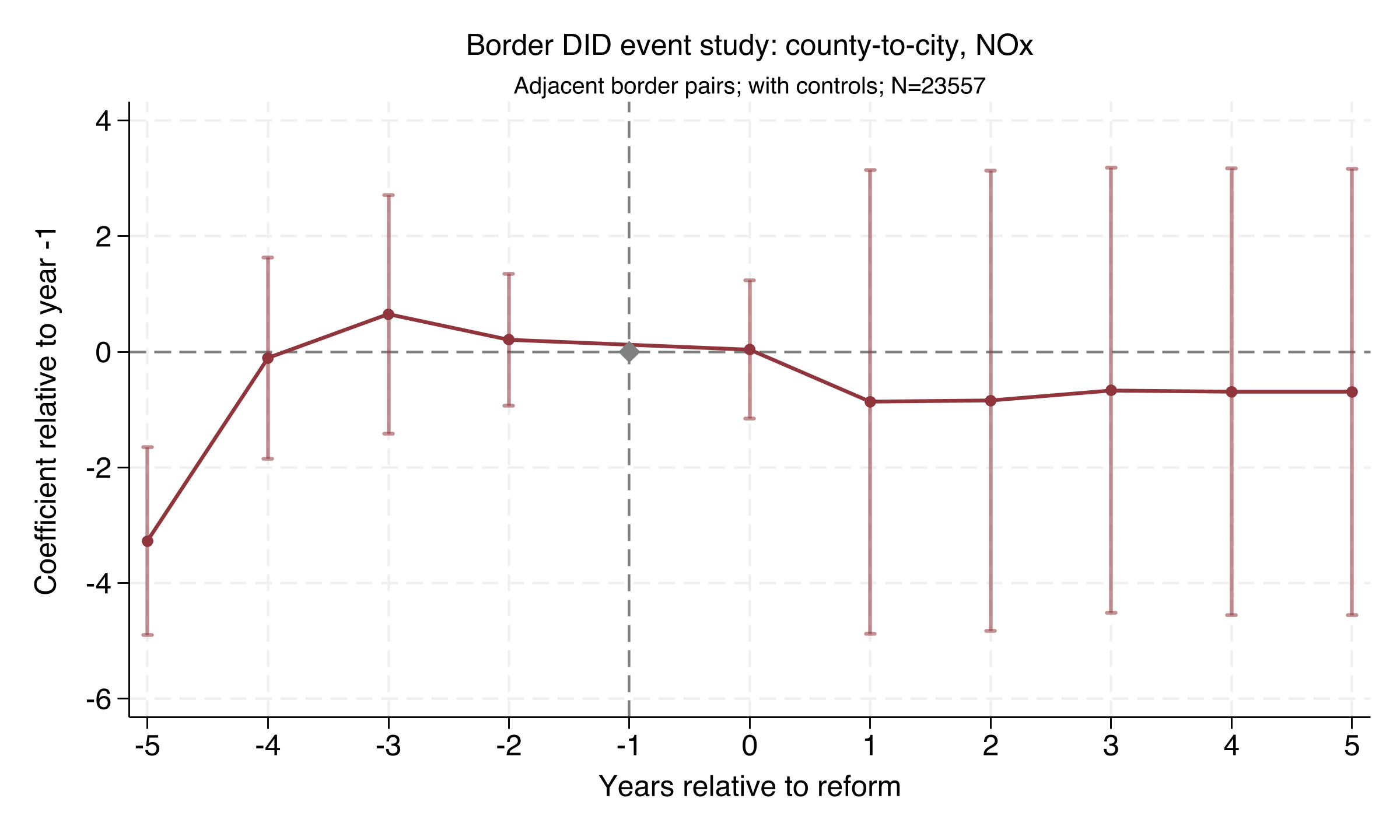}
    \caption{NOx}
  \end{subfigure}\hfill
  \begin{subfigure}[b]{0.32\textwidth}
    \includegraphics[width=\textwidth]{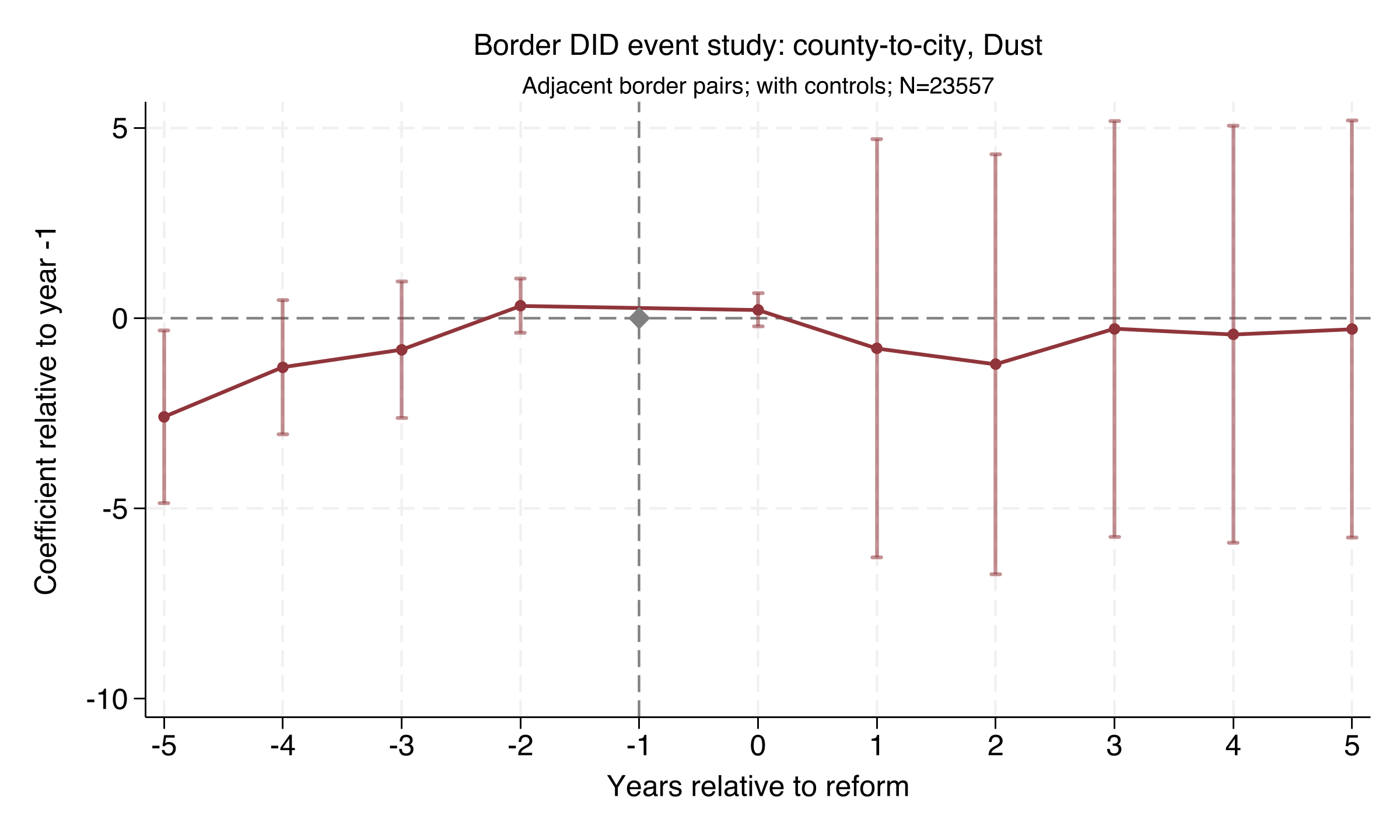}
    \caption{Dust}
  \end{subfigure}\hfill
  \begin{subfigure}[b]{0.32\textwidth}
    \includegraphics[width=\textwidth]{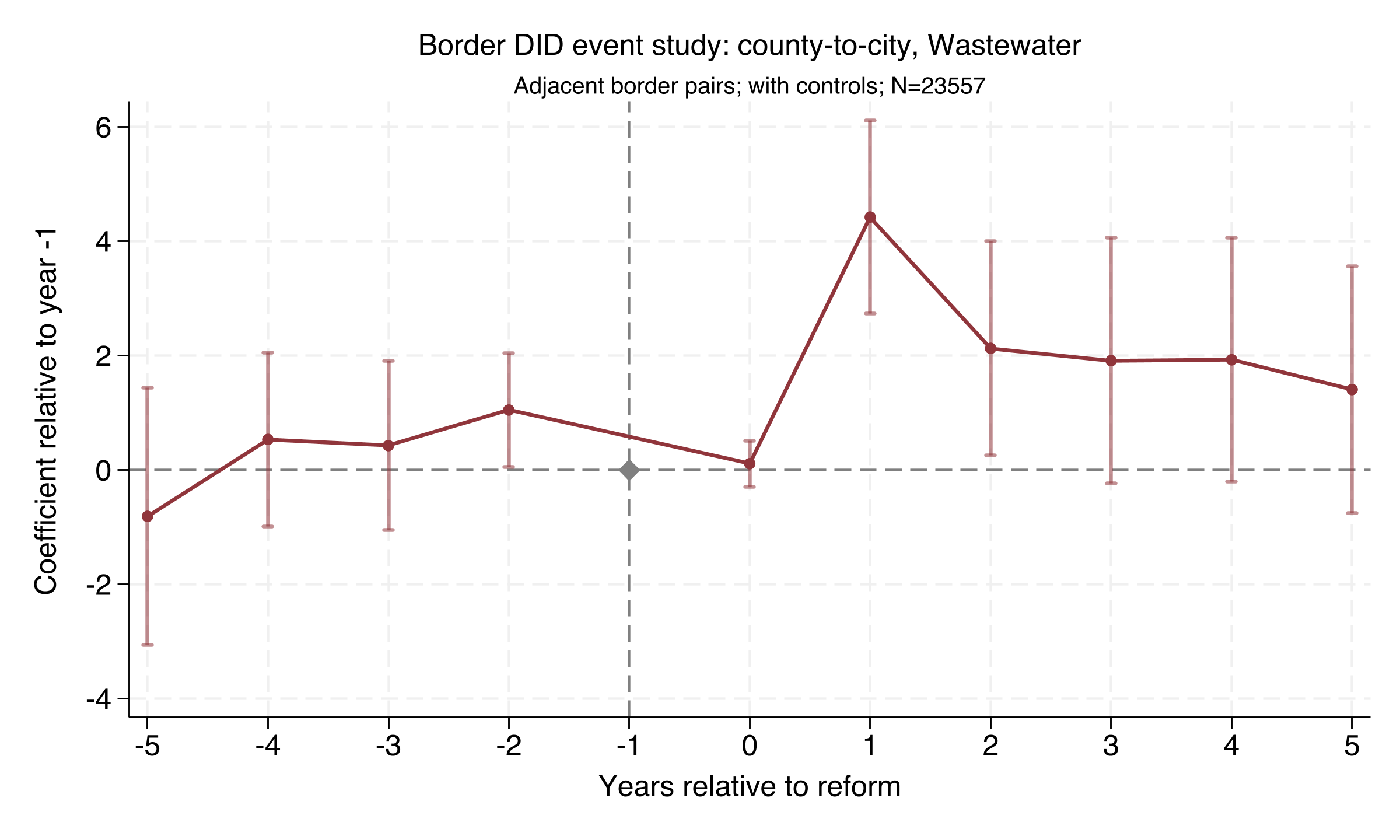}
    \caption{Wastewater}
  \end{subfigure}\hfill
  \begin{subfigure}[b]{0.32\textwidth}
    \includegraphics[width=\textwidth]{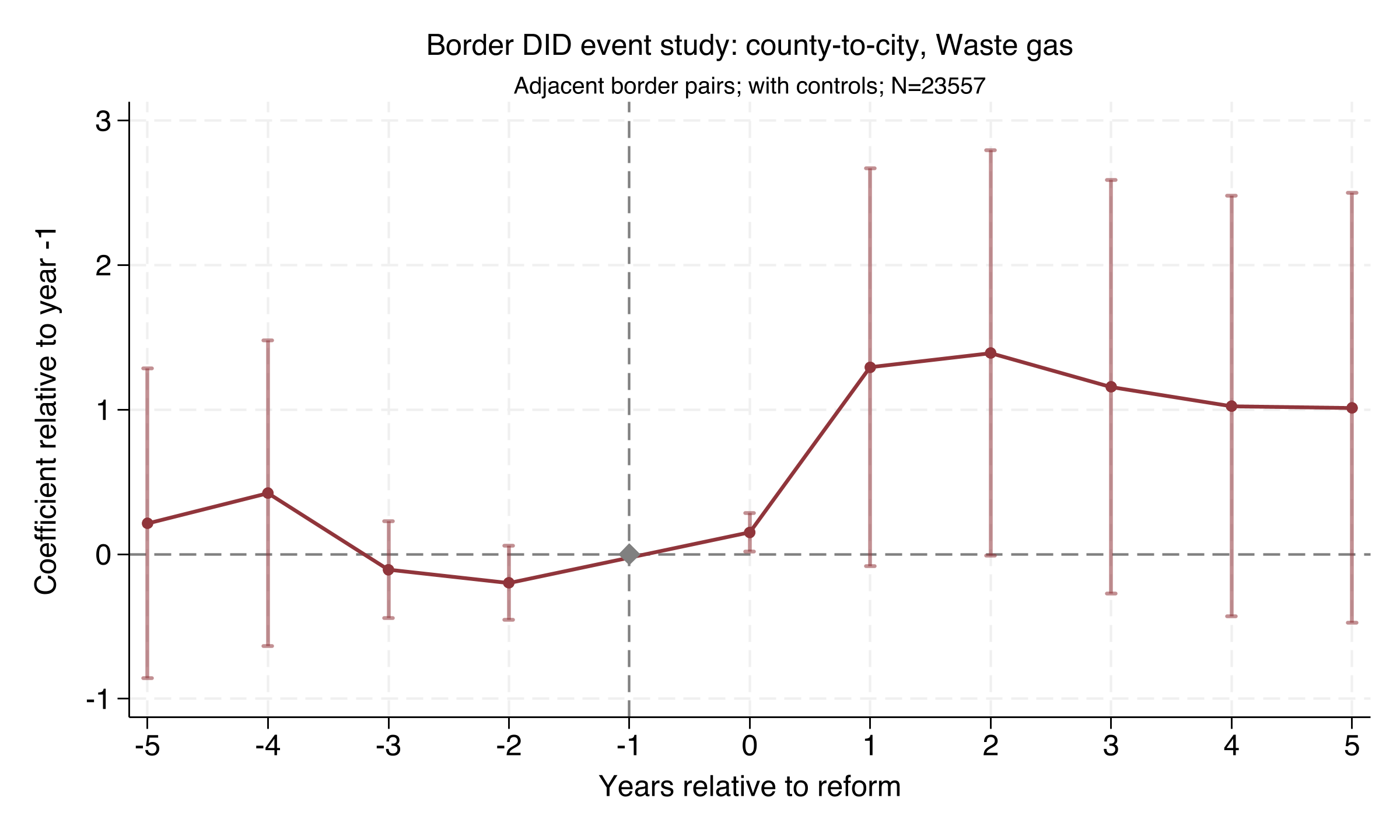}
    \caption{Waste~gas}
  \end{subfigure}\hfill
\end{figure}

\begin{figure}[p]
\caption{County-to-district reform: stacked intensity event studies, per worker}
\label{fig:district_emp}
\centering
  \begin{subfigure}[b]{0.32\textwidth}
    \includegraphics[width=\textwidth]{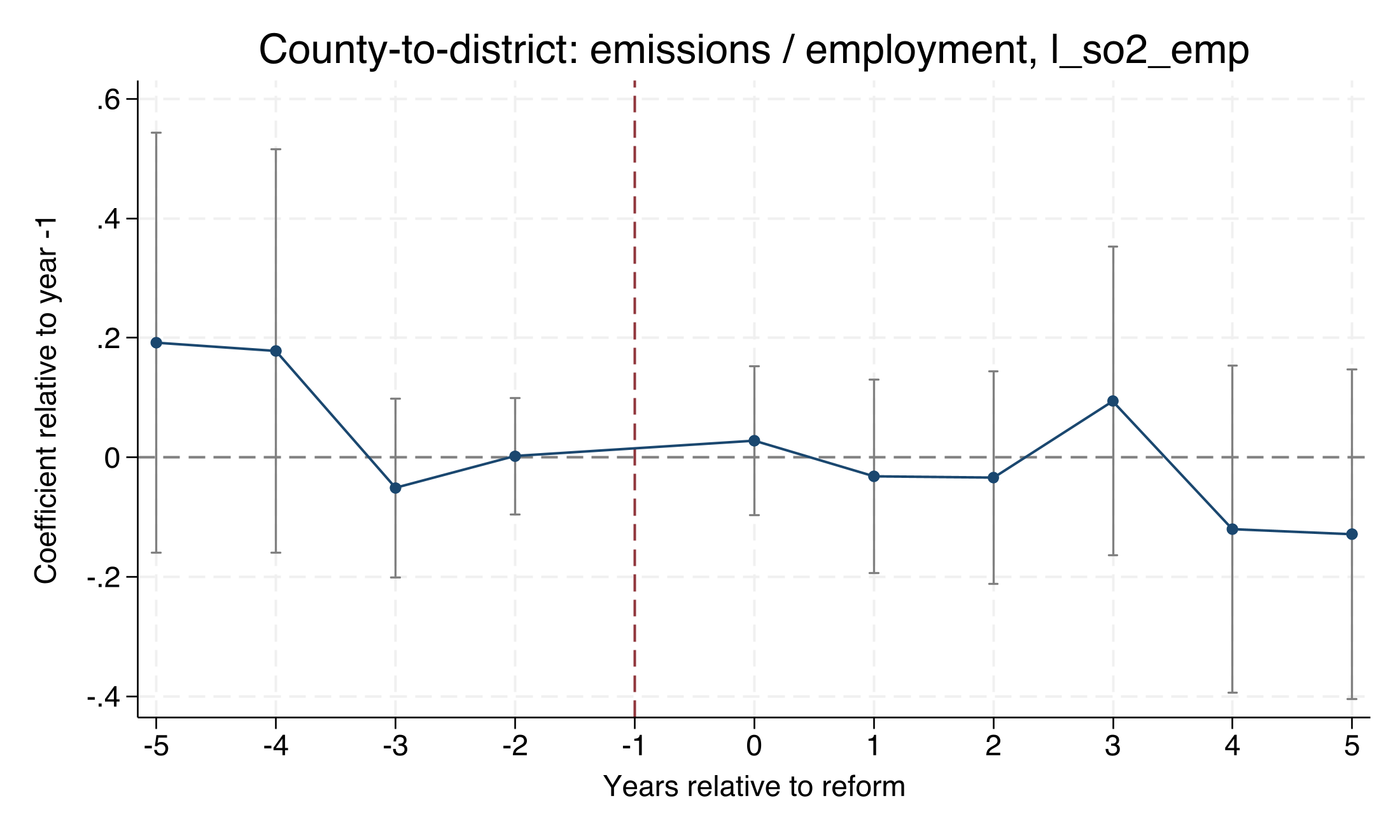}
    \caption{SO$_2$}
  \end{subfigure}\hfill
  \begin{subfigure}[b]{0.32\textwidth}
    \includegraphics[width=\textwidth]{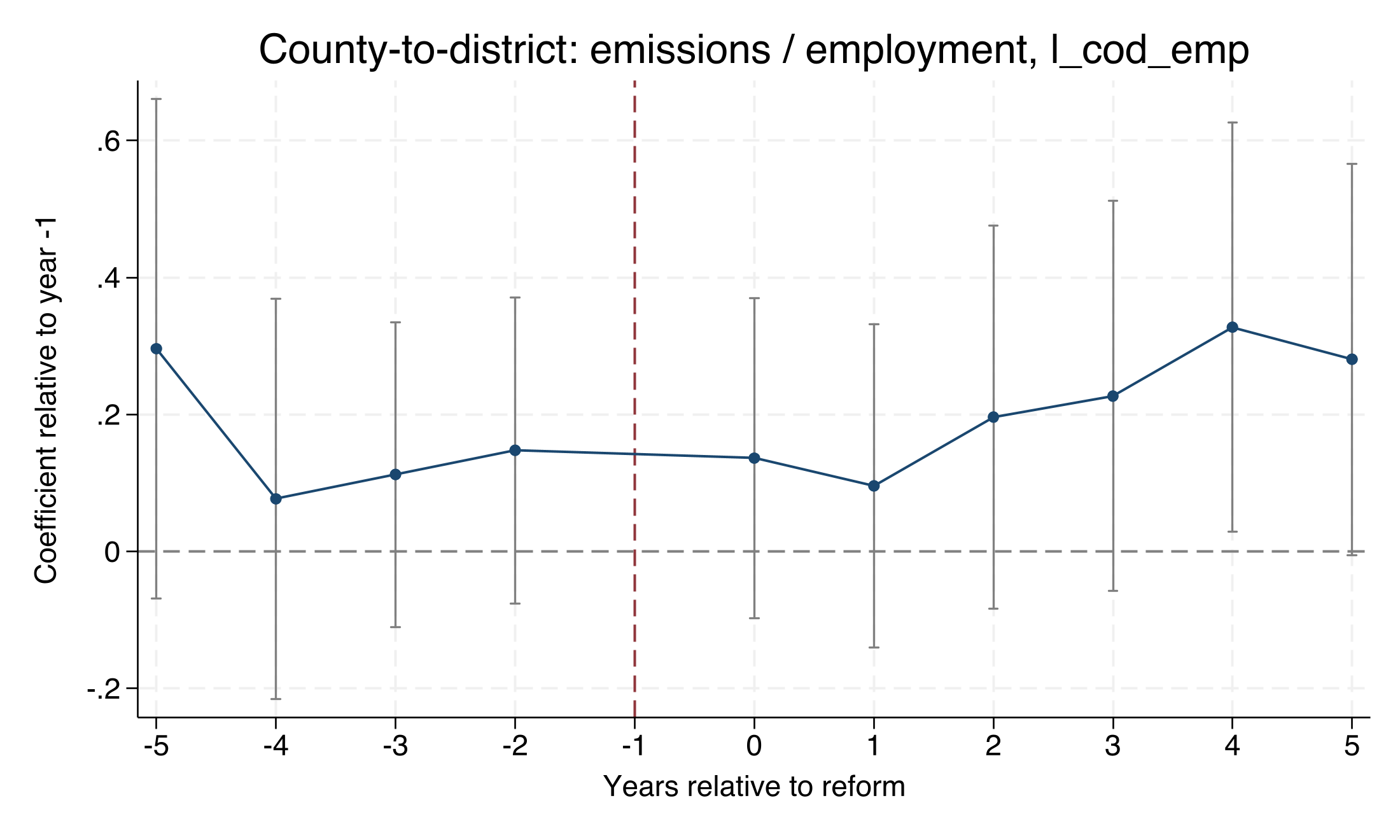}
    \caption{COD}
  \end{subfigure}\hfill
  \begin{subfigure}[b]{0.32\textwidth}
    \includegraphics[width=\textwidth]{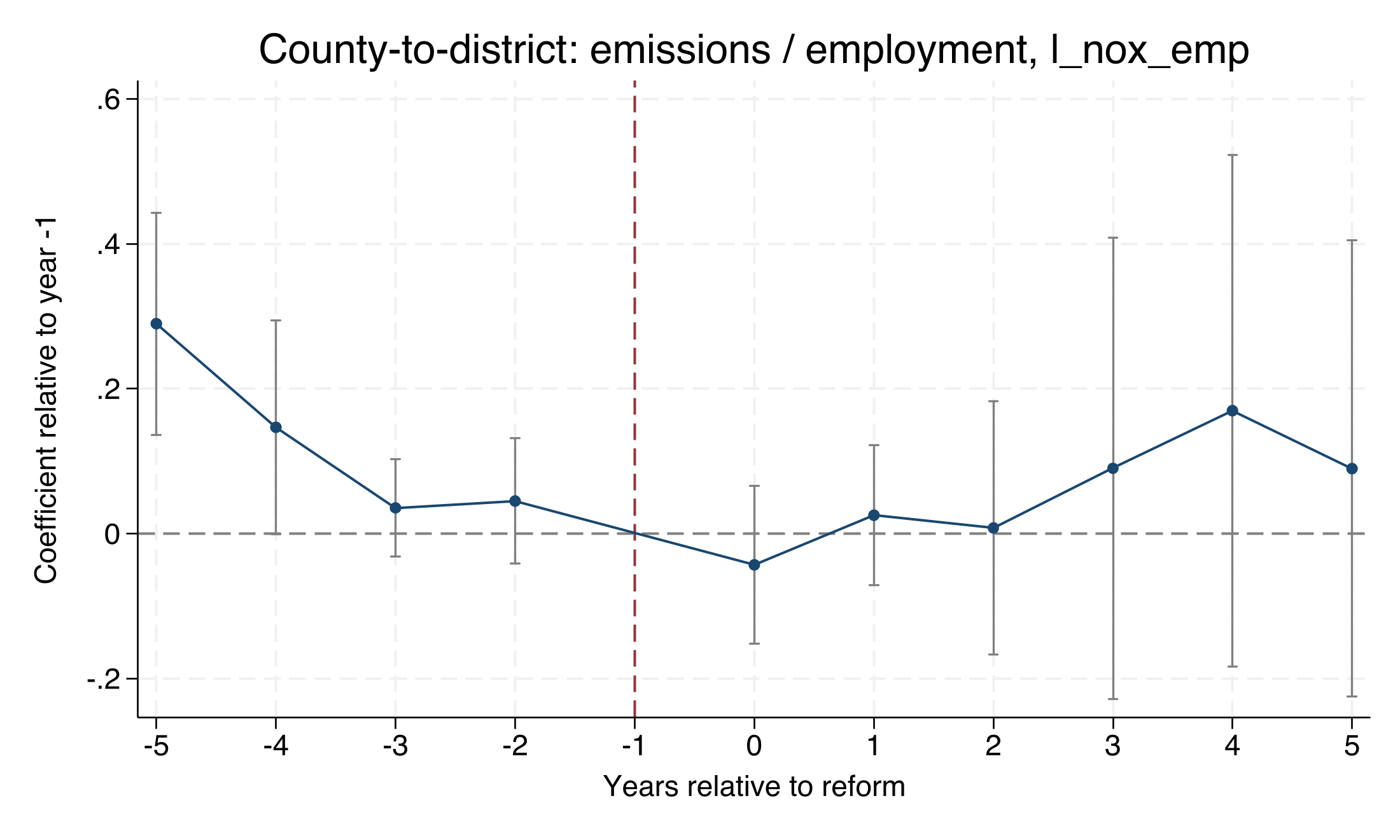}
    \caption{NOx}
  \end{subfigure}\hfill
  \begin{subfigure}[b]{0.32\textwidth}
    \includegraphics[width=\textwidth]{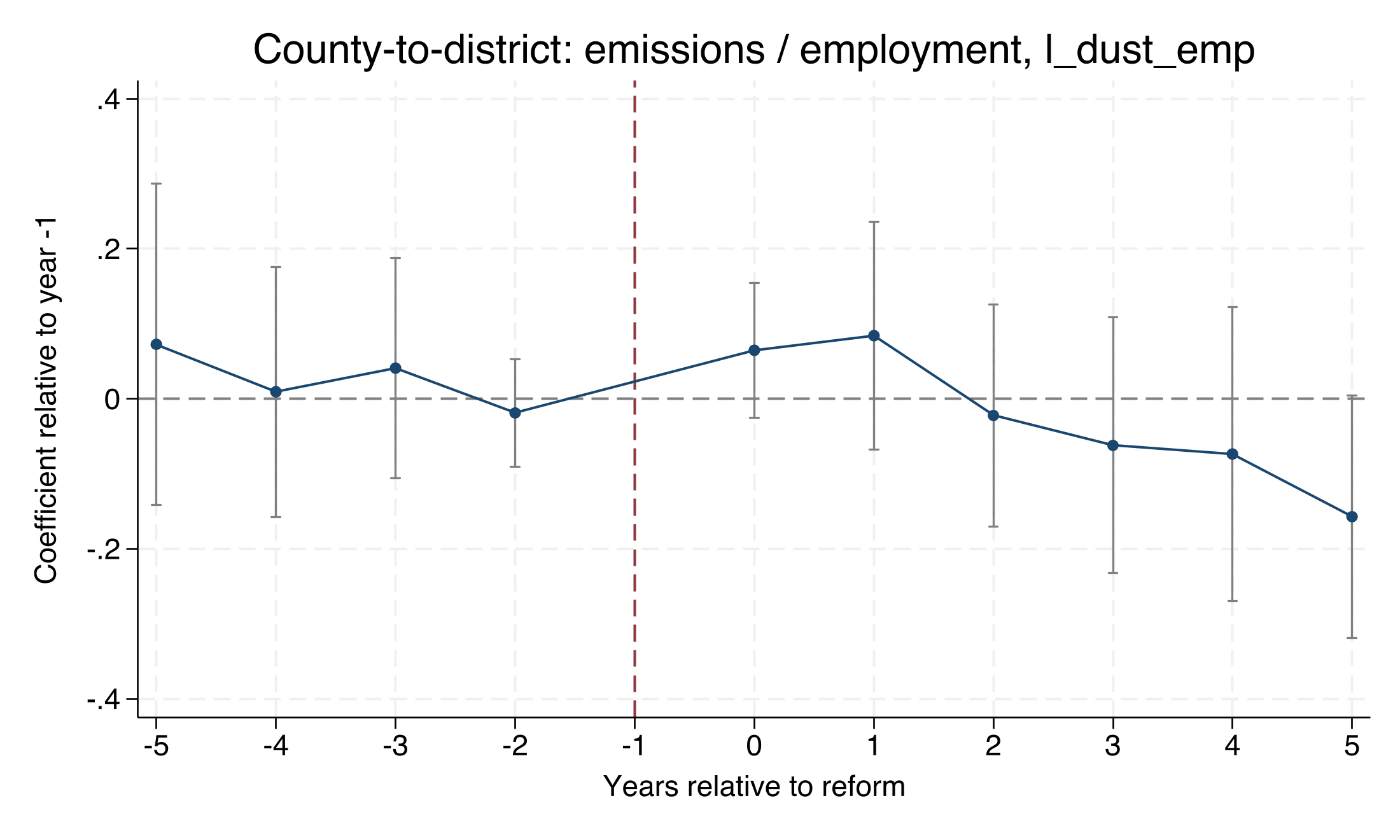}
    \caption{Dust}
  \end{subfigure}\hfill
  \begin{subfigure}[b]{0.32\textwidth}
    \includegraphics[width=\textwidth]{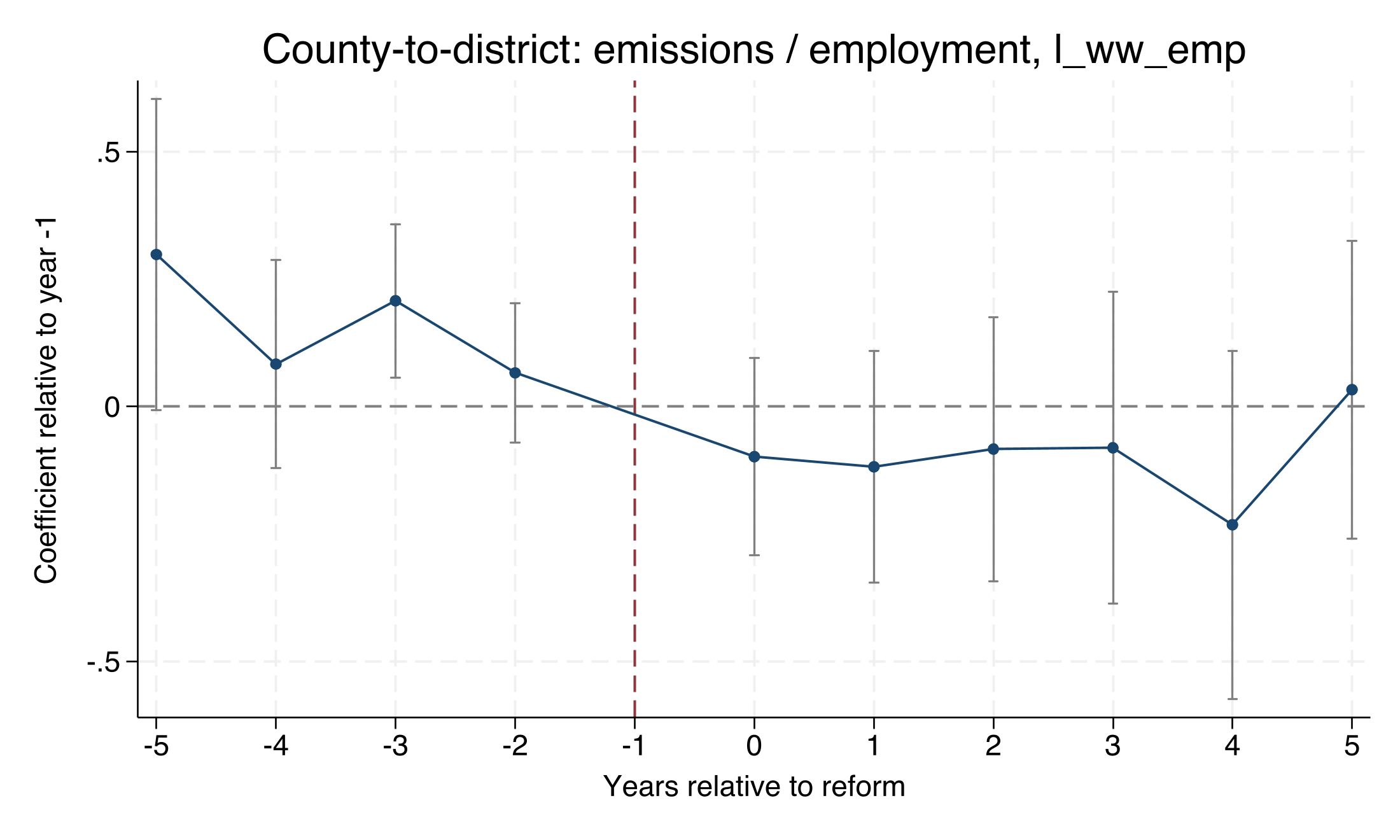}
    \caption{Wastewater}
  \end{subfigure}\hfill
  \begin{subfigure}[b]{0.32\textwidth}
    \includegraphics[width=\textwidth]{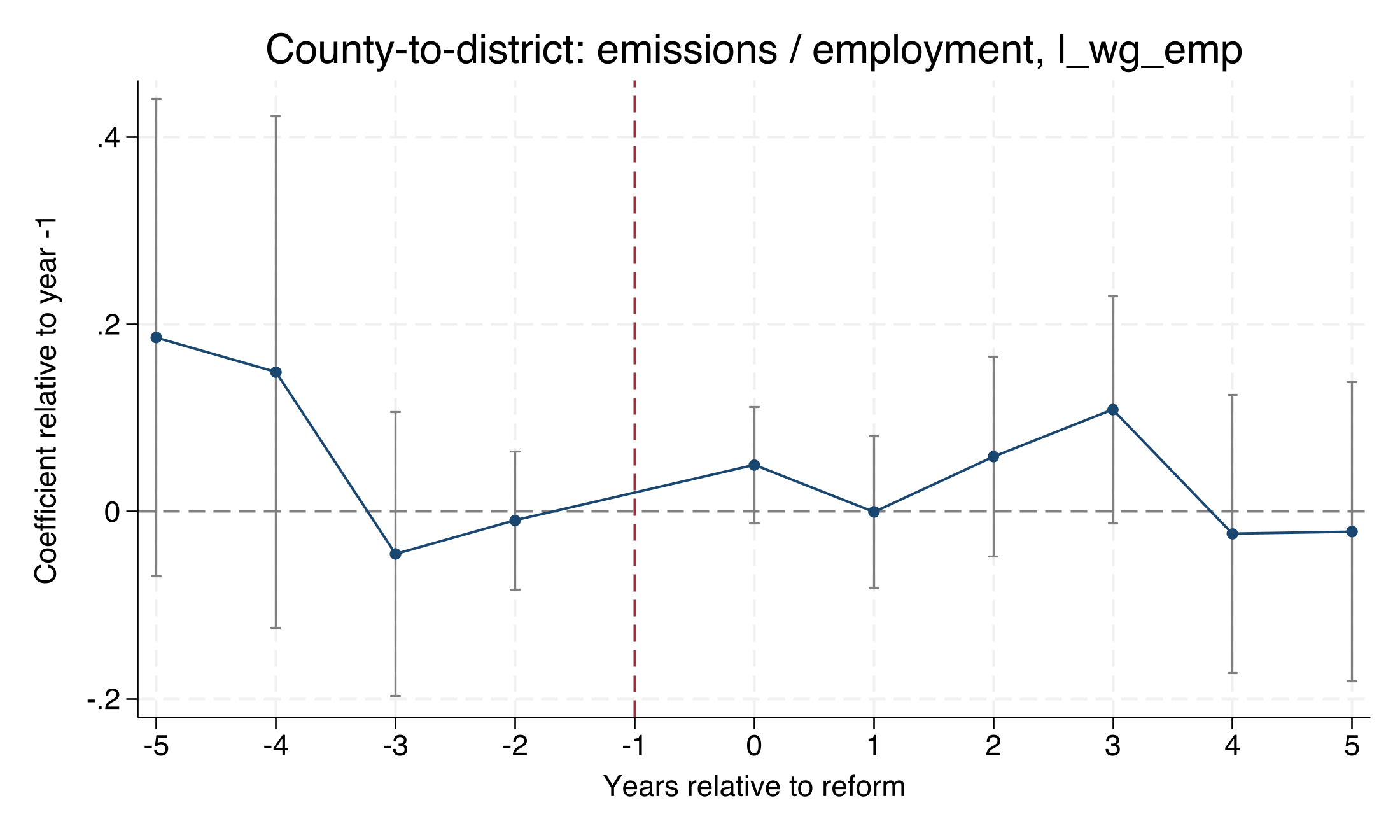}
    \caption{Waste~gas}
  \end{subfigure}\hfill
\end{figure}

\begin{figure}[p]
\caption{County-to-district reform: stacked intensity event studies, per sales}
\label{fig:district_sales}
\centering
  \begin{subfigure}[b]{0.32\textwidth}
    \includegraphics[width=\textwidth]{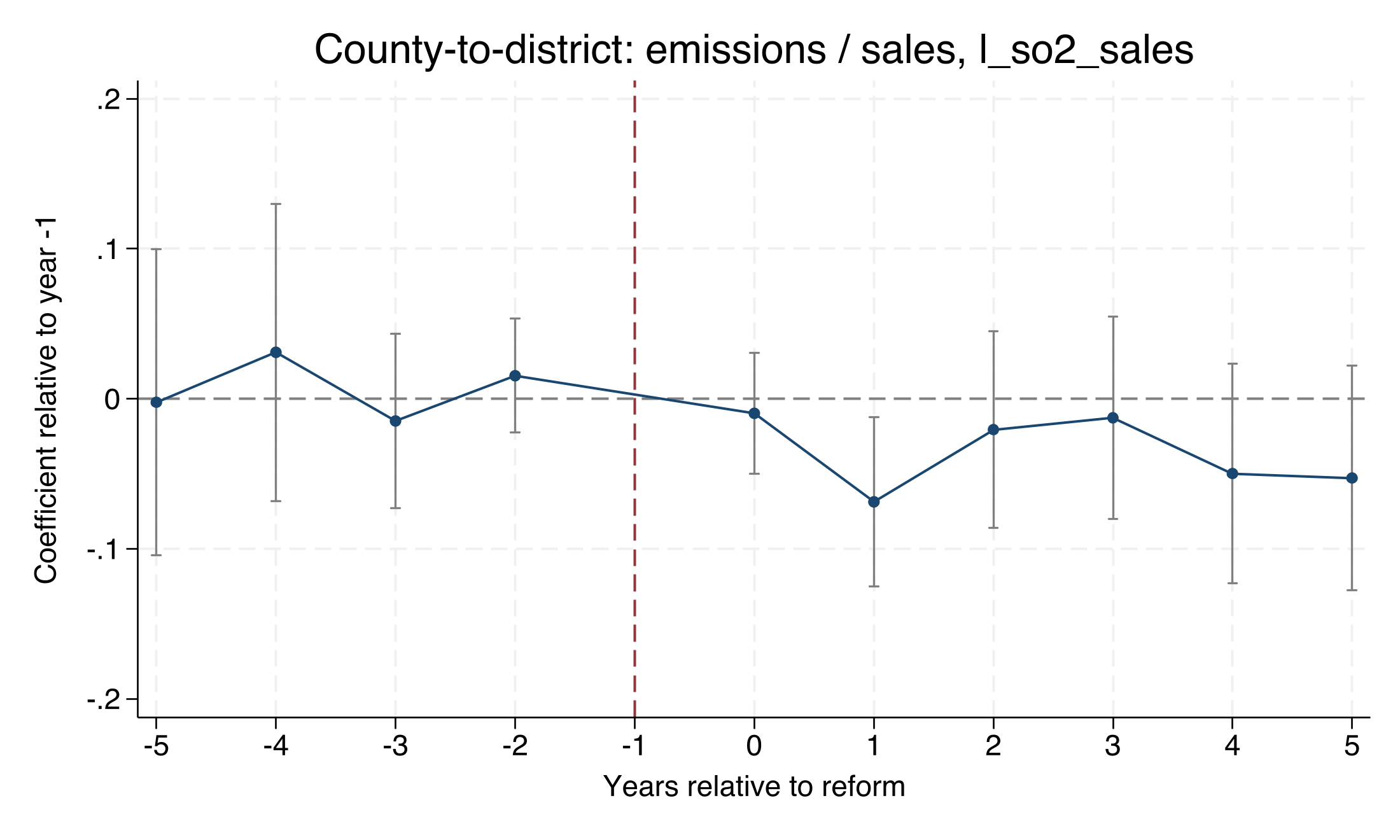}
    \caption{SO$_2$}
  \end{subfigure}\hfill
  \begin{subfigure}[b]{0.32\textwidth}
    \includegraphics[width=\textwidth]{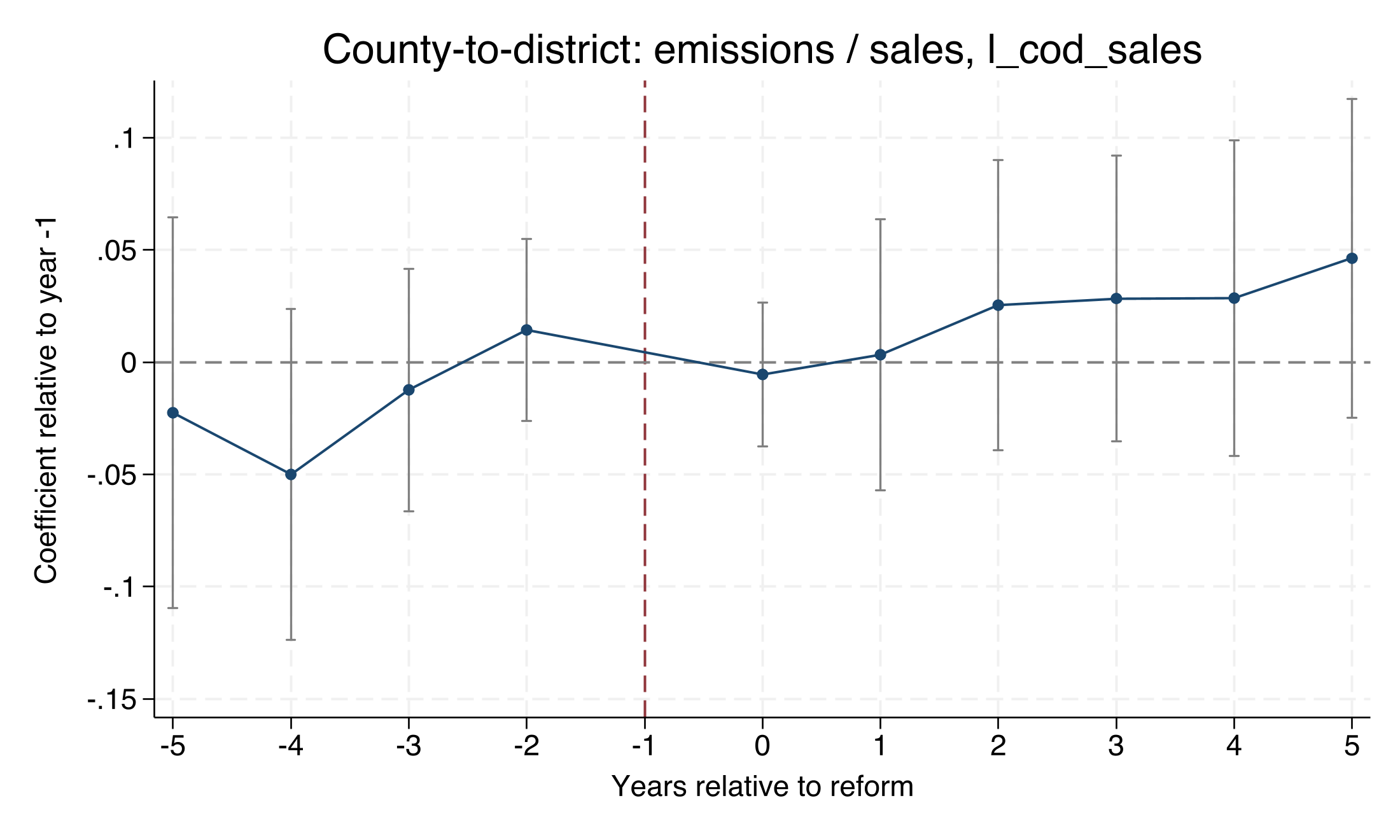}
    \caption{COD}
  \end{subfigure}\hfill
  \begin{subfigure}[b]{0.32\textwidth}
    \includegraphics[width=\textwidth]{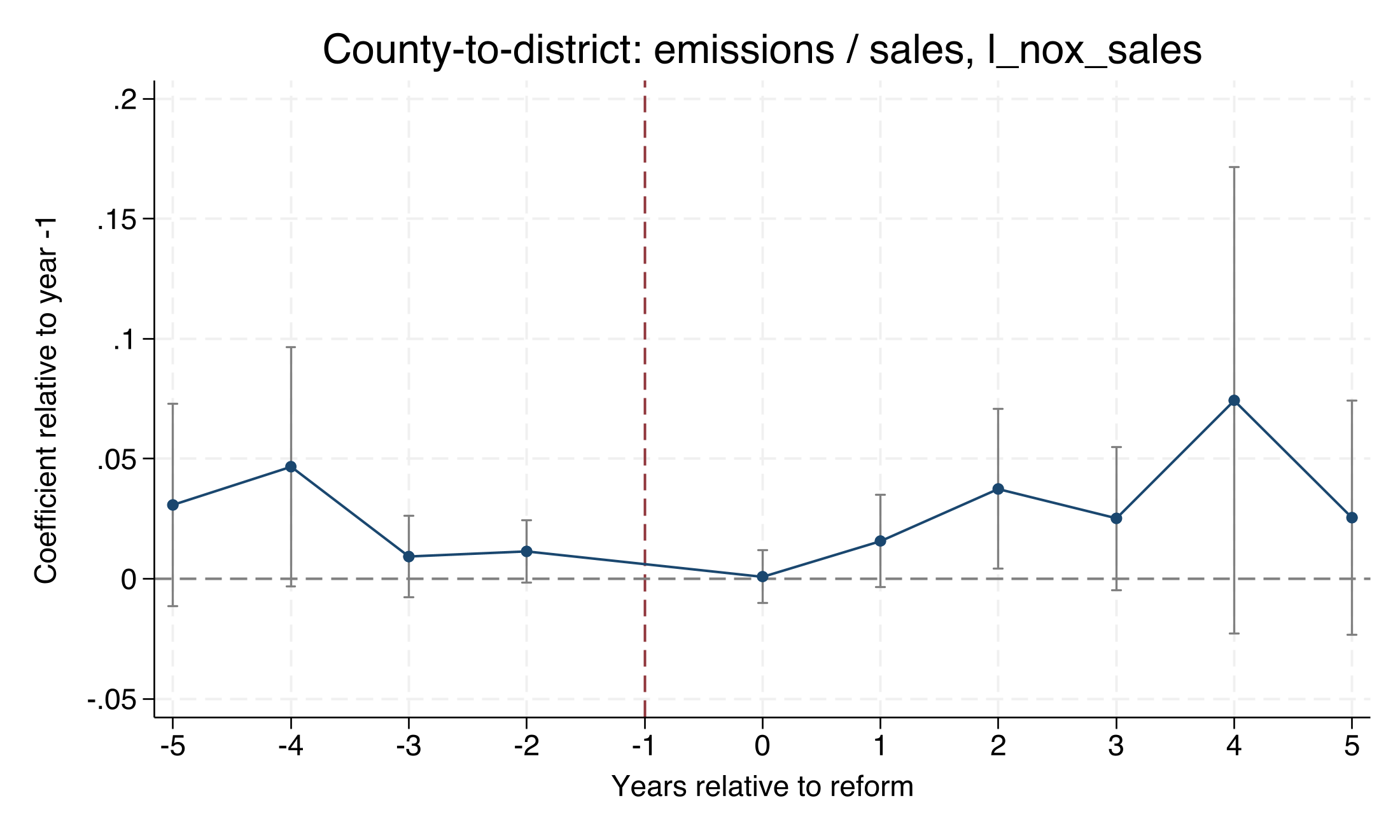}
    \caption{NOx}
  \end{subfigure}\hfill
  \begin{subfigure}[b]{0.32\textwidth}
    \includegraphics[width=\textwidth]{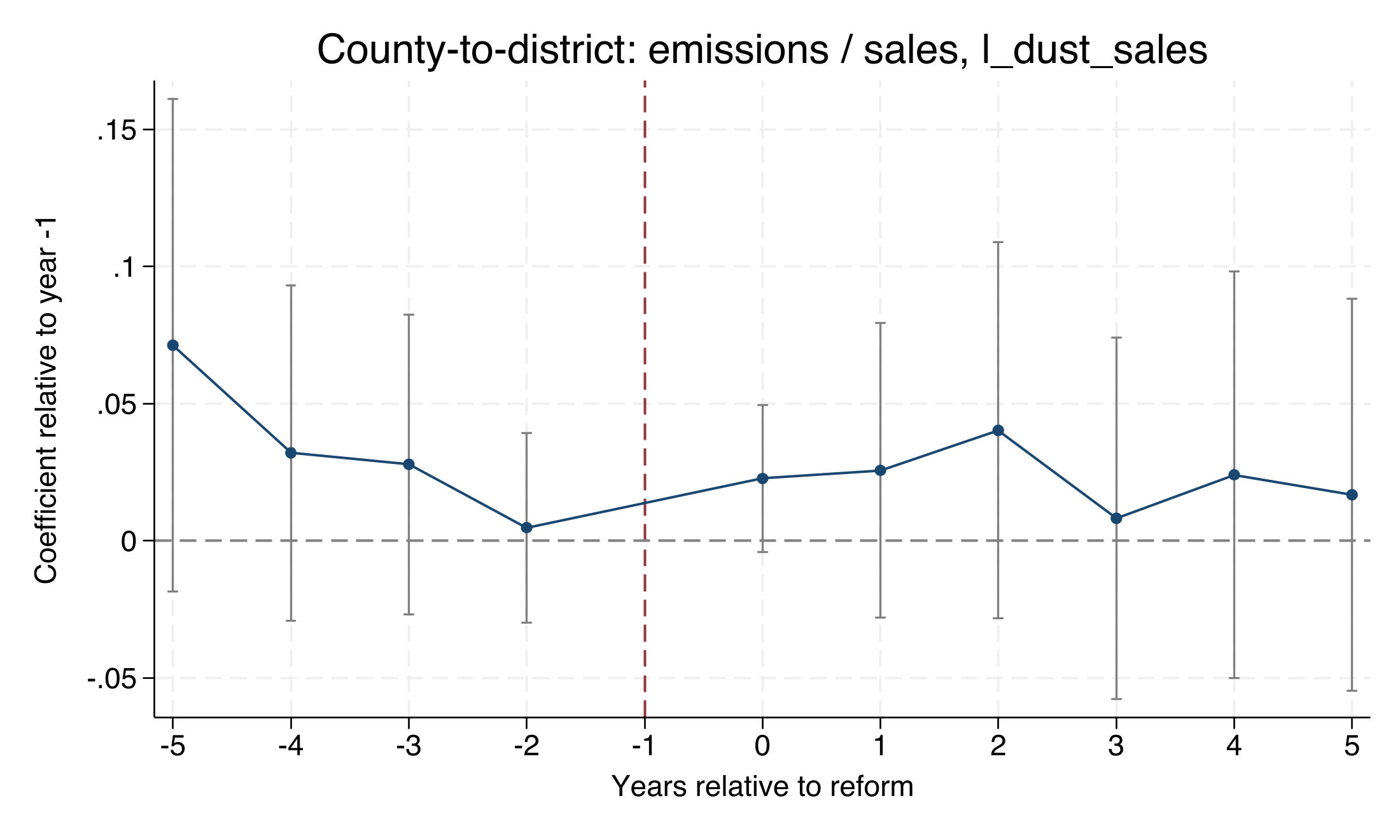}
    \caption{Dust}
  \end{subfigure}\hfill
  \begin{subfigure}[b]{0.32\textwidth}
    \includegraphics[width=\textwidth]{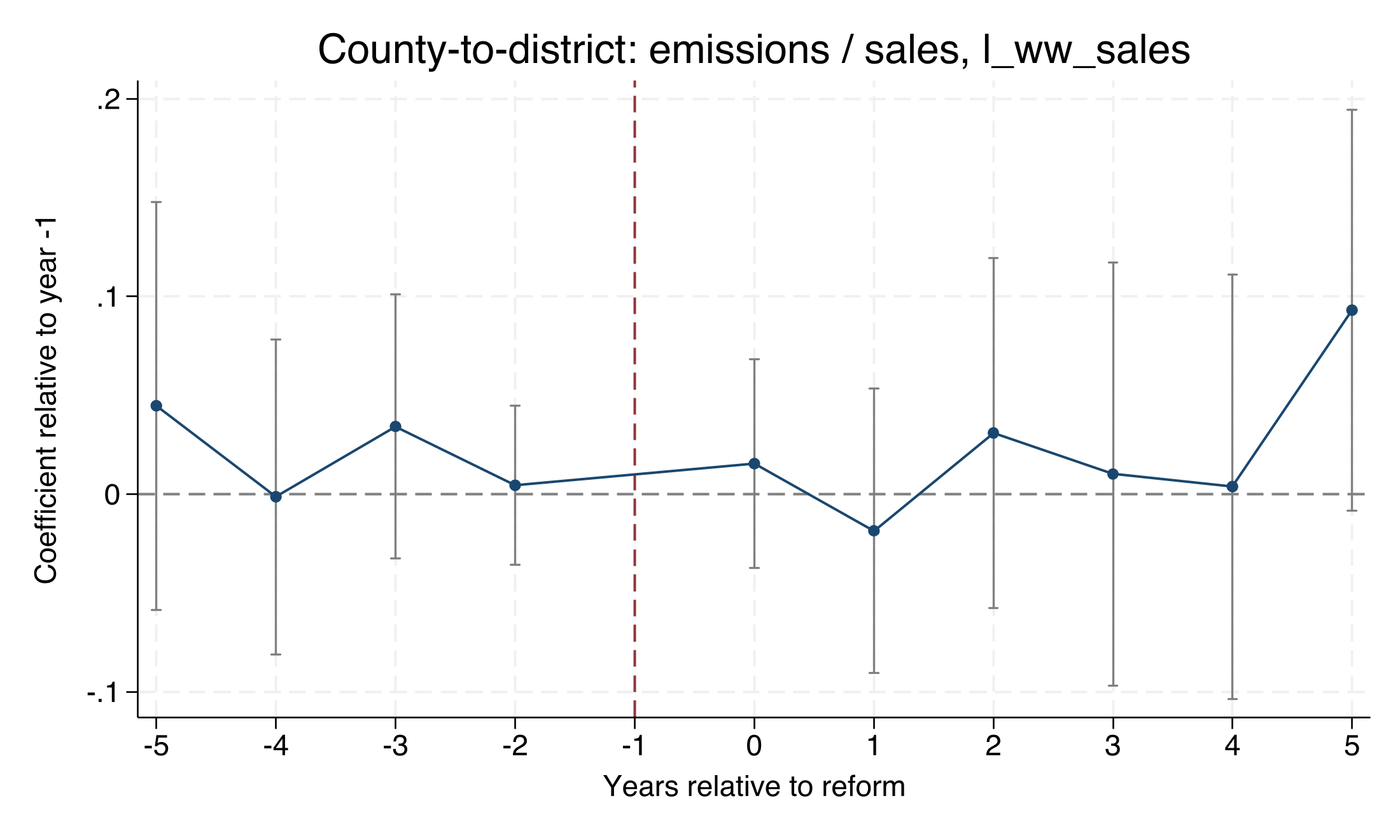}
    \caption{Wastewater}
  \end{subfigure}\hfill
  \begin{subfigure}[b]{0.32\textwidth}
    \includegraphics[width=\textwidth]{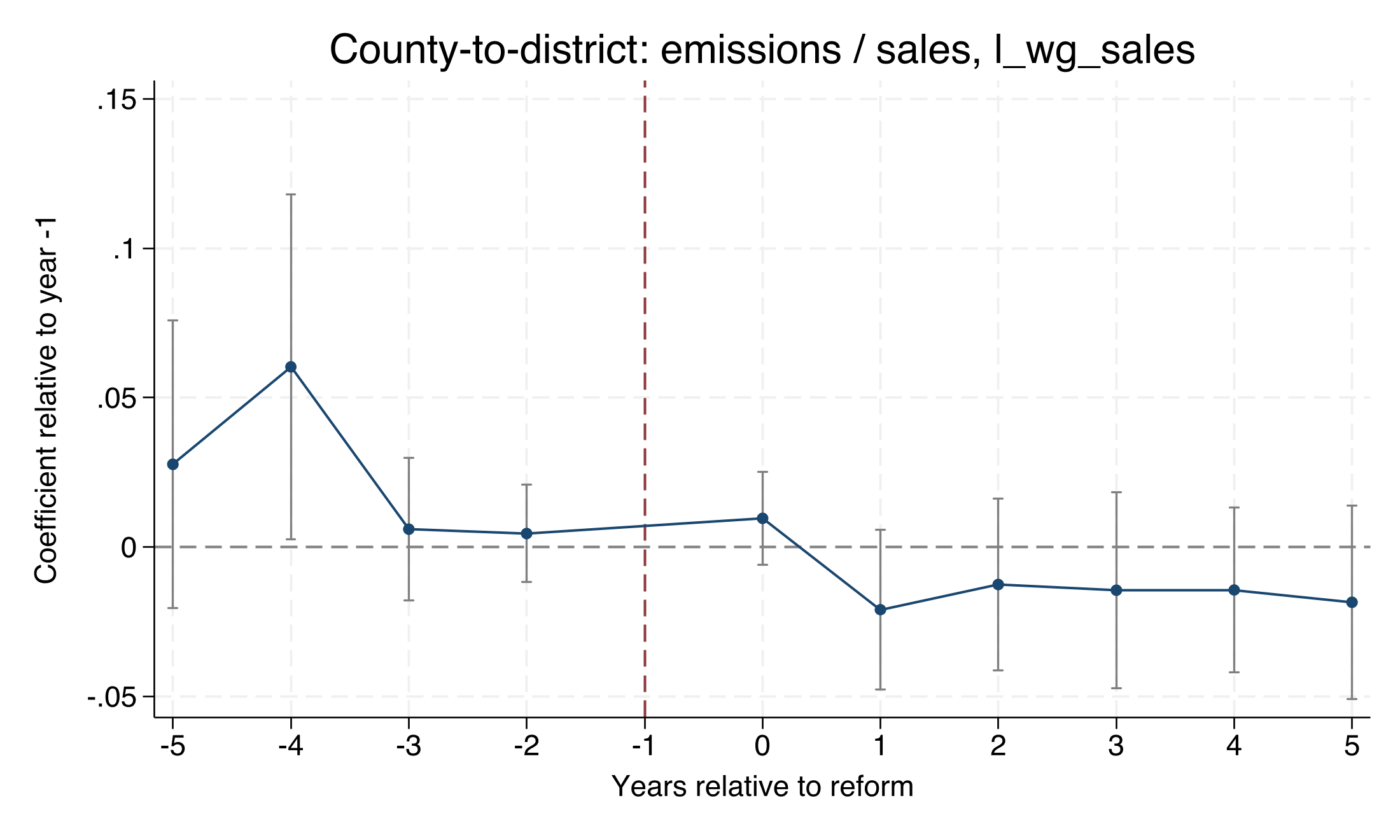}
    \caption{Waste~gas}
  \end{subfigure}\hfill
\end{figure}

\begin{figure}[p]
\caption{County-to-city reform: stacked intensity event studies, per worker}
\label{fig:countycity_emp}
\centering
  \begin{subfigure}[b]{0.32\textwidth}
    \includegraphics[width=\textwidth]{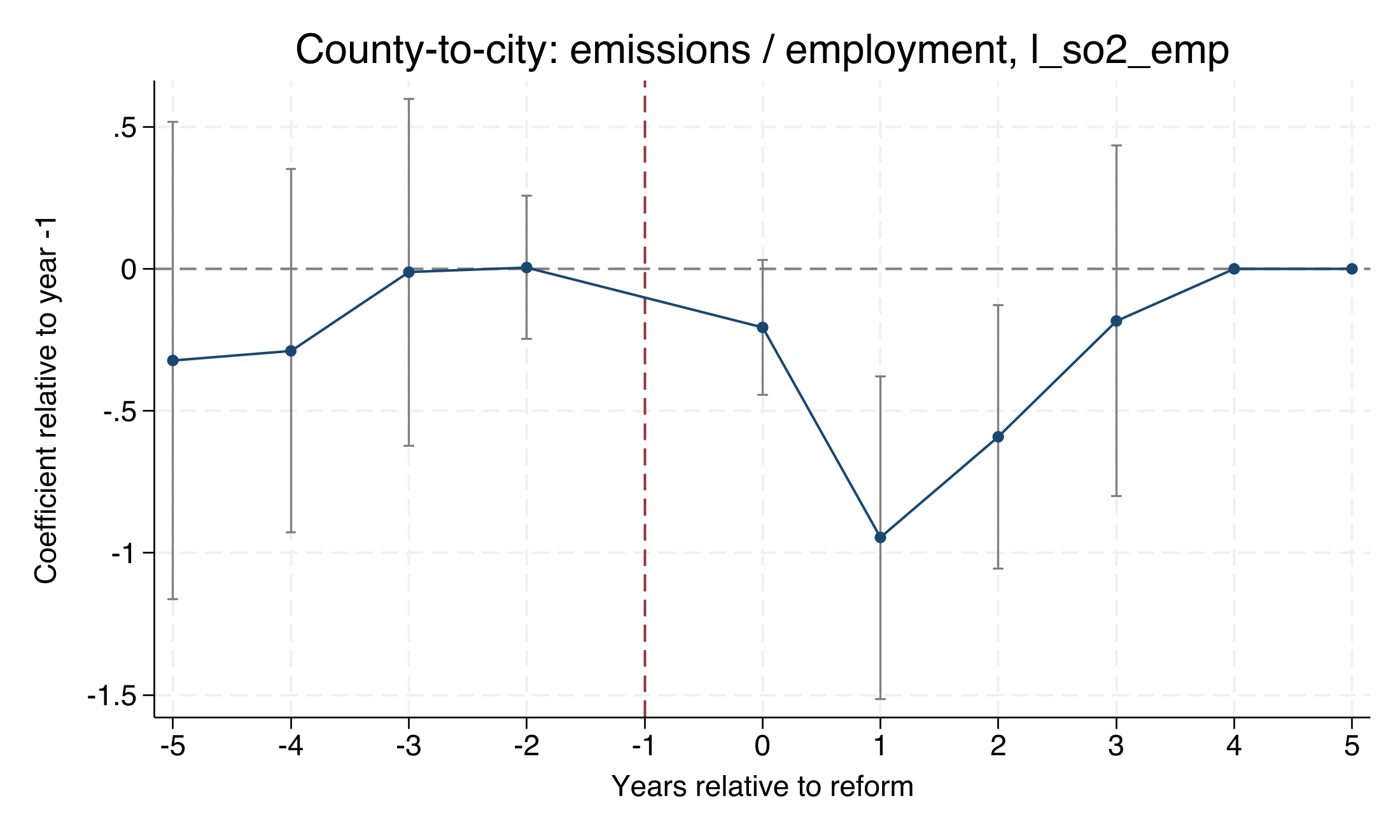}
    \caption{SO$_2$}
  \end{subfigure}\hfill
  \begin{subfigure}[b]{0.32\textwidth}
    \includegraphics[width=\textwidth]{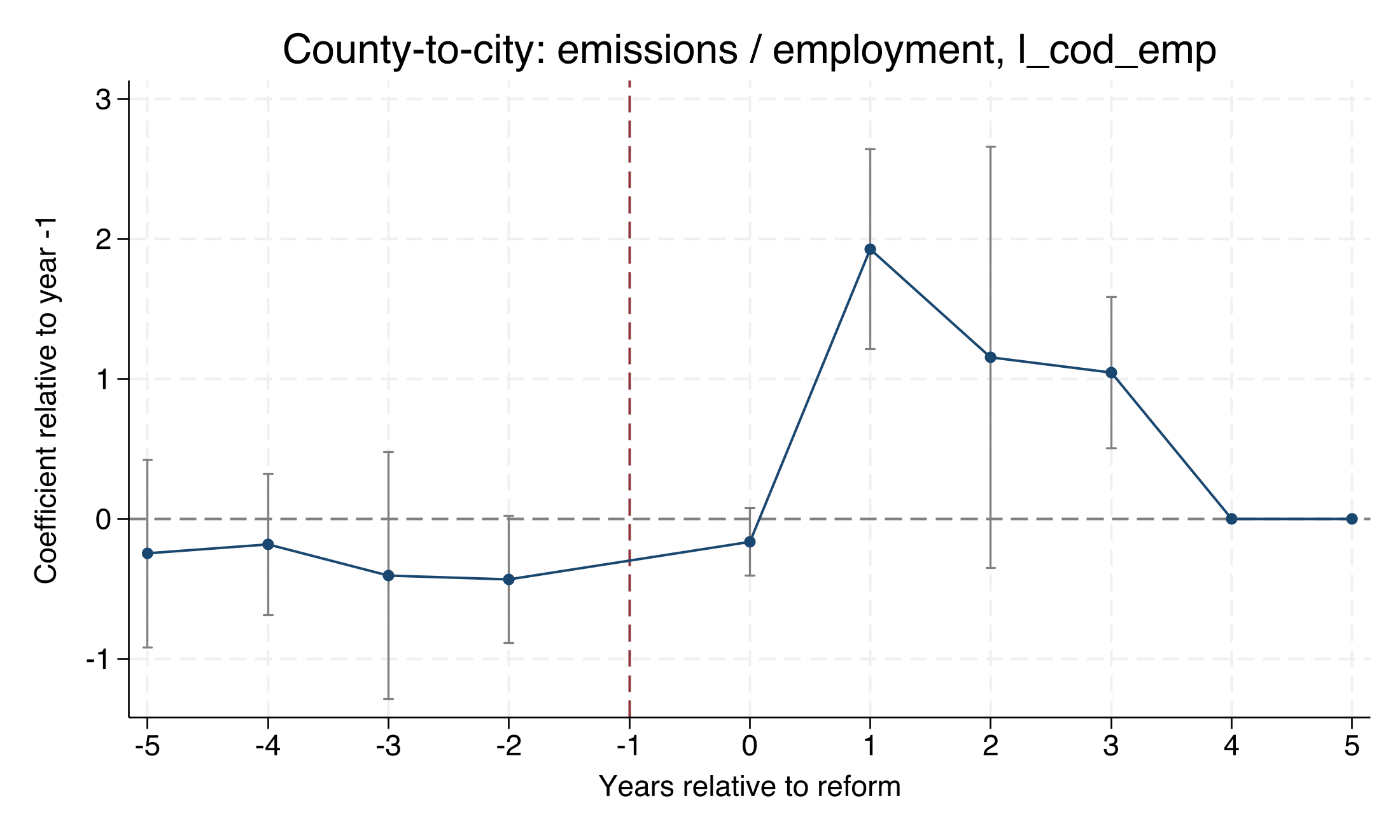}
    \caption{COD}
  \end{subfigure}\hfill
  \begin{subfigure}[b]{0.32\textwidth}
    \includegraphics[width=\textwidth]{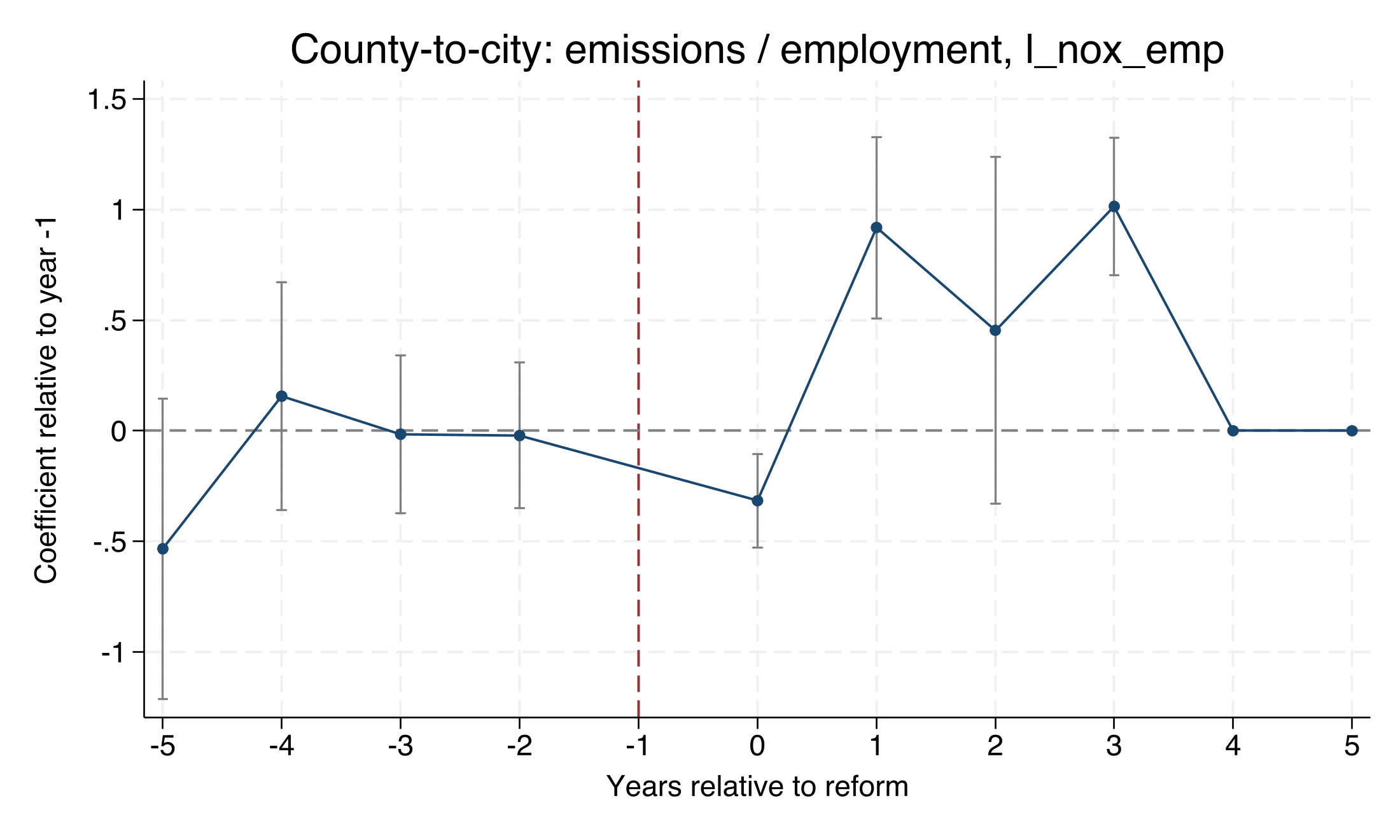}
    \caption{NOx}
  \end{subfigure}\hfill
  \begin{subfigure}[b]{0.32\textwidth}
    \includegraphics[width=\textwidth]{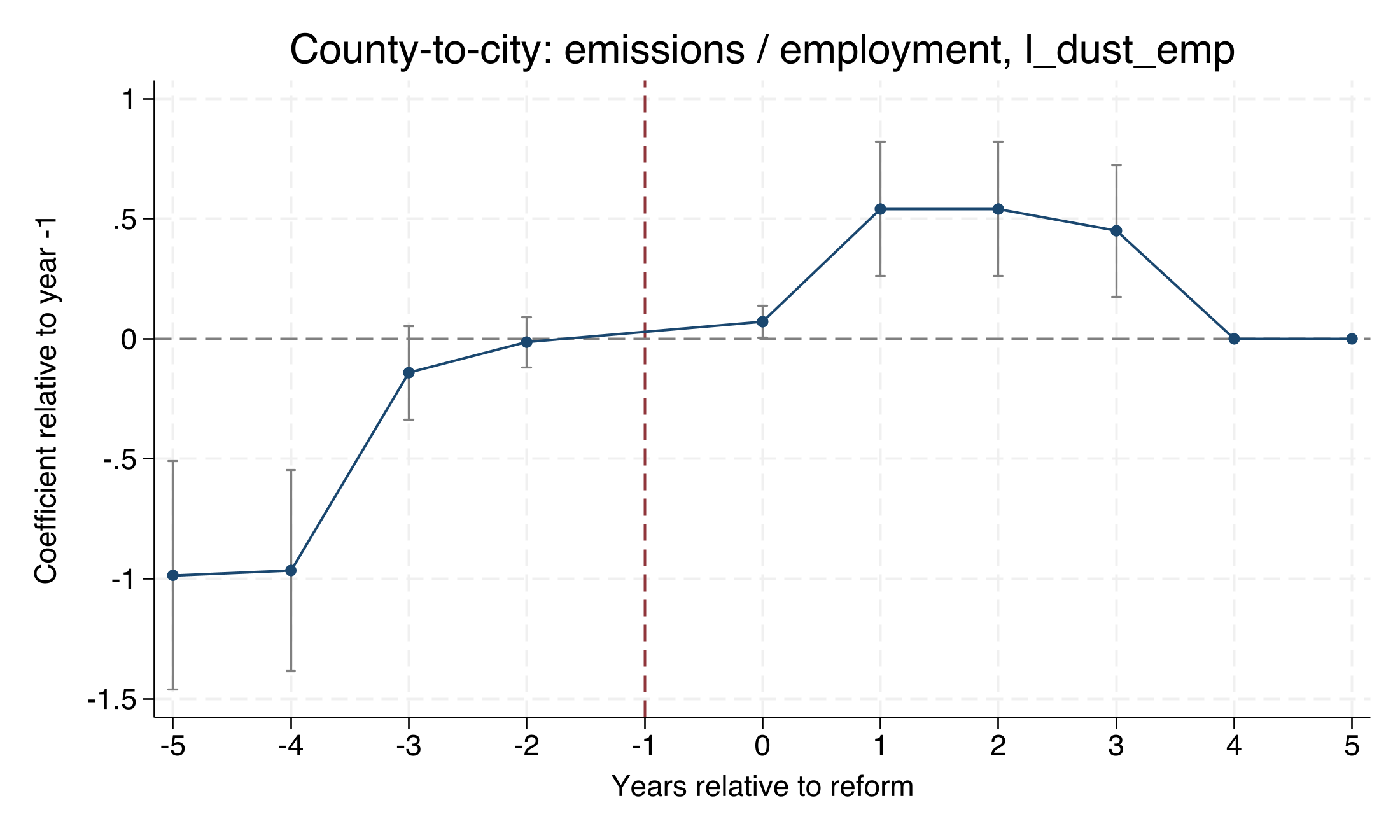}
    \caption{Dust}
  \end{subfigure}\hfill
  \begin{subfigure}[b]{0.32\textwidth}
    \includegraphics[width=\textwidth]{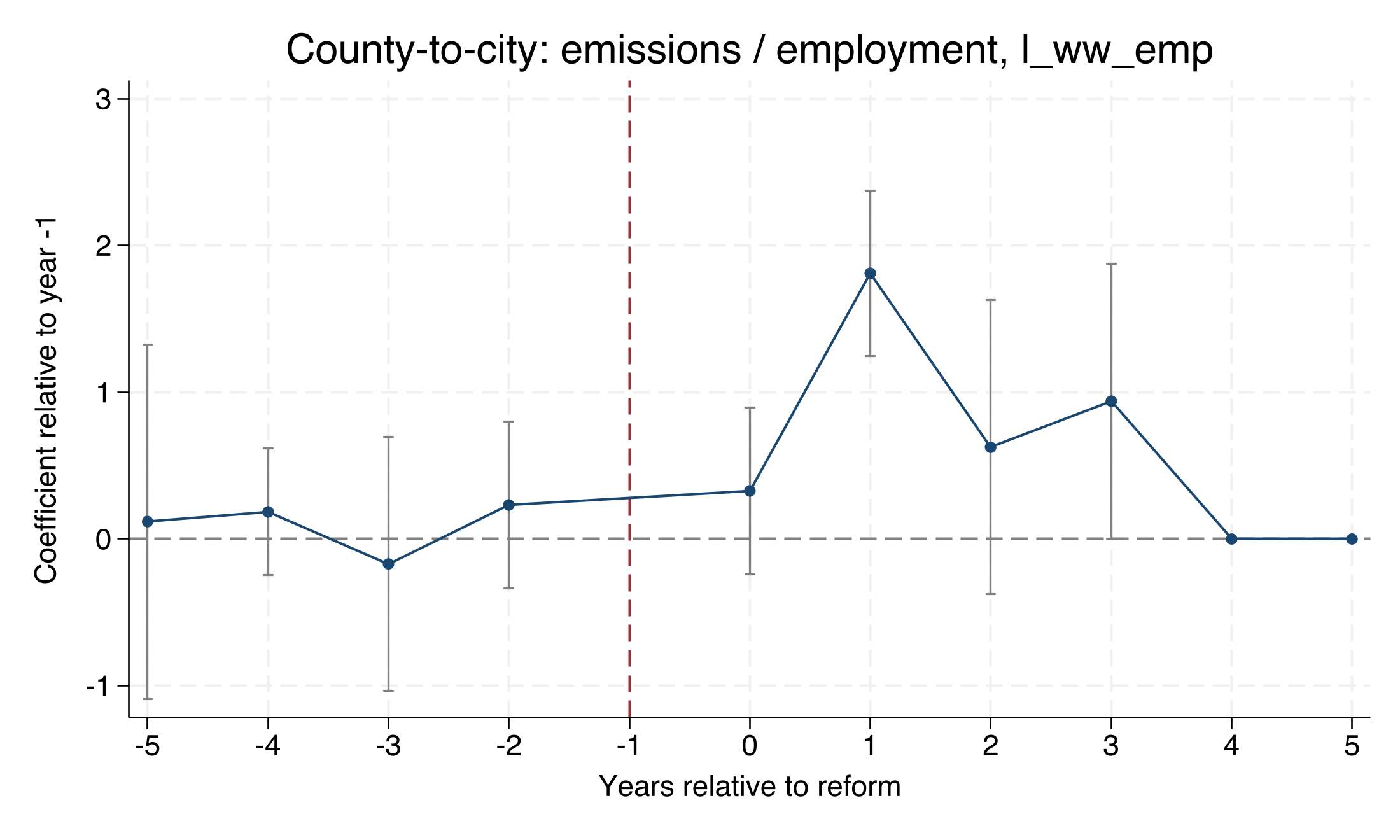}
    \caption{Wastewater}
  \end{subfigure}\hfill
  \begin{subfigure}[b]{0.32\textwidth}
    \includegraphics[width=\textwidth]{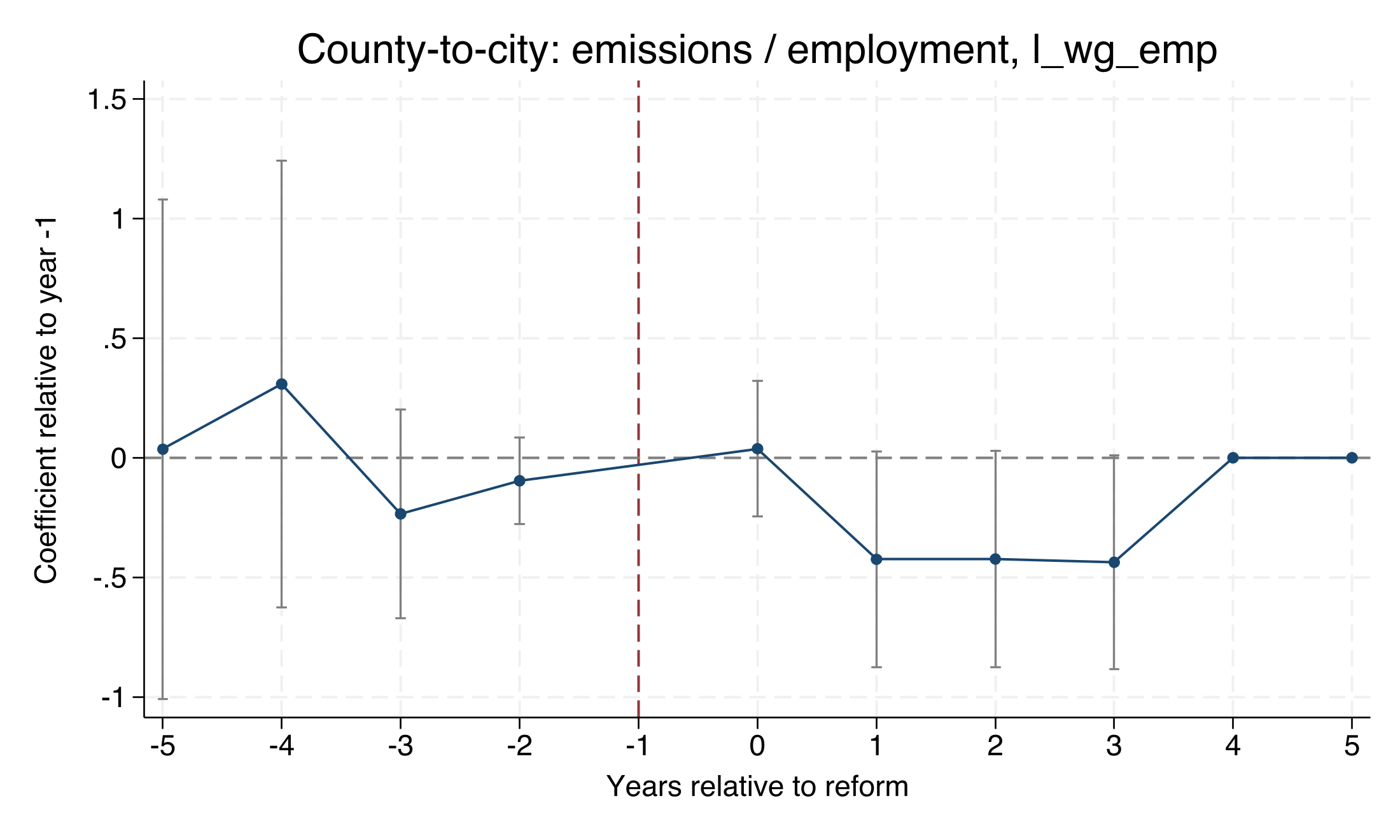}
    \caption{Waste~gas}
  \end{subfigure}\hfill
\end{figure}

\begin{figure}[p]
\caption{County-to-city reform: stacked intensity event studies, per sales}
\label{fig:countycity_sales}
\centering
  \begin{subfigure}[b]{0.32\textwidth}
    \includegraphics[width=\textwidth]{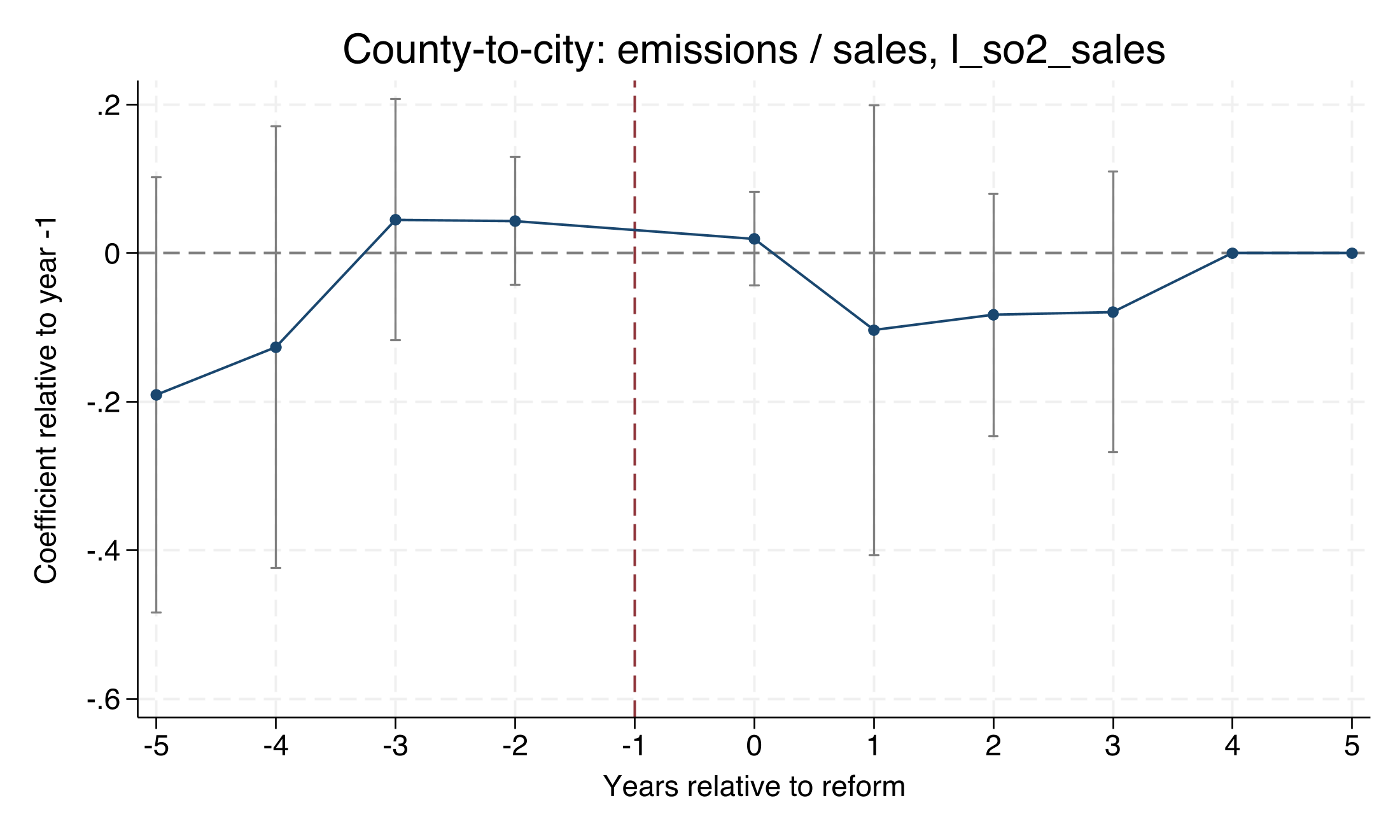}
    \caption{SO$_2$}
  \end{subfigure}\hfill
  \begin{subfigure}[b]{0.32\textwidth}
    \includegraphics[width=\textwidth]{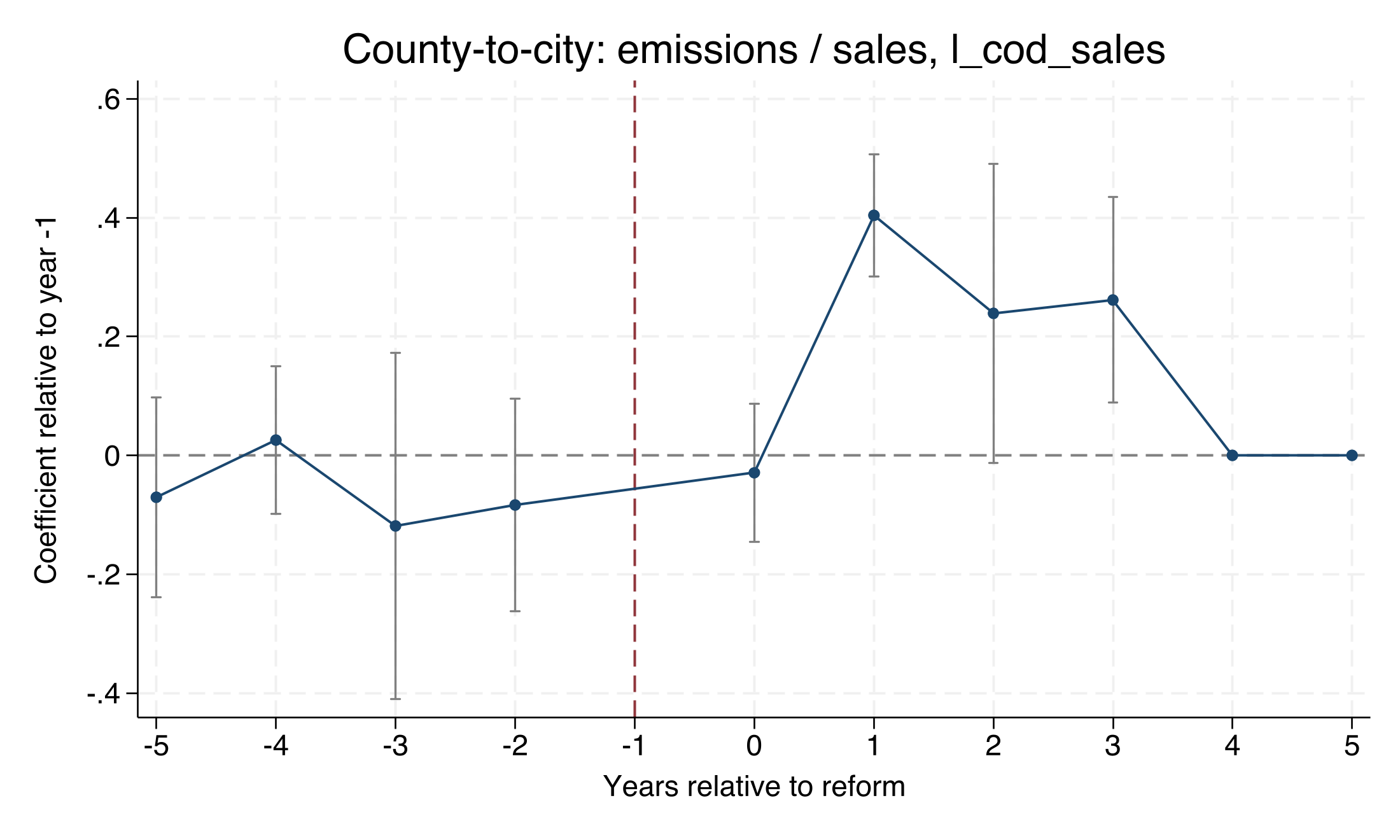}
    \caption{COD}
  \end{subfigure}\hfill
  \begin{subfigure}[b]{0.32\textwidth}
    \includegraphics[width=\textwidth]{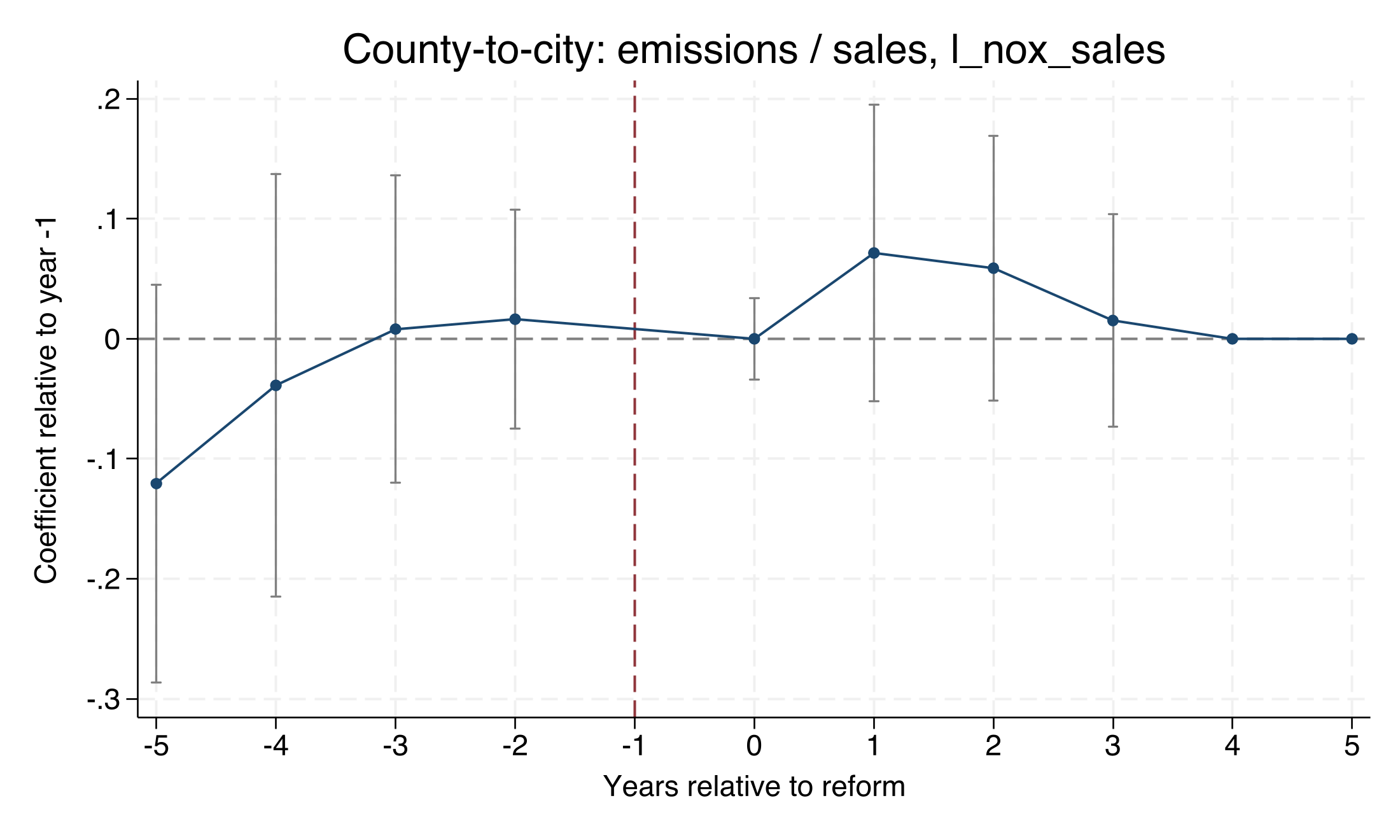}
    \caption{NOx}
  \end{subfigure}\hfill
  \begin{subfigure}[b]{0.32\textwidth}
    \includegraphics[width=\textwidth]{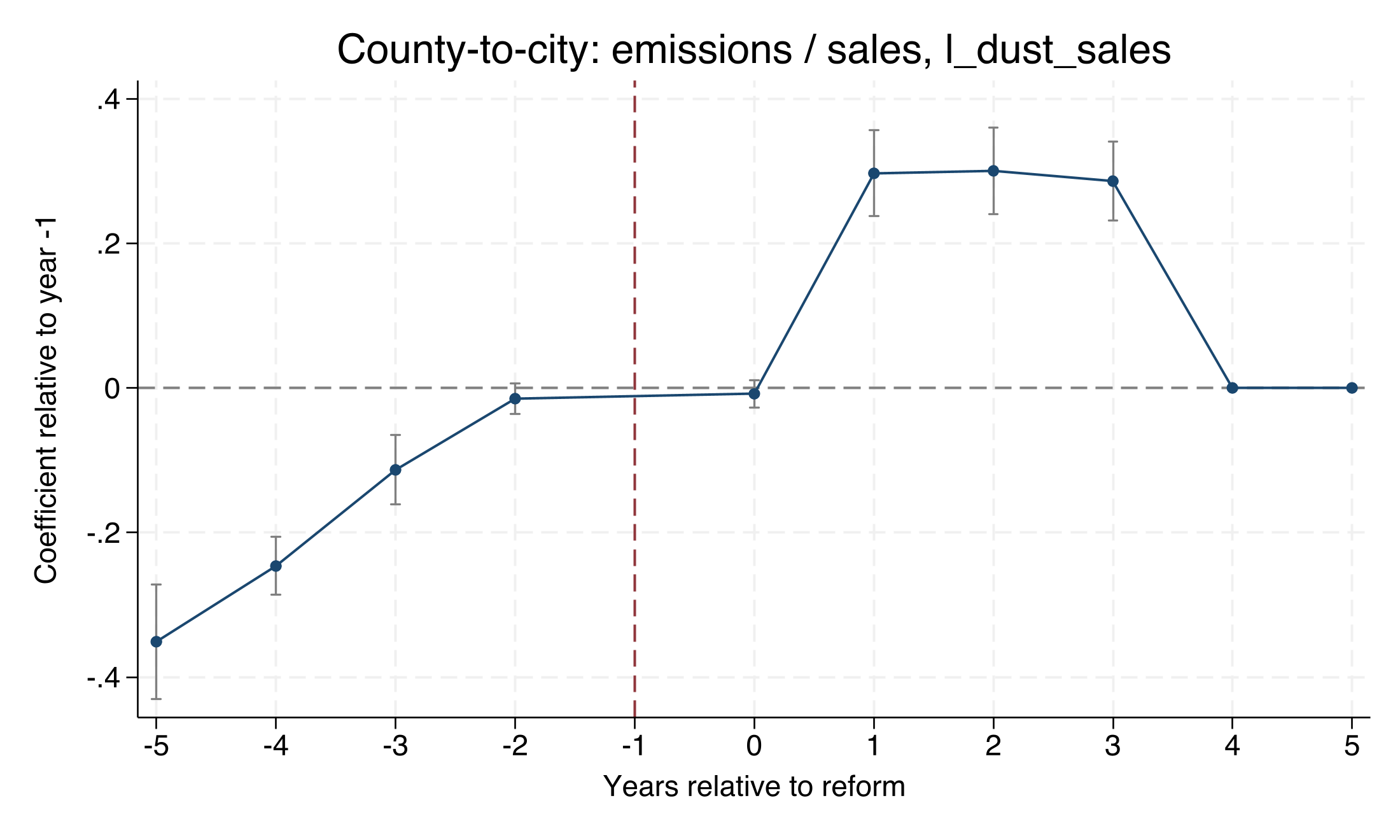}
    \caption{Dust}
  \end{subfigure}\hfill
  \begin{subfigure}[b]{0.32\textwidth}
    \includegraphics[width=\textwidth]{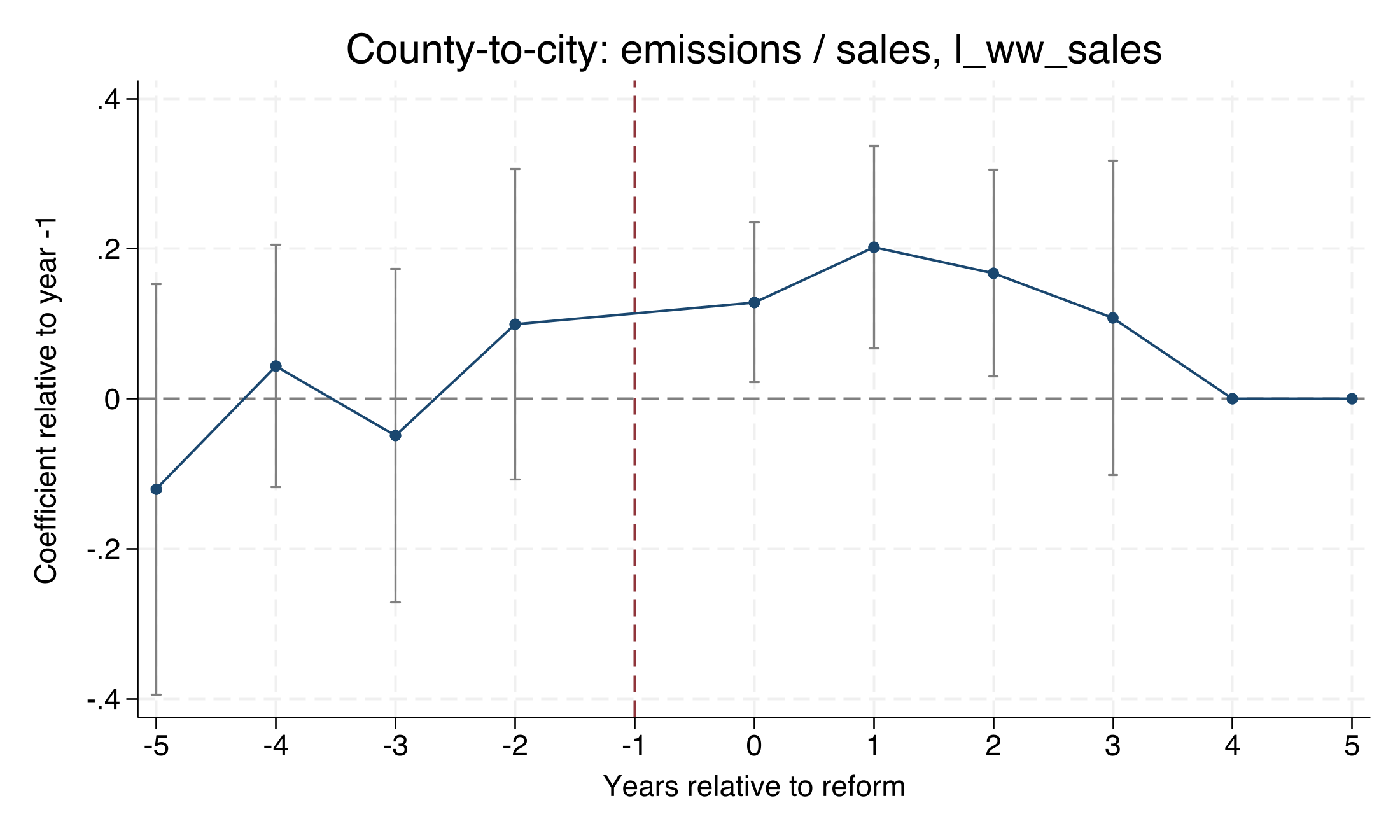}
    \caption{Wastewater}
  \end{subfigure}\hfill
  \begin{subfigure}[b]{0.32\textwidth}
    \includegraphics[width=\textwidth]{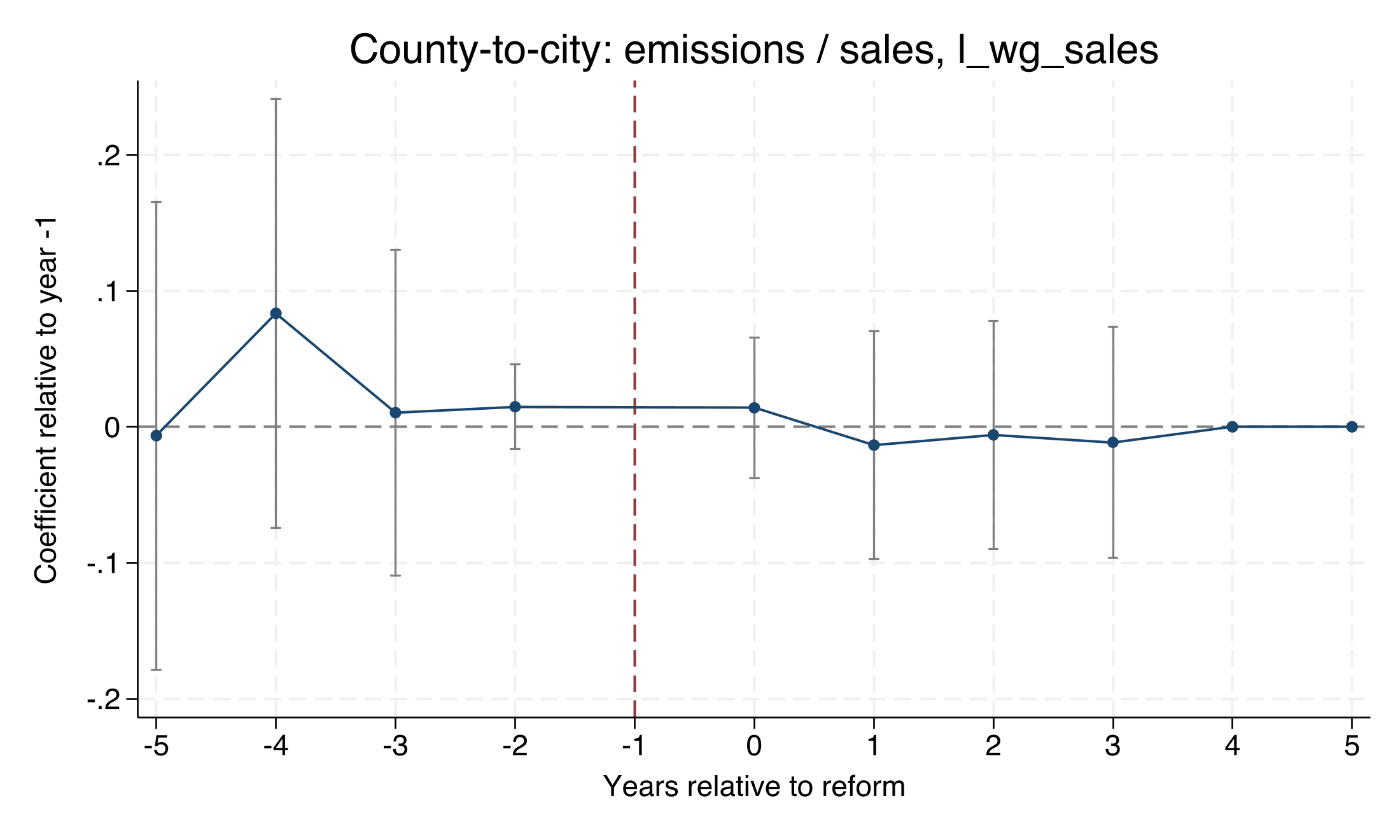}
    \caption{Waste~gas}
  \end{subfigure}\hfill
\end{figure}

\begin{figure}[p]
\caption{County-to-district reform: stacked DID trends, per worker}
\label{fig:district_trends_emp}
\centering
  \begin{subfigure}[b]{0.32\textwidth}
    \includegraphics[width=\textwidth]{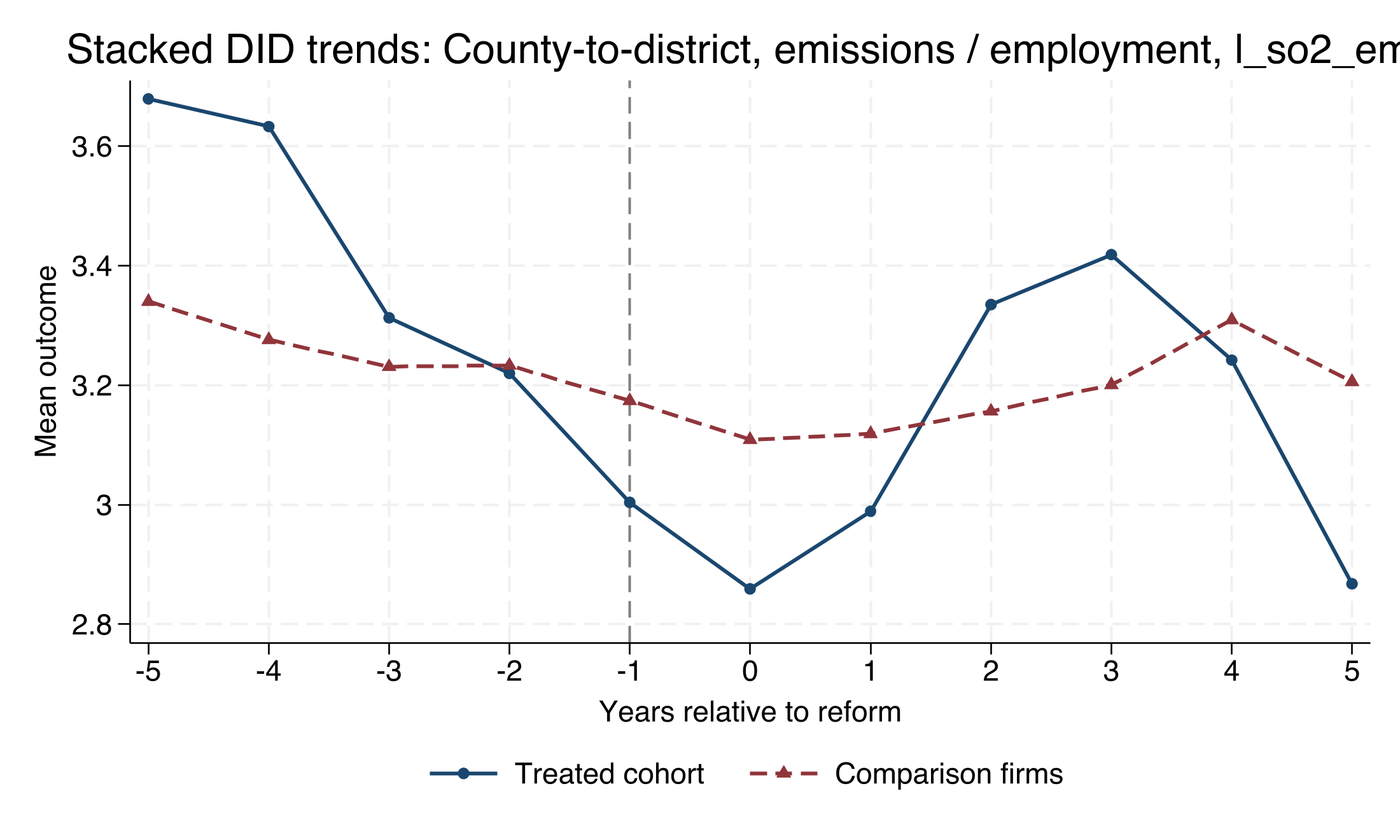}
    \caption{SO$_2$}
  \end{subfigure}\hfill
  \begin{subfigure}[b]{0.32\textwidth}
    \includegraphics[width=\textwidth]{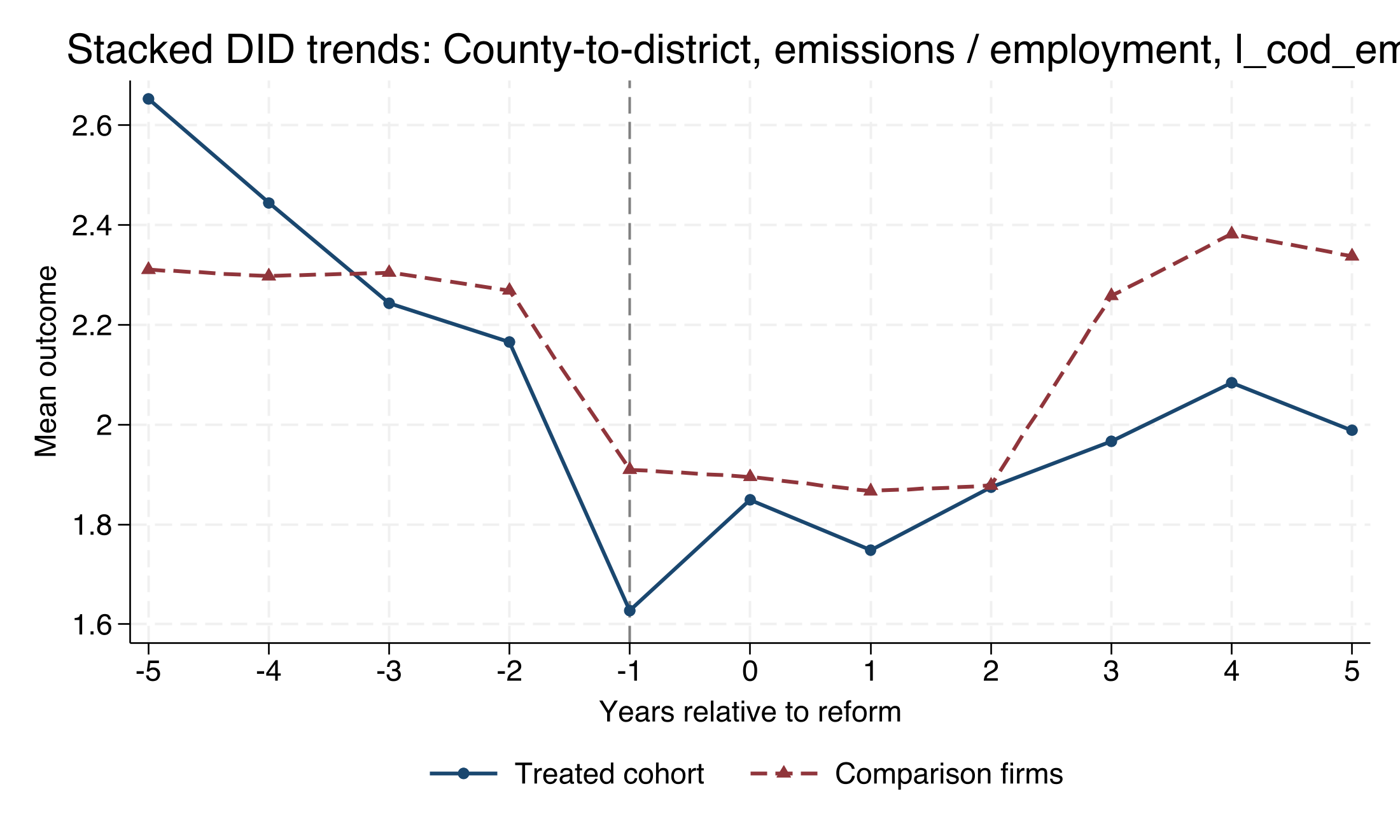}
    \caption{COD}
  \end{subfigure}\hfill
  \begin{subfigure}[b]{0.32\textwidth}
    \includegraphics[width=\textwidth]{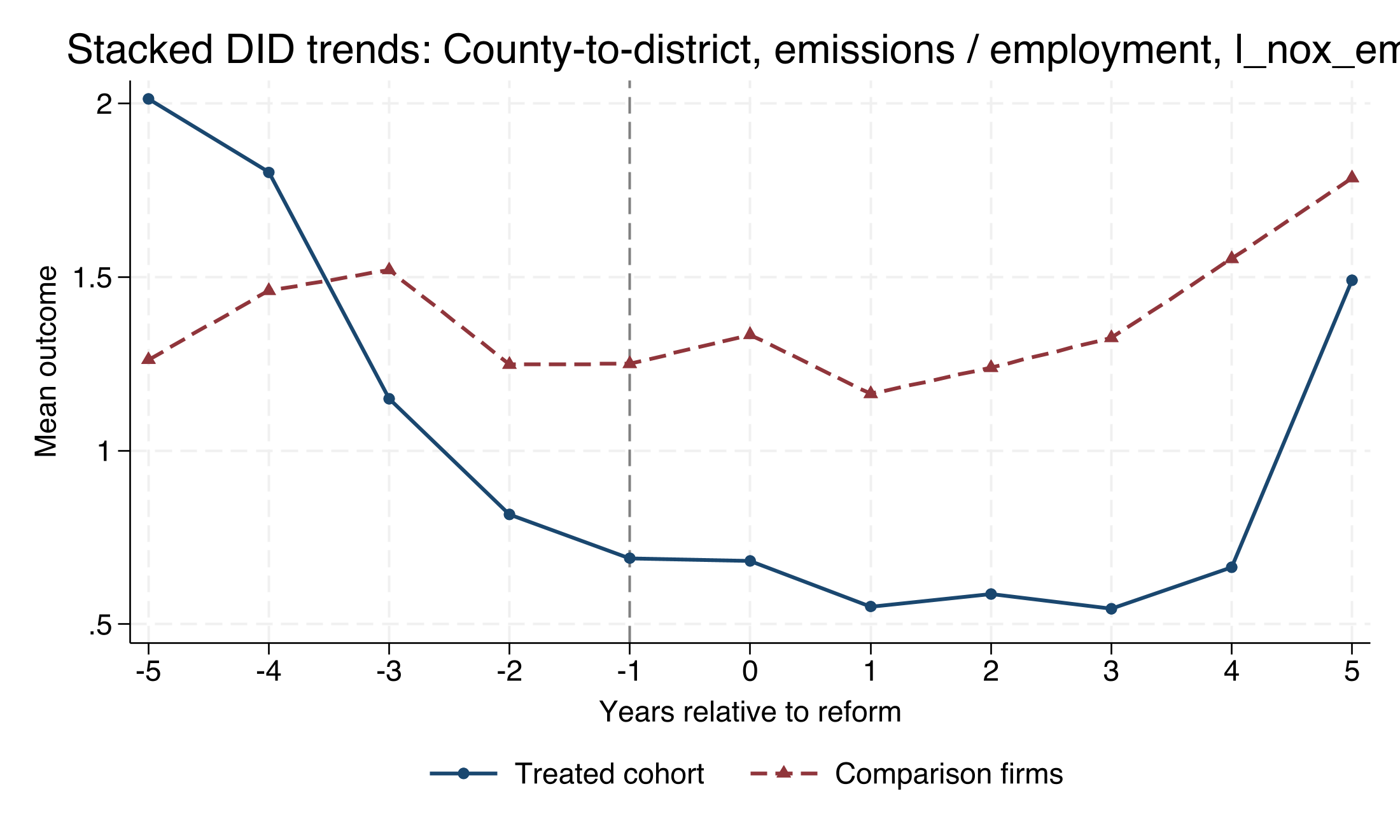}
    \caption{NOx}
  \end{subfigure}\hfill
  \begin{subfigure}[b]{0.32\textwidth}
    \includegraphics[width=\textwidth]{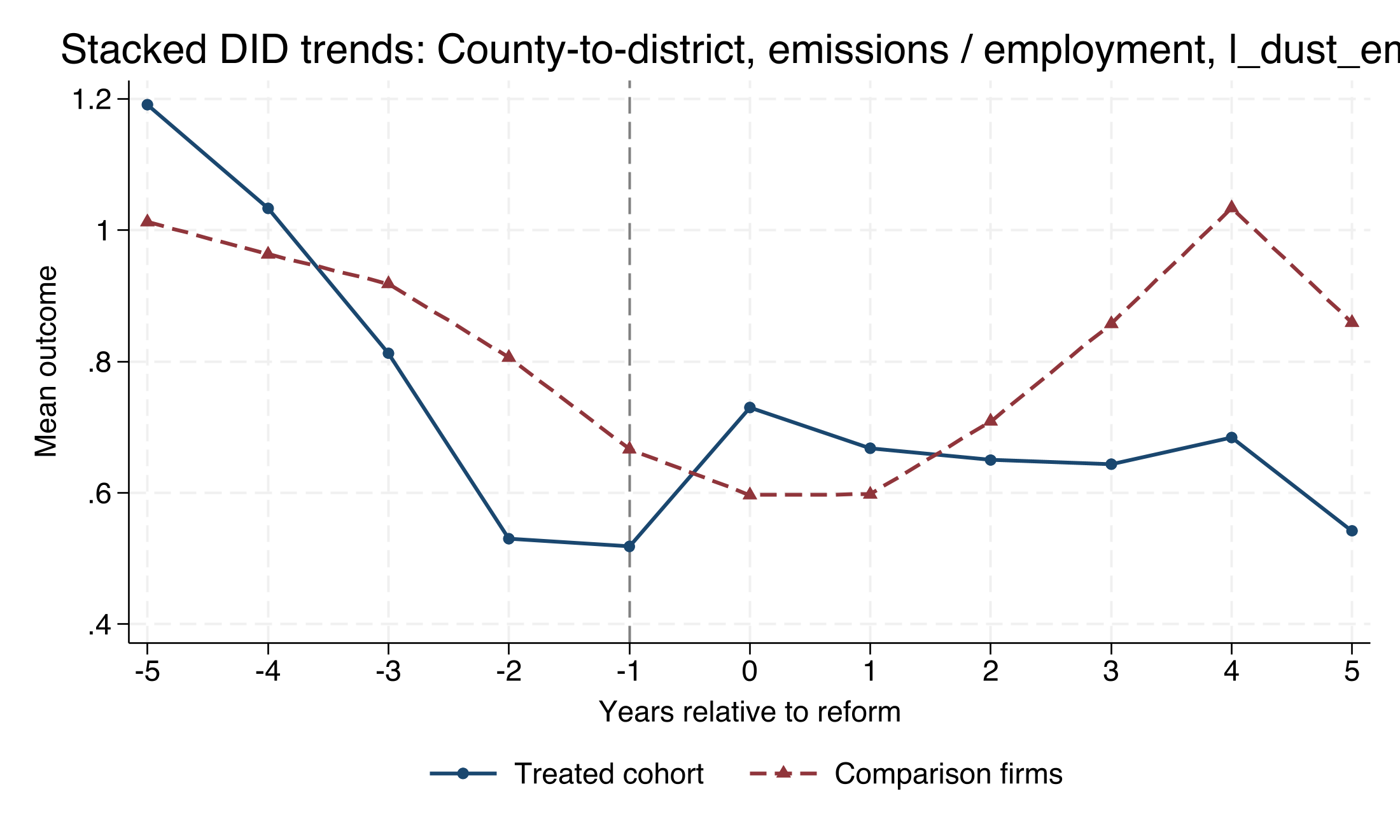}
    \caption{Dust}
  \end{subfigure}\hfill
  \begin{subfigure}[b]{0.32\textwidth}
    \includegraphics[width=\textwidth]{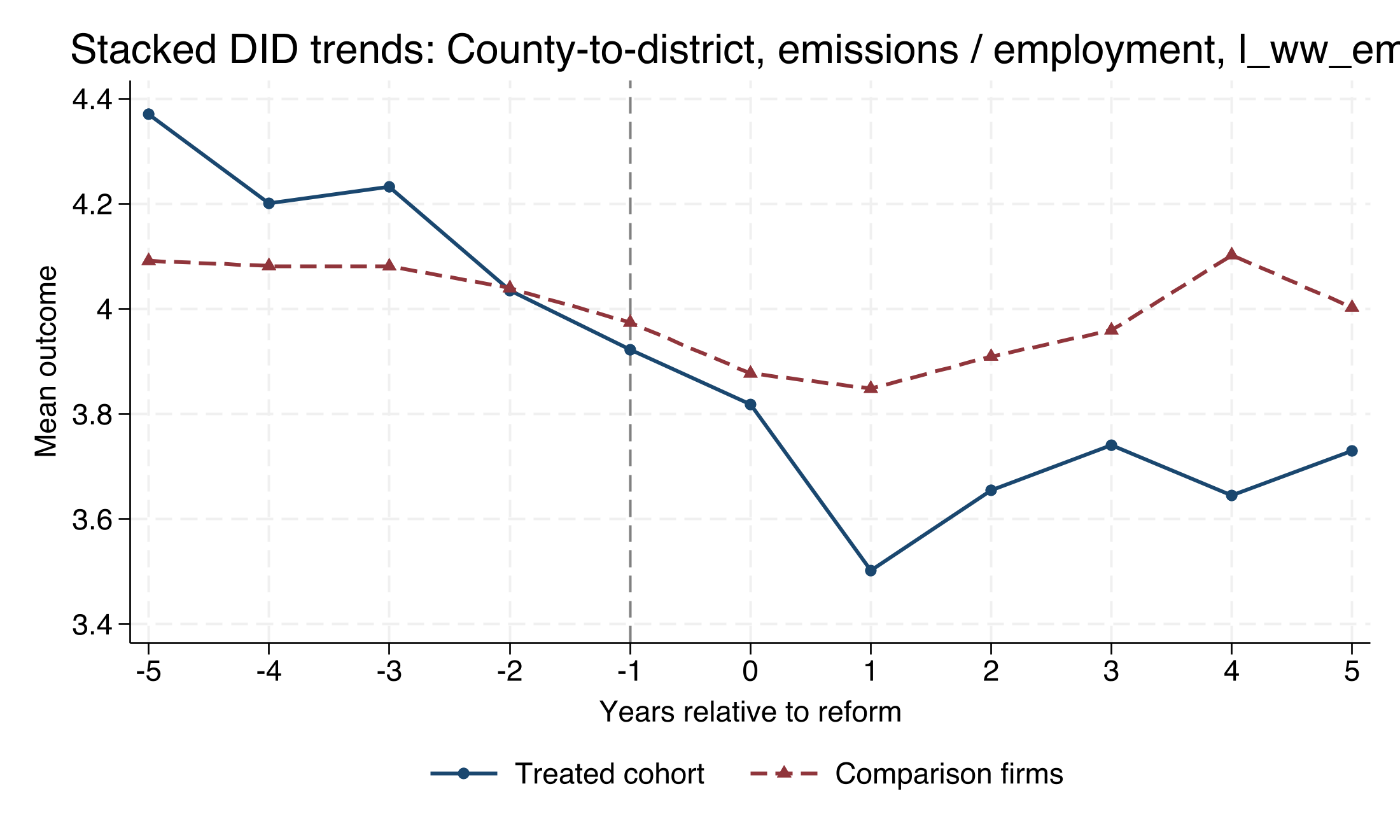}
    \caption{Wastewater}
  \end{subfigure}\hfill
  \begin{subfigure}[b]{0.32\textwidth}
    \includegraphics[width=\textwidth]{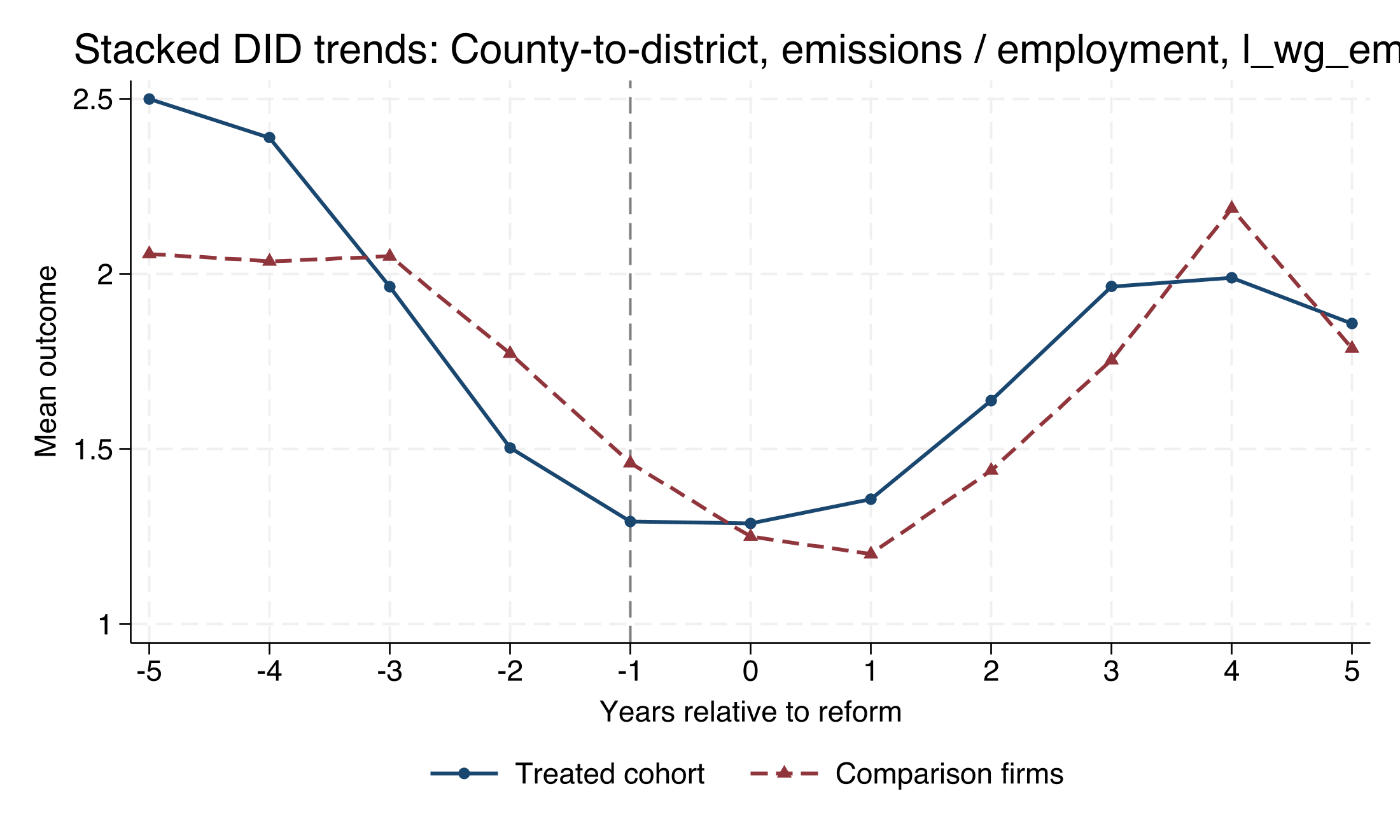}
    \caption{Waste~gas}
  \end{subfigure}\hfill
\end{figure}

\begin{figure}[p]
\caption{County-to-district reform: stacked DID trends, per sales}
\label{fig:district_trends_sales}
\centering
  \begin{subfigure}[b]{0.32\textwidth}
    \includegraphics[width=\textwidth]{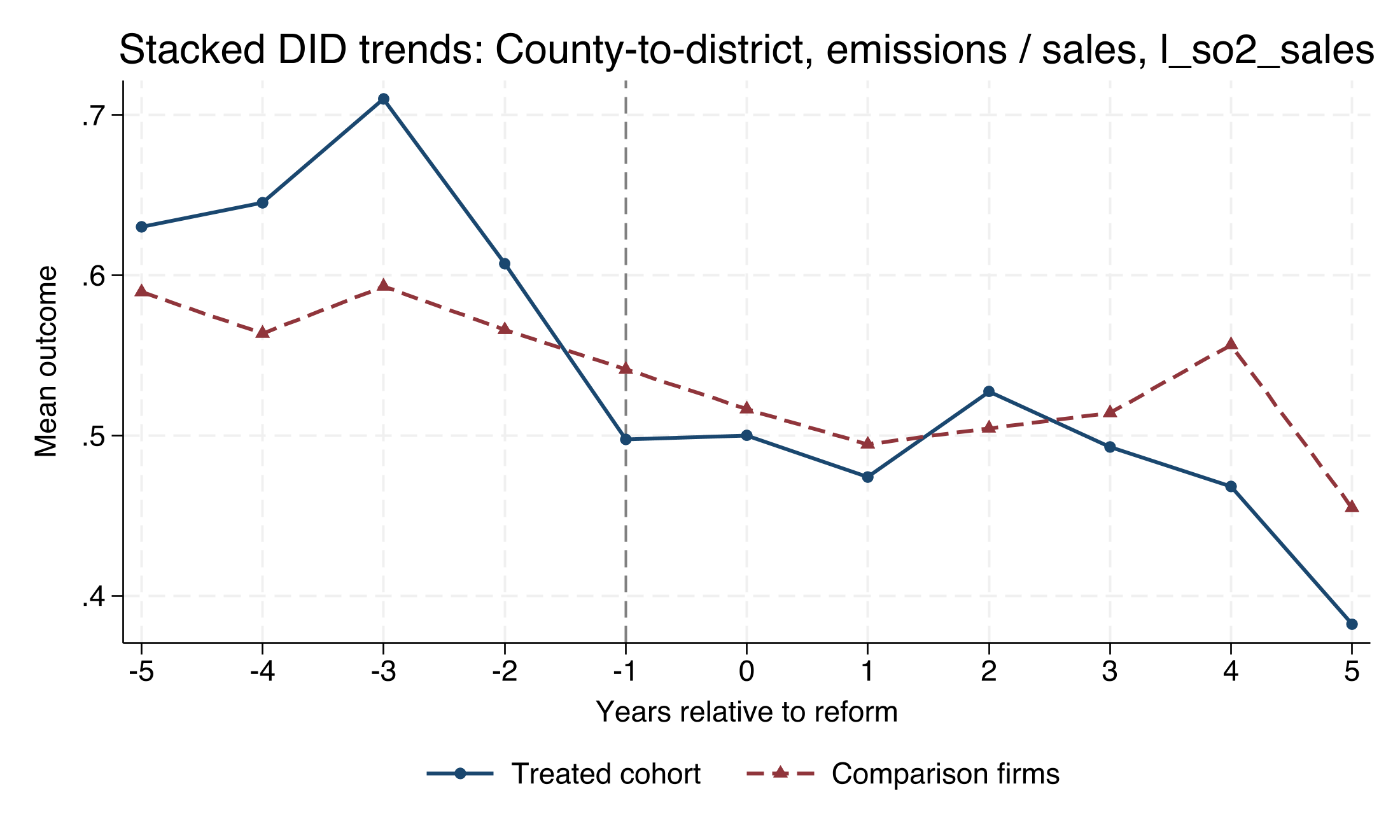}
    \caption{SO$_2$}
  \end{subfigure}\hfill
  \begin{subfigure}[b]{0.32\textwidth}
    \includegraphics[width=\textwidth]{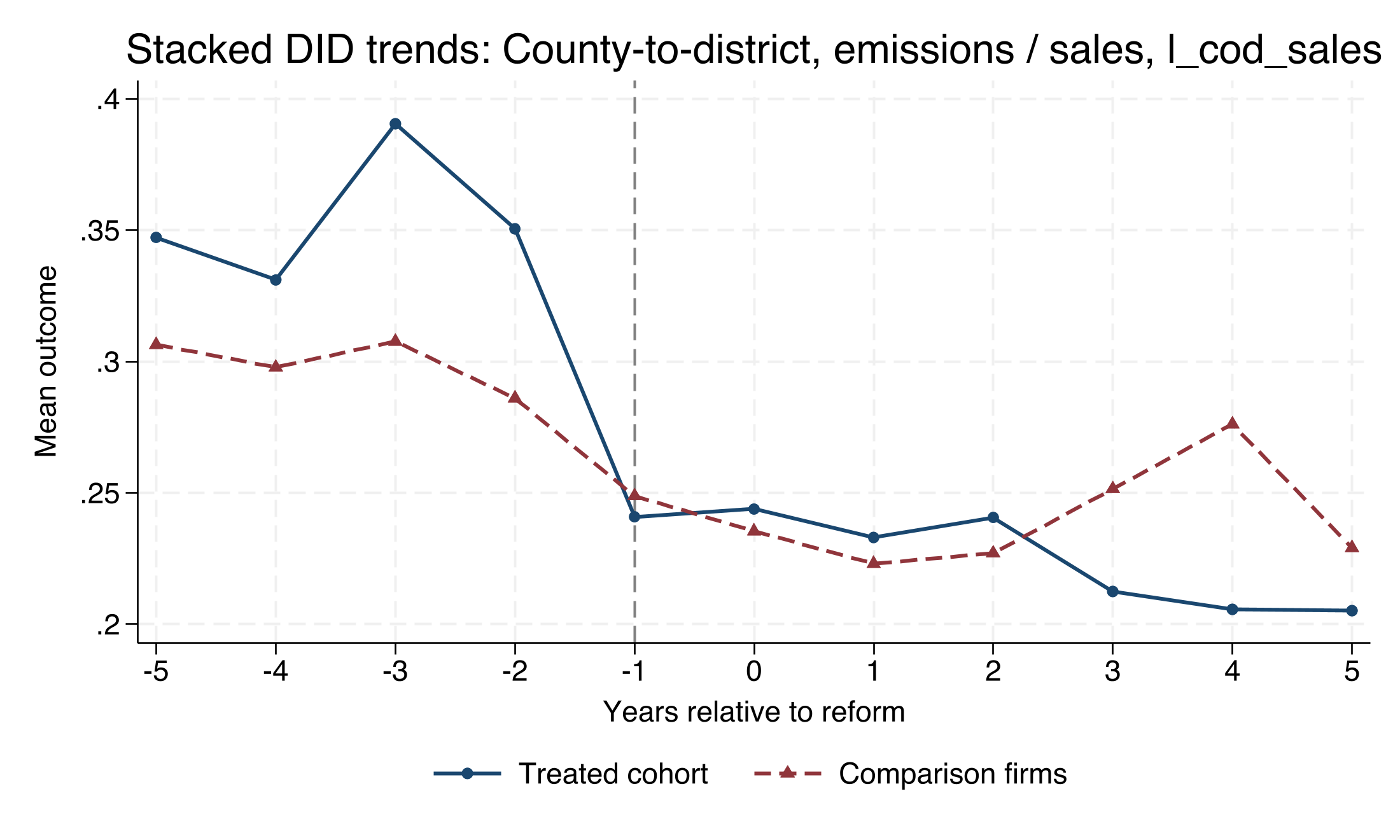}
    \caption{COD}
  \end{subfigure}\hfill
  \begin{subfigure}[b]{0.32\textwidth}
    \includegraphics[width=\textwidth]{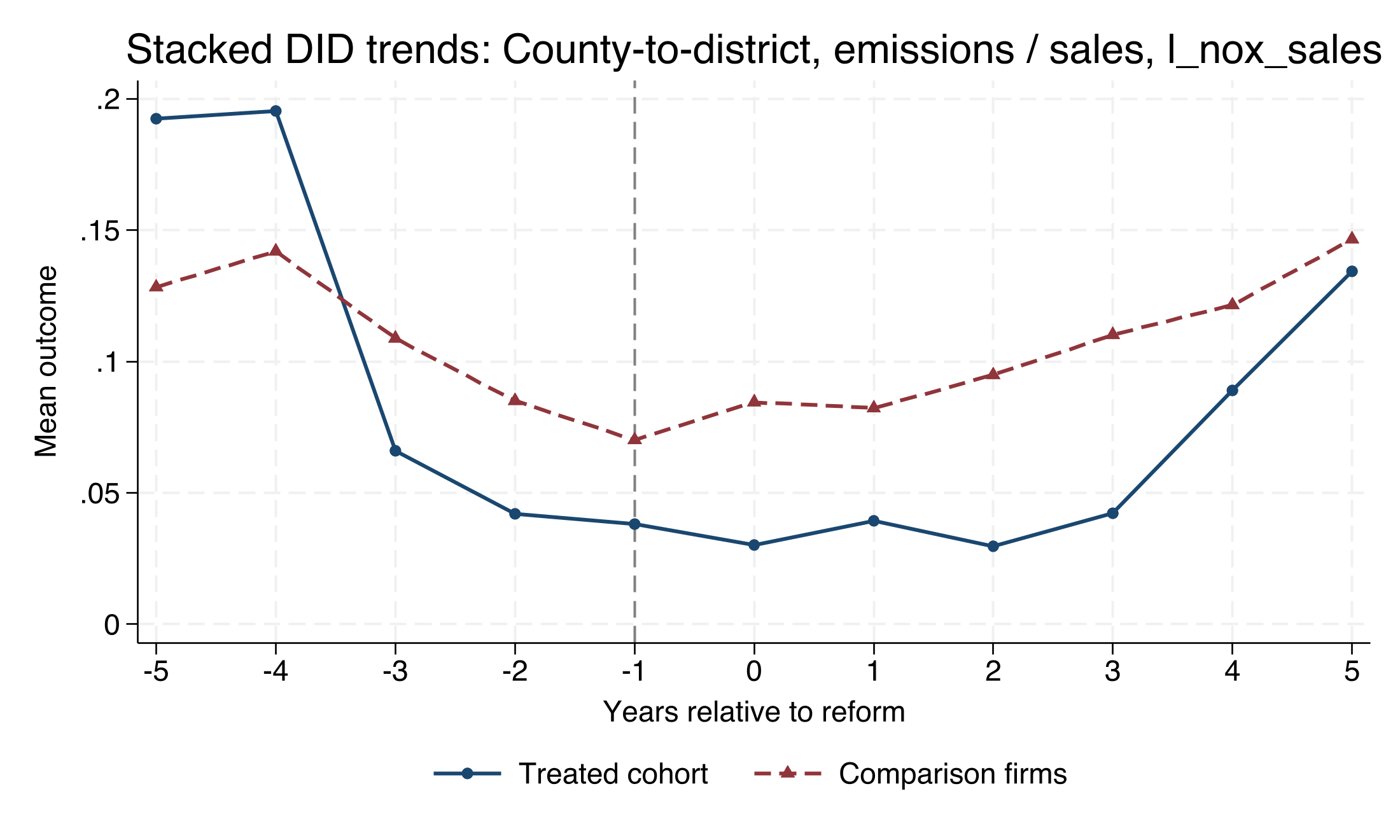}
    \caption{NOx}
  \end{subfigure}\hfill
  \begin{subfigure}[b]{0.32\textwidth}
    \includegraphics[width=\textwidth]{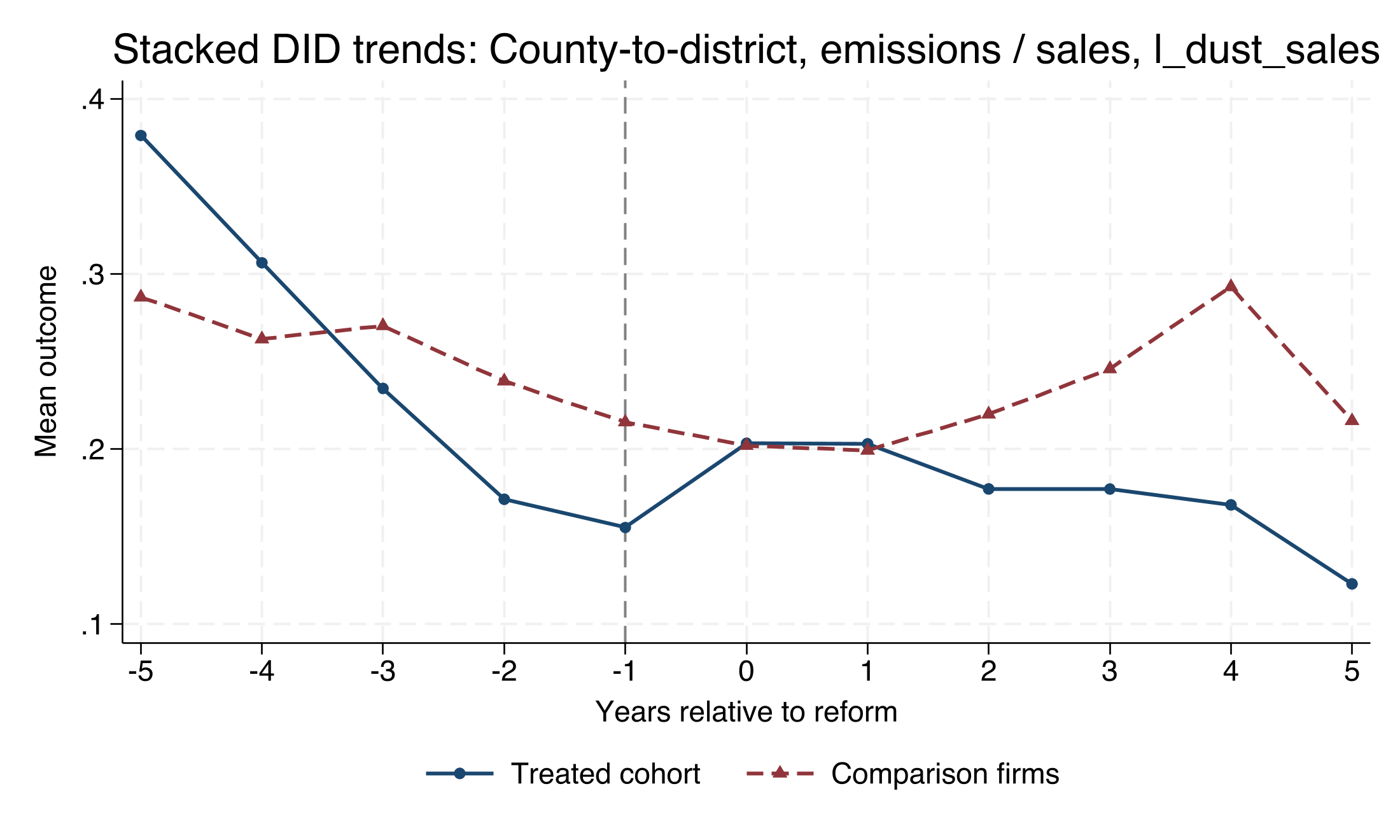}
    \caption{Dust}
  \end{subfigure}\hfill
  \begin{subfigure}[b]{0.32\textwidth}
    \includegraphics[width=\textwidth]{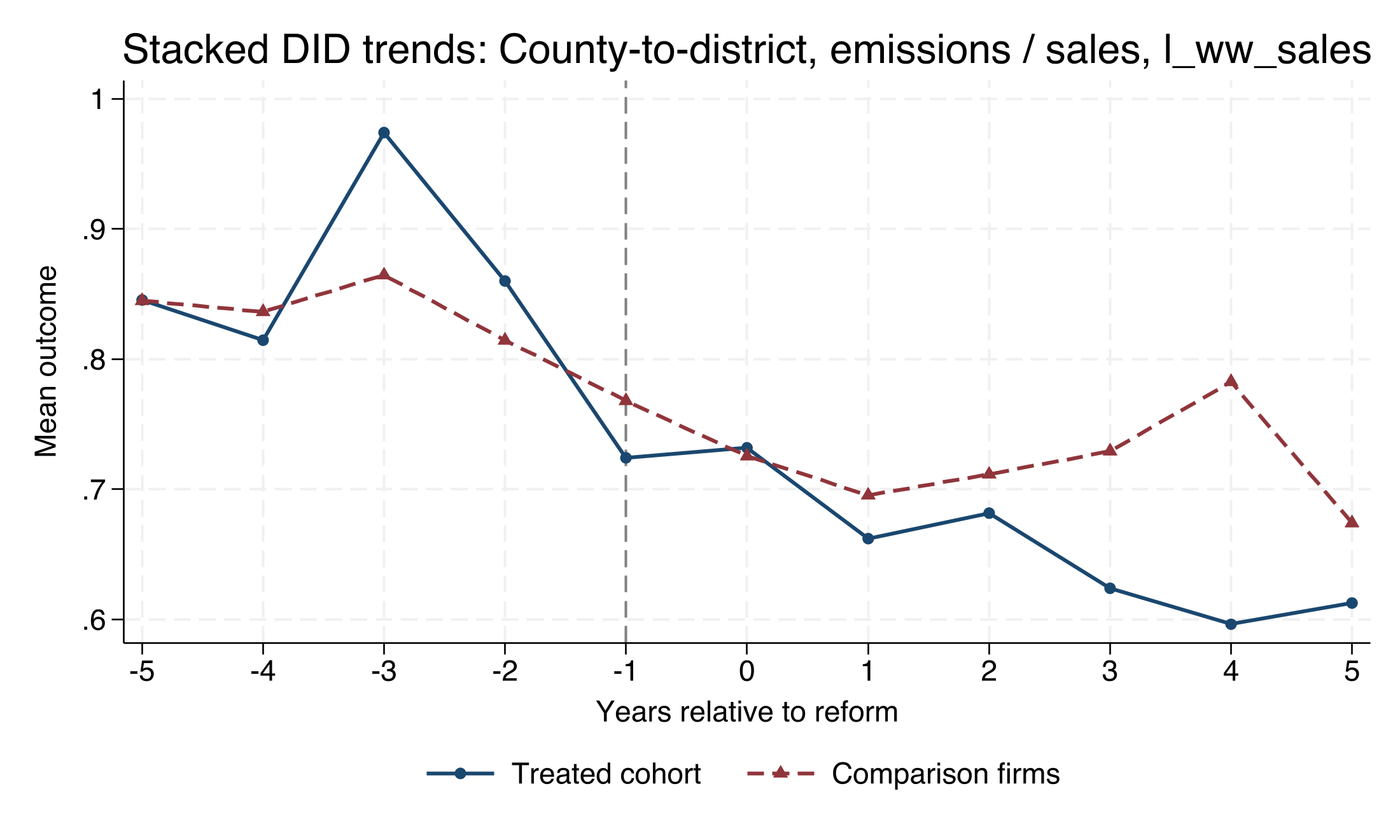}
    \caption{Wastewater}
  \end{subfigure}\hfill
  \begin{subfigure}[b]{0.32\textwidth}
    \includegraphics[width=\textwidth]{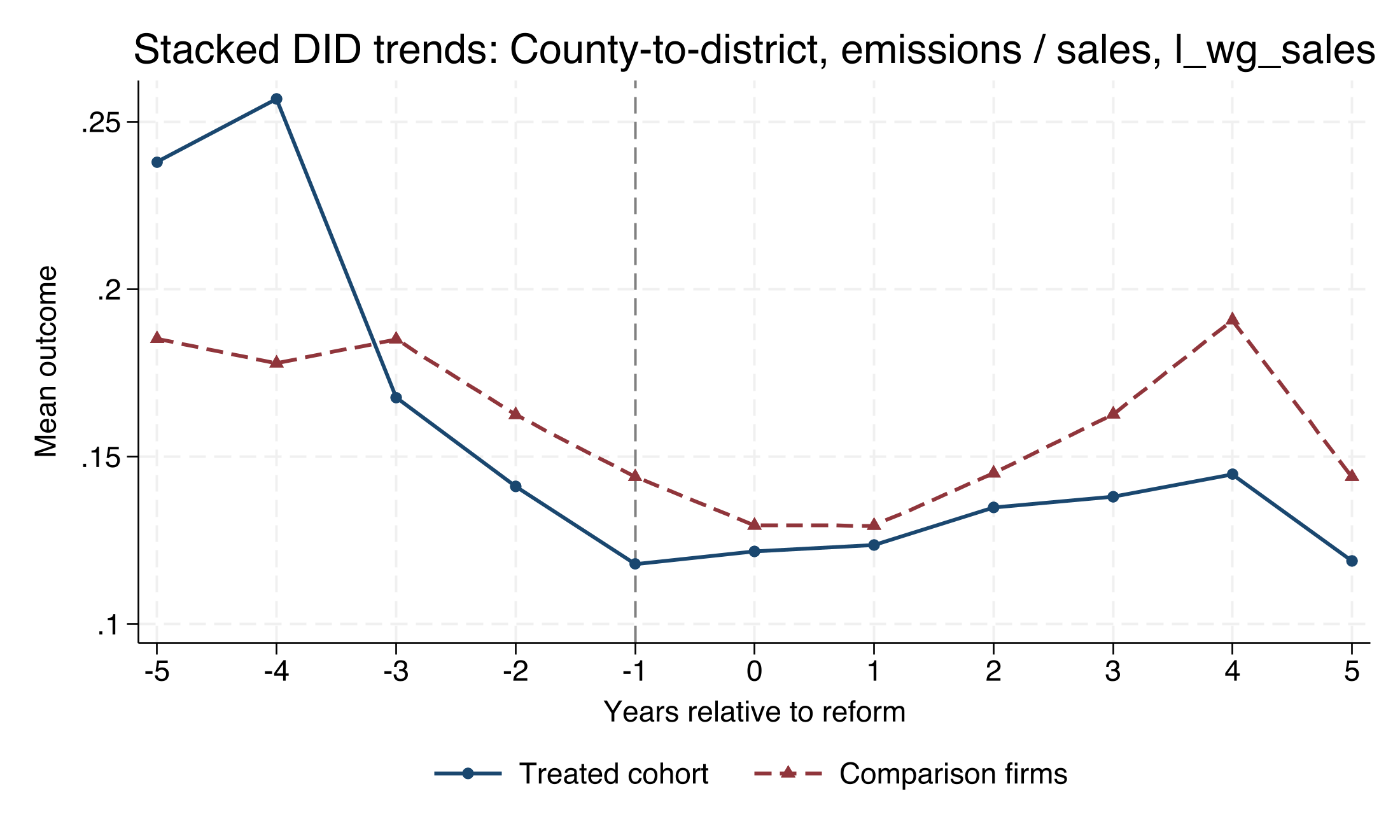}
    \caption{Waste~gas}
  \end{subfigure}\hfill
\end{figure}

\begin{figure}[p]
\caption{County-to-city reform: stacked DID trends, per worker}
\label{fig:countycity_trends_emp}
\centering
  \begin{subfigure}[b]{0.32\textwidth}
    \includegraphics[width=\textwidth]{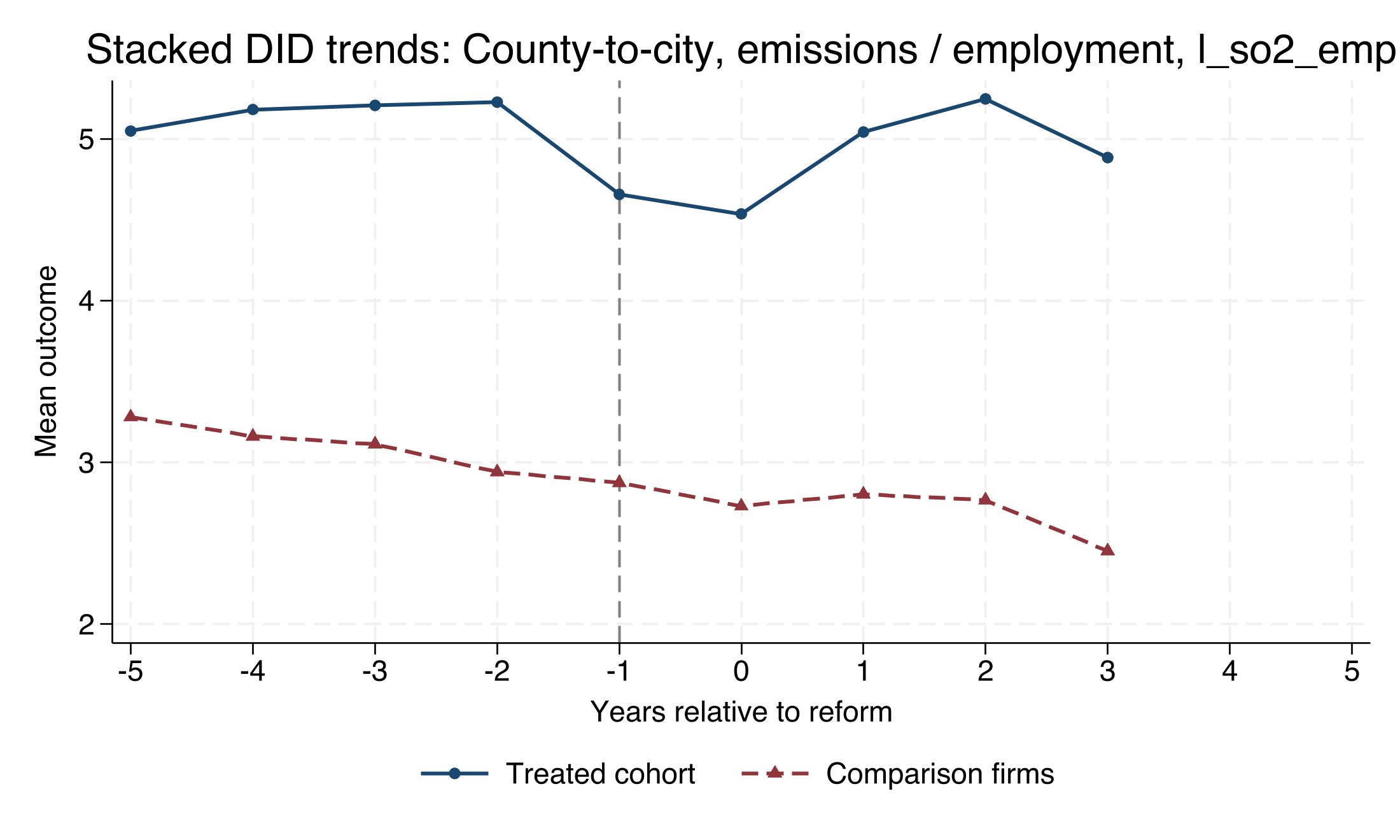}
    \caption{SO$_2$}
  \end{subfigure}\hfill
  \begin{subfigure}[b]{0.32\textwidth}
    \includegraphics[width=\textwidth]{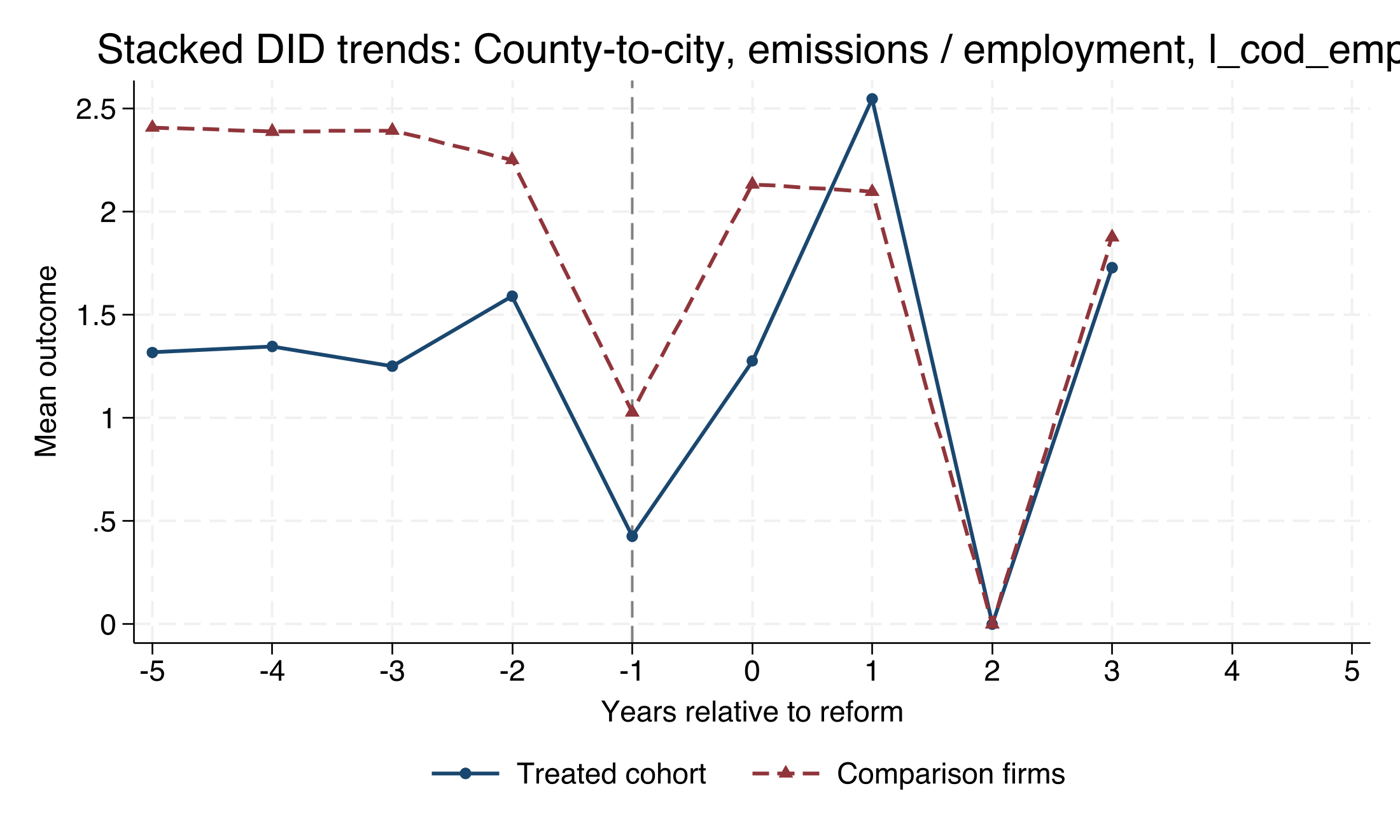}
    \caption{COD}
  \end{subfigure}\hfill
  \begin{subfigure}[b]{0.32\textwidth}
    \includegraphics[width=\textwidth]{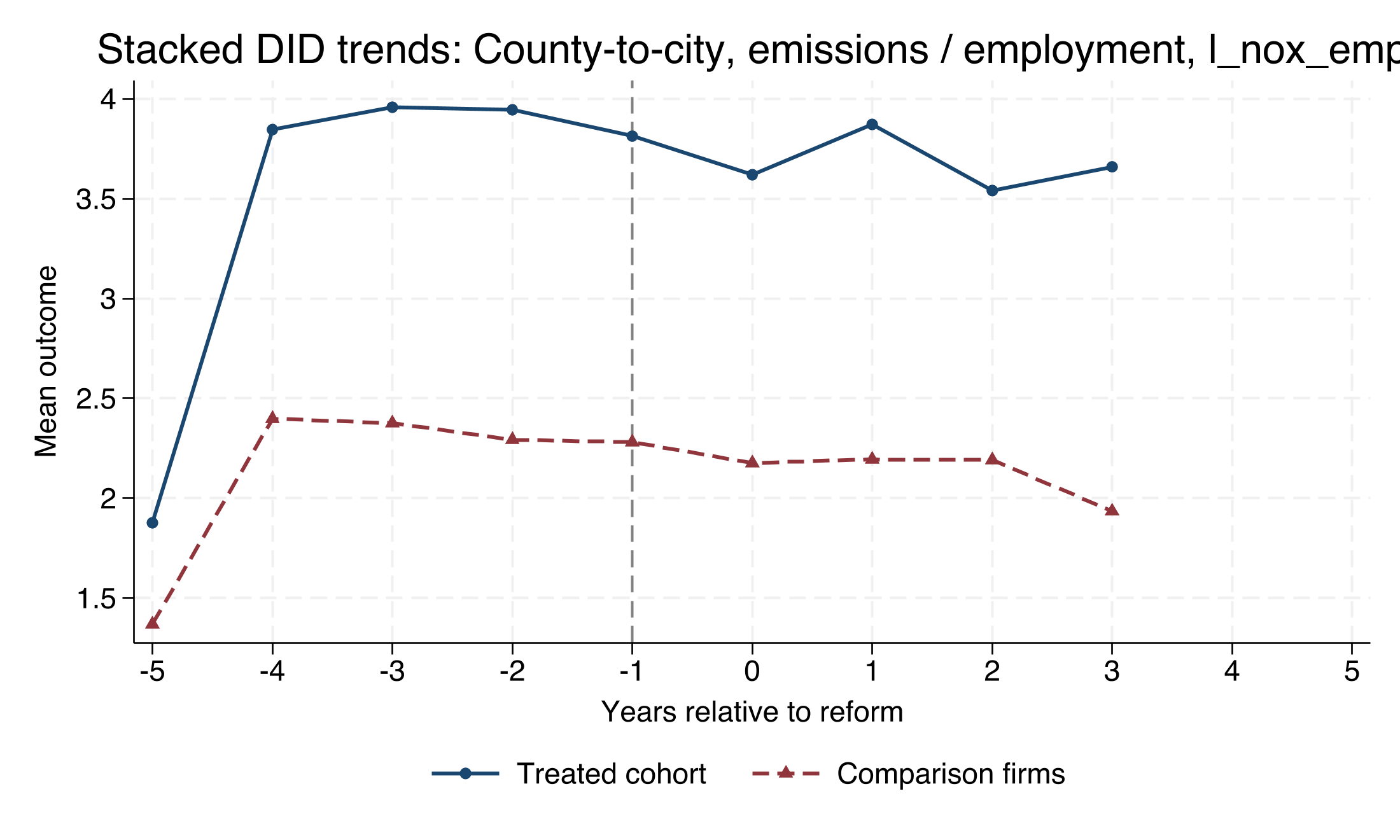}
    \caption{NOx}
  \end{subfigure}\hfill
  \begin{subfigure}[b]{0.32\textwidth}
    \includegraphics[width=\textwidth]{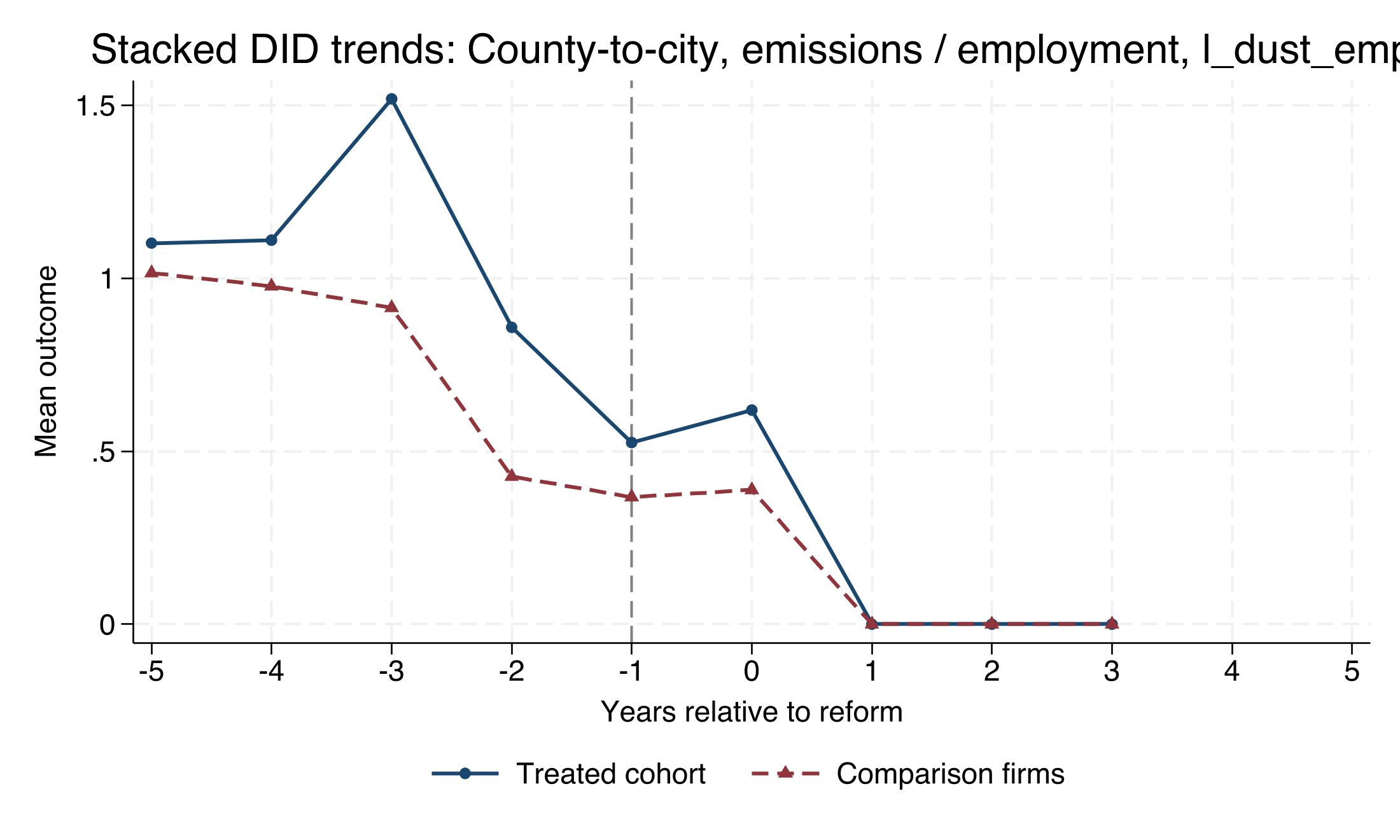}
    \caption{Dust}
  \end{subfigure}\hfill
  \begin{subfigure}[b]{0.32\textwidth}
    \includegraphics[width=\textwidth]{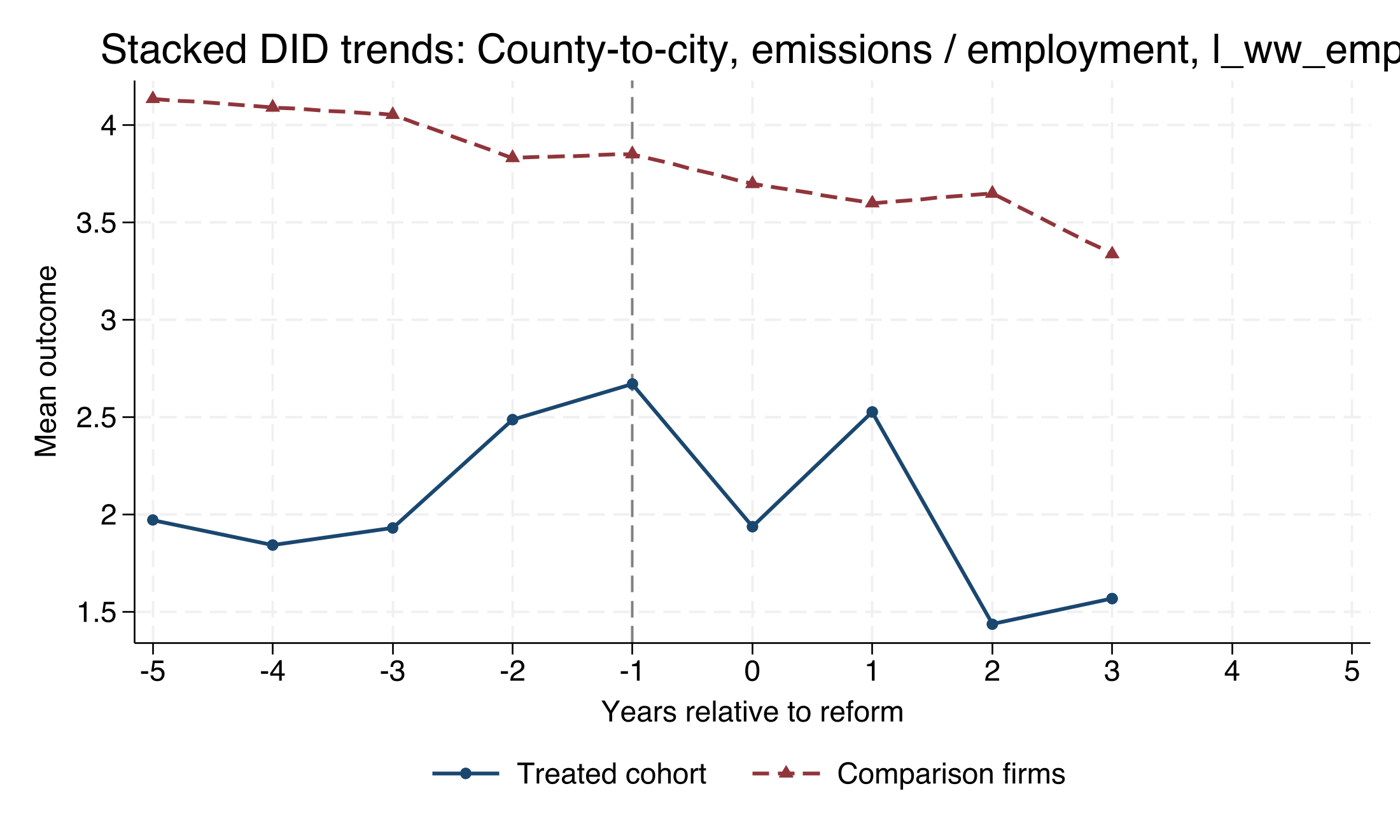}
    \caption{Wastewater}
  \end{subfigure}\hfill
  \begin{subfigure}[b]{0.32\textwidth}
    \includegraphics[width=\textwidth]{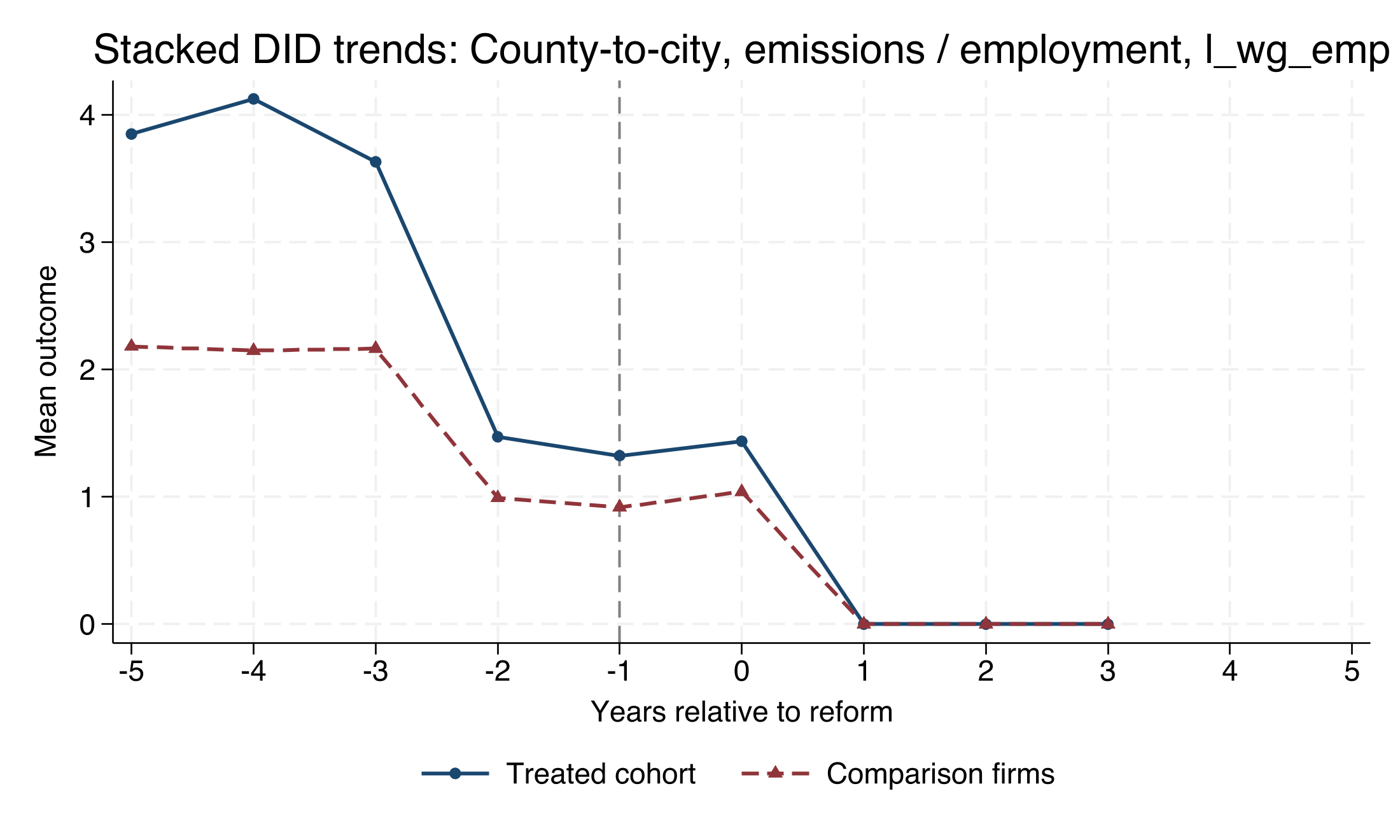}
    \caption{Waste~gas}
  \end{subfigure}\hfill
\end{figure}

\begin{figure}[p]
\caption{County-to-city reform: stacked DID trends, per sales}
\label{fig:countycity_trends_sales}
\centering
  \begin{subfigure}[b]{0.32\textwidth}
    \includegraphics[width=\textwidth]{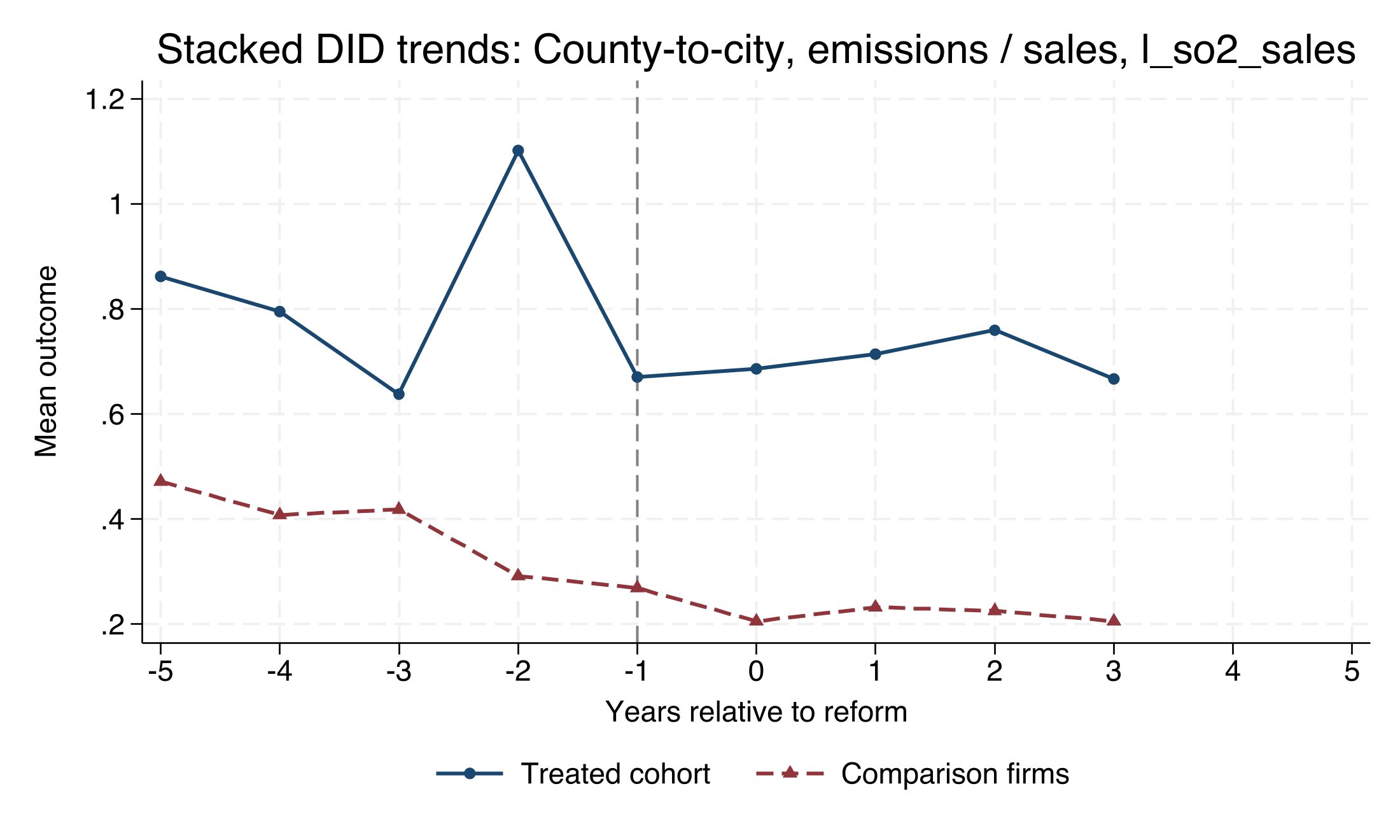}
    \caption{SO$_2$}
  \end{subfigure}\hfill
  \begin{subfigure}[b]{0.32\textwidth}
    \includegraphics[width=\textwidth]{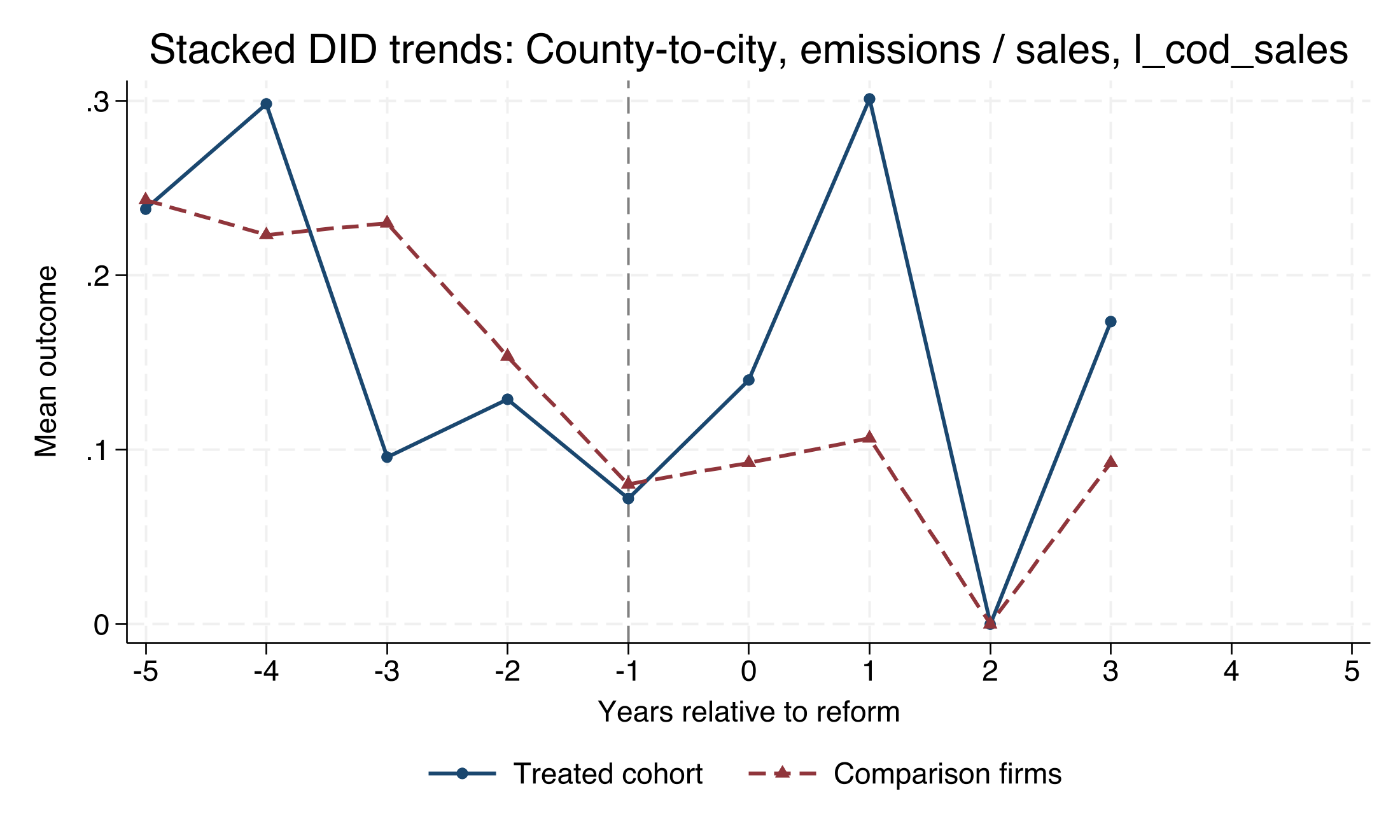}
    \caption{COD}
  \end{subfigure}\hfill
  \begin{subfigure}[b]{0.32\textwidth}
    \includegraphics[width=\textwidth]{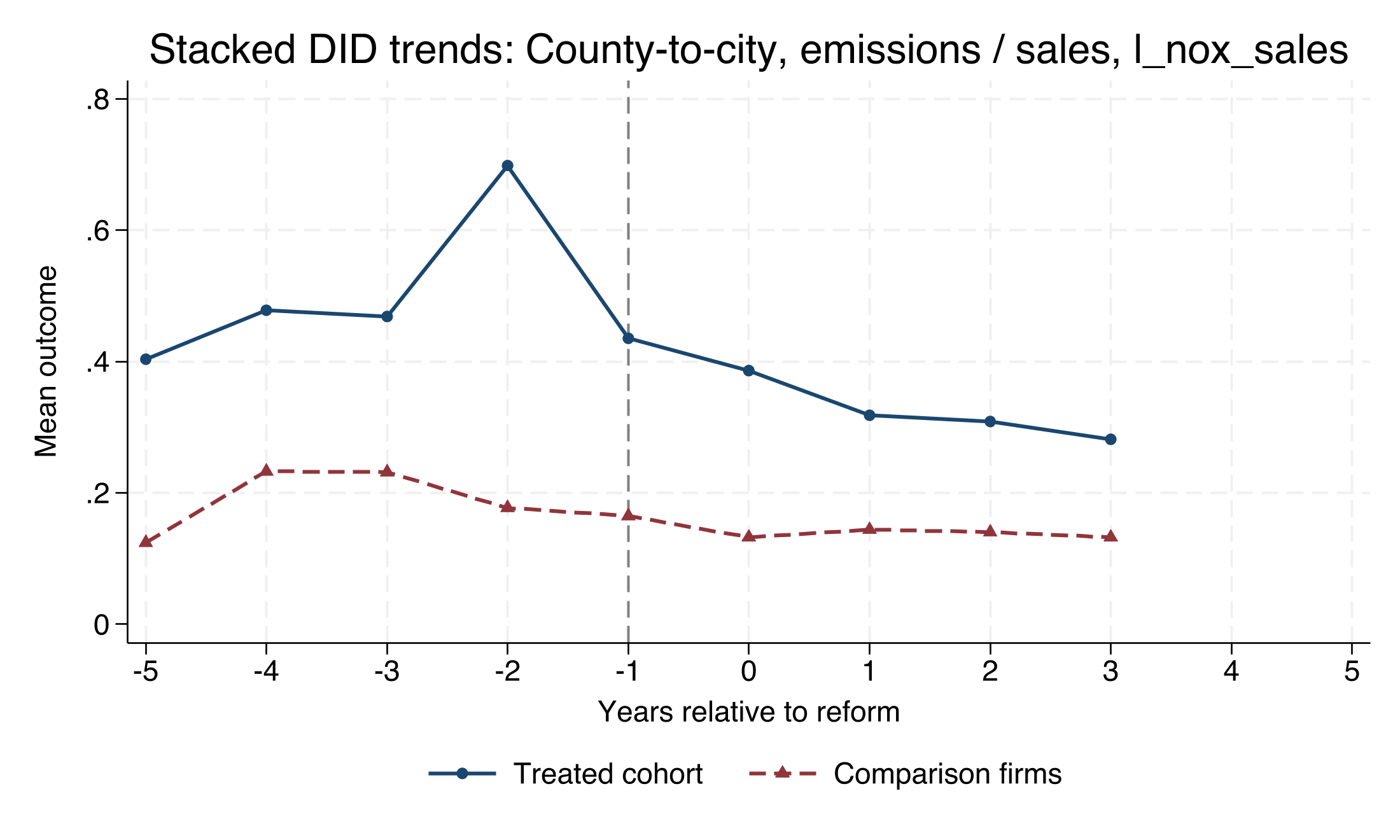}
    \caption{NOx}
  \end{subfigure}\hfill
  \begin{subfigure}[b]{0.32\textwidth}
    \includegraphics[width=\textwidth]{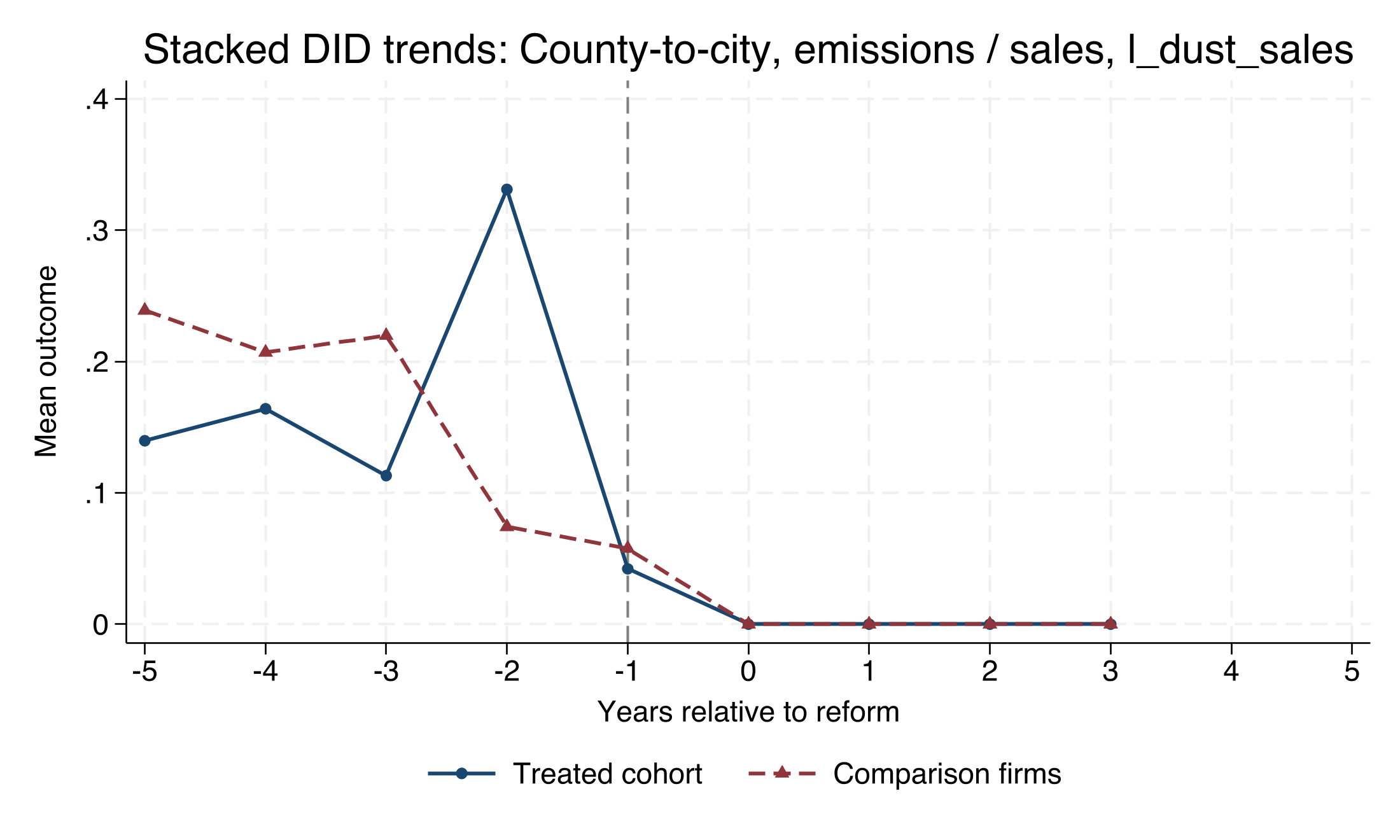}
    \caption{Dust}
  \end{subfigure}\hfill
  \begin{subfigure}[b]{0.32\textwidth}
    \includegraphics[width=\textwidth]{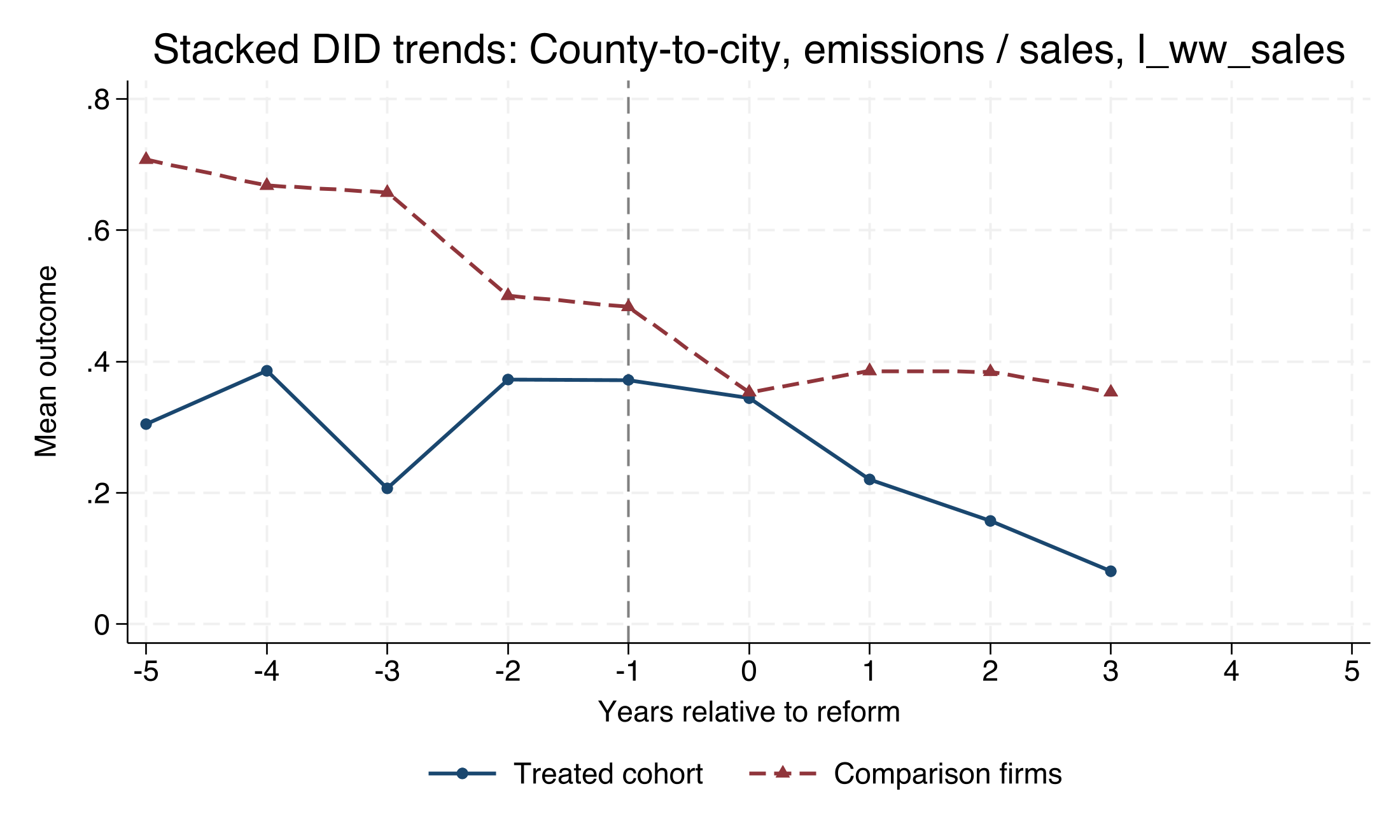}
    \caption{Wastewater}
  \end{subfigure}\hfill
  \begin{subfigure}[b]{0.32\textwidth}
    \includegraphics[width=\textwidth]{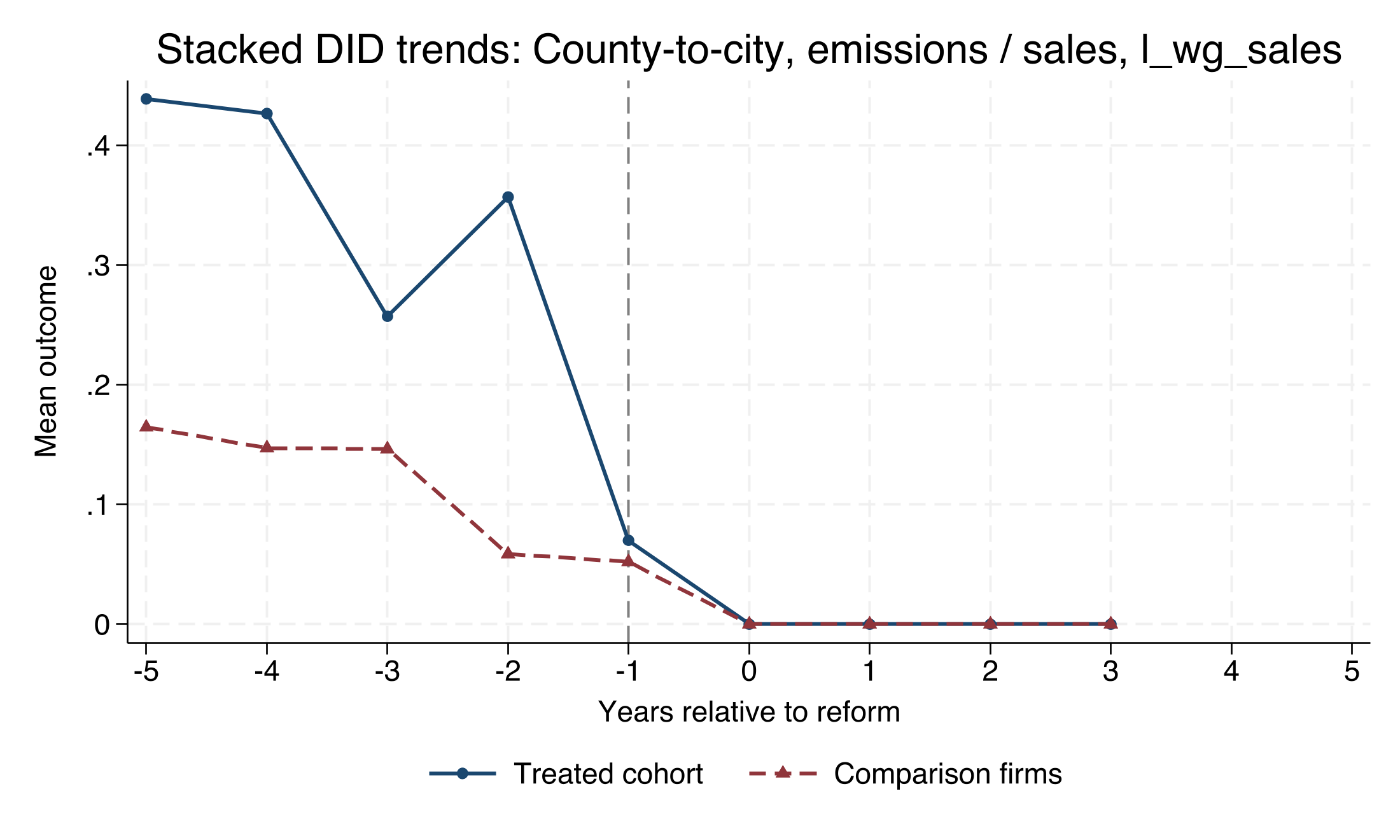}
    \caption{Waste~gas}
  \end{subfigure}\hfill
\end{figure}

\end{document}